\documentclass[12pt]{article}
\usepackage{amsfonts,amssymb,epsfig,amsmath}
\usepackage{color}

\usepackage[vcentermath]{youngtab}
\renewcommand{\baselinestretch}{1.2}

\def\det{{\rm det}}

\newcommand{\be}{\begin{eqnarray}}
\newcommand{\ee}{\end{eqnarray}}

\newcommand{\bn}{\begin{enumerate}}
\newcommand{\en}{\end{enumerate}}

\begin{document}

\makeatletter \@addtoreset{equation}{section} \makeatother
\renewcommand{\theequation}{\thesection.\arabic{equation}}
\renewcommand{\thefootnote}{\alph{footnote}}

\begin{titlepage}

\begin{center}
\hfill {\tt KIAS-P13039}\\
\hfill {\tt SNUTP13-003}\\

\vspace{2cm}

{\Large\bf The general M5-brane superconformal index}

\vspace{2cm}

\renewcommand{\thefootnote}{\alph{footnote}}

{\large Hee-Cheol Kim$^1$, %Joonho Kim$^2$,
Seok Kim$^{2,3}$, Sung-Soo Kim$^1$ and Kimyeong Lee$^1$}

\vspace{1cm}

\textit{$^1$School of Physics, Korea Institute for Advanced Study,
Seoul 130-722, Korea.}

\textit{$^2$Department of Physics and Astronomy \& Center for
Theoretical Physics,\\
Seoul National University, Seoul 151-747, Korea.}

\textit{$^3$Perimeter Institute for Theoretical Physics,
Waterloo, Ontario, Canada N2L 2Y5.}

%\textit{$^3$Center for Quantum Spacetime, Sogang University, Seoul 121-742, Korea.}\\

\vspace{0.7cm}

E-mails: {\tt heecheol1@gmail.com, skim@phya.snu.ac.kr,\\
sungsoo.kim@kias.re.kr, klee@kias.re.kr
}

\end{center}

\vspace{1.5cm}

\begin{abstract}

We calculate and study the general superconformal index for the 6d $U(N)$ $(2,0)$
theory with four chemical potentials, from the indices of gauge theories on
$\mathbb{CP}^2\times\mathbb{R}$. Our index agrees with the large $N$ supergravity
index on $AdS_7\times S^4$ at low energies, and also yields the   negative  `Casimir
energy' with an $N^3$ scaling which was recently calculated from a QFT on $S^5$.
Our approach also   suggests  a systematic study of the $(1,0)$ superconformal indices.

\end{abstract}

\end{titlepage}

\renewcommand{\thefootnote}{\arabic{footnote}}

\setcounter{footnote}{0}

\renewcommand{\baselinestretch}{1}

\tableofcontents

\renewcommand{\baselinestretch}{1.2}

\section{Introduction}

Recently, some progress has been made in calculating and understanding supersymmetric
observables of the M5-brane worldvolume theories, or more generally the 6d $(2,0)$
superconformal field theories \cite{Witten:1995zh,Seiberg:1996vs}. One interesting
observable is the superconformal index, or the BPS partition function on
$S^5\times S^1$ \cite{Kinney:2005ej,Bhattacharya:2008zy}. This
index counts local BPS operators of the 6d theory, or equivalently the BPS states
in the radially quantized CFT.

This index was calculated from the partition function of supersymmetric gauge theories
on $S^5$, obtained by a dimensional reduction of the 6d theory on $S^1$ \cite{Kim:2012av}.
The 5d gauge coupling $g_{YM}^2$ is identified with the temperature-like
chemical potential (the radius $r_1$ of $S^1$), which is a natural M5-brane
version of the relation between the M-theory circle radius and the type IIA dilaton
\cite{Hull:1994ys}. The usage of this (apparently) non-renormalizable 5d SYM to study the
6d $(2,0)$ theory has been considered in \cite{Douglas:2010iu,Bern:2012di}.\footnote{Since
the path integral in our context is supersymmetric and does not suffer from serious
UV divergences, it might be alright to view the 5d SYM just as a low energy effective
description for our purpose.}
Roughly speaking, the 5d observable is supposed to probe interesting 6d
physics if its path integral acquires contribution from `instantonic' configurations. As the instantons' classical action depends on $g_{YM}^2=4\pi^2 r_1$, the 6d physics
depending on $r_1$ is probed by these configurations. The partition
function on $S^5$ acquires contribution from instantonic loops which wrap suitable
circle factors of $S^5$, satisfying this condition. The appearance of $r_1$ this way
makes it possible for the $S^5$ partition function to be a 6d index, with the
chemical potential corresponding to $r_1$.
There are many works over the past couple
of years on this partition function. See \cite{Hosomichi:2012ek} for studies on
the off-shell classical field theory on round $S^5$, \cite{Imamura:2012xg} for the
studies of the theories on squashed $S^5$,
\cite{Kallen:2012cs,Kallen:2012va,Lockhart:2012vp,Imamura:2012bm,Kim:2012qf} for
the studies on perturbative part of the partition function,
\cite{Kim:2012av,Lockhart:2012vp,Kim:2012qf} for the studies on the non-perturbative
corrections,
\cite{Kim:2012av,Kallen:2012zn} for $N^3$ scalings of the index version of 6d Casimir
energies, \cite{Kim:2012qf,Minahan:2013jwa} for BPS Wilson loops on $S^5$ which are
interpreted as Wilson surfaces in 6d.\footnote{
6d $(2,0)$ partition function on other manifolds with $S^1$ factors was also studied
from 5d SYM. An example is the study of 5d gauge theory on Omega-deformed
$\mathbb{R}^4\times S^1$,
as the partition function of the 6d $(2,0)$ theory on $\mathbb{R}^4\times T^2$
\cite{Kim:2011mv}. Other examples on nontrivial curved spaces focus on the usage of
the 6d theory as a tool
to engineer and study interesting QFT's in lower dimensions \cite{Gaiotto:2009we}.
They discuss topologically twisted 6d theories on product spaces, especially on the
relations between QFT's in $n$ dimensions and SCFT's in $6-n$ dimensions. From the 5d
SYM approach, \cite{Kawano:2012up} studied the 2d TQFT partition function on Riemann
surfaces as the 4d superconformal index \cite{Gadde:2011ik}. \cite{Yagi:2013fda}
studied the relation between 3d Chern-Simons partition functions on a 3-manifolds,
and the $S^3$ partition function \cite{Dimofte:2011ju} or the 3d superconformal index \cite{Dimofte:2011py}. The 5d SYM for the circle compactified $(2,0)$ theory was also
studied from various different viewpoints: for instance, from the self-dual string junctions \cite{Lee:2006gqa}, instanton partons \cite{Collie:2009iz}, DLCQ M5-branes \cite{Aharony:1997th,Aharony:1997an,Kim:2011mv,Lambert:2011gb}, deconstruction
\cite{Lambert:2012qy}, and so on.}

The expression for the partition function on $S^5$ was obtained in the weak-coupling
expansion in the inverse-temperature-like chemical potential
$\beta=\frac{g_{YM}^2}{4\pi^2r}\ll 1$, where $g_{YM}$ is the 5d gauge coupling constant
and $r$ is the radius scale of (squashed) $S^5$.
On the other hand, to study the 6d index, one should have
a strong-coupling expansion with the fugacity $e^{-\beta}\ll 1$ at
large $\beta$, since the indices reveal the information on the integer degeneracies after
this expansion. So far, such strong coupling expansions were explicitly done only in
some special cases.\footnote{The studies of \cite{vafa} could enable such an expansion,
which wrote the instanton partition function on $\mathbb{R}^4\times S^1$
as BPS self-dual strings' indices, making its S-duality transformation
properties manifest \cite{Tachikawa:2011ch} .}

In this paper, following a rather different route, we obtain an expression for
the 6d $(2,0)$ index, which takes a manifest index form. We only discuss the index
for the $A_n$ type (more precisely, $U(N)$) theory. Following and expanding the idea
of \cite{Kim:2012tr},
we start by constructing an infinite class of supersymmetric 5d SYM theories on $\mathbb{CP}^2\times\mathbb{R}$ by reducing the 6d theory on
$S^5/\mathbb{Z}_K\times\mathbb{R}$ along the Hopf fiber of $S^5$. The
supersymmetric $\mathbb{Z}_K$ quotient acts on the 6d fields, which makes a $1/K$
fractional identification of the Hopf fiber angle on $S^5$ together with a
certain internal rotation. The latter internal rotation, or `twisting,'
is needed to have some SUSY to survive the $\mathbb{Z}_K$ quotient for general $K$.
Two such theories are constructed in \cite{Kim:2012tr},
which preserve $4$ or $12$ Hermitian supercharges manifestly in 5d classical field theory.
We extend the first QFT of \cite{Kim:2012tr} into an infinite class of QFT's, by considering
all possible twistings with the $SO(5)_R$ internal symmetry which secure the $2$ SUSY
which we want to use to define our 6d superconformal index. The states with nonzero
Kaluza-Klein momentum along the Hopf fiber circle is visible in the 5d theory as Yang-Mills
instantons. The coupling $\tilde{g}_{YM}^2$ of the 5d field theory is given by
$\tilde{g}_{YM}^2=\frac{4\pi^2r}{K}$. So these KK states have masses of order
$\frac{4\pi^2}{\tilde{g}_{YM}^2}=\frac{K}{r}$, and are heavy
for $K\gg 1$ compared to the mass gap $\frac{1}{r}$ of this theory. The 5d SYM's obtained
in \cite{Kim:2012tr} and in this paper are thus low energy effective action in the
range $  E\ll\frac{K}{r}$.

The most interesting case with $K=1$, without any orbifold on the 6d theory,
is strongly interacting at the length scale of $\mathbb{CP}^2$ radius $r$. Still,
inclusion of the non-perturbative instantons could allow us to obtain the 6d BPS
spectrum appearing in the index, as the possible issues of UV divergences are
much milder in BPS sectors. We calculate the index of this strongly interacting theory
using supersymmetric localization, which again acquires contribution from Yang-Mills
instanton states on $\mathbb{CP}^2$. We interpret it as the 6d superconformal index
\cite{Bhattacharya:2008zy}. Unlike the $S^5$ partition function, the BPS partition
function on $\mathbb{CP}^2\times S^1$ is manifestly an index due to the explicit presence
of the Euclidean time circle in 5d. So for the purpose of studying the 6d index,
our approach from $\mathbb{CP}^2\times S^1$ could be more efficient. However, there seem
to be different benefits from the two approaches, one from SYM on $S^5$ and another from
SYM on $\mathbb{CP}^2\times S^1$. We comment on it in section 5.

For the detailed study of the non-Abelian index, we mostly focus on a 5d theory with
a particular twisting. The final result is given by a contour integral, with the
integration measure acquiring
contributions from various instantons on $\mathbb{CP}^2$. The latter is essentially
factorized into $3$ contributions from self-dual instantons localized on fixed points of
$\mathbb{CP}^2$ under $U(1)^2\subset SU(3)$ isometry. Each of the $3$
contributions is identified as the instanton partition function on Omega deformed
$\mathbb{R}^4\times S^1$ \cite{Nekrasov:2002qd,Nekrasov:2003rj}.

Our index agrees with the results in special cases known from different approaches.
Firstly, we reproduce the index for the
Abelian 6d theory. This can be directly computed from the free 6d theory \cite{Bhattacharya:2008zy}, and was also obtained using the $S^5$ partition
function \cite{Lockhart:2012vp,Kim:2012qf}. Secondly, we find that the correct
unrefined index (turning off all but one chemical potentials) is obtained,
which was first obtained in \cite{Kim:2012av,Kim:2012qf} from the $S^5$ partition
function. Finally, we find that our general index, with all $4$ chemical potentials
turned on, agrees with the gravity dual index on $AdS_7\times S^4$ at
low enough energies. We show this by suitably expanding the index up to some order
in fugacities. We first show that our finite $N$ indices agree
with gravity up to `energies' of order $N$ (more precisely, for $k\leq N$ where $k$ is
the instanton number), for $N\leq 3$. Of course, the index beyond the supergravity regime
at $k>N$ is a prediction of our study, which we partially explore as well. We also
show that the strict large $N$ index agrees with the gravity dual for $k\leq 2$. We hope
to go to higher orders in both tests in the near future.

From the QFT's on $\mathbb{CP}^2\times\mathbb{R}$,
one could hopefully study 6d $(1,0)$ superconformal index more efficiently.
In particular, having a manifest index form should make a systematic analysis
of the 6d BPS spectrum possible. In section 5.3, we briefly comment on the indices
for the 6d $(1,0)$ SCFT's from our approach, and possible new issues there.

The partition function of $\mathbb{CP}^2\times\mathbb{R}$ QFT
reproduces the `Casimir energy' which exhibits an $N^3$ scaling at large $N$,
which was first obtained from the QFT on $S^5$ \cite{Kim:2012av,Kallen:2012zn}.
We make comments on this quantity in section 5.1, also emphasizing some subtleties.

Finally, as we are suggesting that $S^5$ and $\mathbb{CP}^2\times S^1$ partition
functions are the same 6d index. A closely related question is the S-duality of the
5d $\mathcal{N}=1^\ast$ SYM on $\mathbb{R}^4\times S^1$ \cite{Tachikawa:2011ch}, as the
important building block of both partition functions on $S^5$ and $\mathbb{CP}^2\times S^1$
is the partition function on $\mathbb{R}^4\times S^1$ in Omega background.
We comment on some aspects of this duality in section 5.2, having in mind an important
progress which was recently made in \cite{vafa}.

The remaining part of this paper is organized as follows. In section 2, after
explaining the 5d QFT's on $\mathbb{CP}^2\times\mathbb{R}$, we use supersymmetric
localization to calculate the index. In section 3, we use it
to study the 6d $(2,0)$ theory. Section 4 explains some aspects of the 5d QFT at general
orbifold with $K>1$, emphasizing the strong coupling nature of the vacuum energy with
$N^3$ scaling. Section 5 is devoted to discussions
on various issues. Appendix A derives the S-duality of the Abelian instanton partition
function on Omega-deformed $\mathbb{R}^4\times S^1$.
Appendix B explains the details of the spinors that we use in this paper.
Appendix C derives the vector multiplet part of our off-shell action from supergravity
methods \cite{Festuccia:2011ws}.

\section{The 5d QFT and the index}

We would like to study the index partition function of 6d $(2,0)$ theory on
$S^5\times S^1$, where the time direction $\mathbb{R}$ is compactified to $S^1$ by
a temperature-like chemical potential. When the manifold contains a factor of circle,
preferably
with a small radius compared to the other length scales in the QFT, one can
dimensionally reduce the 6d theory to 5d SYM theories. The reduction would naively
seem to be forgetting the information on the 6d physics by losing the Kaluza-Klein
modes along the circle. The hope of this approach is
that non-perturbative effects in the 5d theory would let one to restore the naively
lost 6d information, similar to the way that non-perturbative sector of the type IIA
string theory contains the information on the 11 dimensional physics of M-theory.
Despite the recent studies \cite{Douglas:2010iu}, the precise logical background of
this approach does not seem to be very solid, as the 5d theory appears to be non-renormalizable. See also \cite{Bern:2012di} for some related discussions.
We shall be assuming that at least the BPS partition functions
can be calculated from the 5d SYM theory.\footnote{At least for calculations,
supersymmetric path integrals enjoy the property of localization, which means that
such integrals are secretly Gaussian and almost free of UV divergence issues.
Still, one should understand what are the possible higher derivative terms
preserving the required symmetries, especially the maximal $16$ SUSY in the flat
space limit. If all these terms are $Q$-exact, not affecting our BPS partition
function, our studies could be independent of the proposal of \cite{Douglas:2010iu}.
If not, our works should somehow rely on \cite{Douglas:2010iu}.
We do not know the answer at the moment.}

For the convenience of engineering the 5d SYM theory via dimensional reduction,
we generalize the 6d theory by considering the theory on $S^5/\mathbb{Z}_K\times\mathbb{R}$ \cite{Kim:2012tr} (which we shall eventually remove by setting $K=1$).
Among others, the $\mathbb{Z}_K$ orbifold acts on the Hopf fiber of $S^5$, and shifts its
angle by $\frac{2\pi}{K}$. To make a supersymmetric orbifold at generic $K$, one has to
twist this shift with suitable internal rotations.
The twists are explained below shortly. The idea is to first obtain a 5d
effective theory at $K\gg 1$ when the KK energy scale $\frac{K}{r}$ along the Hopf fiber
is much larger than the mass gap $\frac{1}{r}$ of the 6d theory on $S^5\times\mathbb{R}$.
The effective theory on $\mathbb{CP}^2\times\mathbb{R}$ is valid for the energy
$  E\ll\frac{K}{r}$. (The spectrum for  $Er$ is expected to be    discrete.)
At $K\sim 1$, especially at $K=1$ that we are mostly interested in, there is no regime
in which one can regard the 5d QFT as an effective theory. Speaking differently, the
Kaluza-Klein states with energies of order $\frac{K}{r}$ will not be heavier than the
5d degrees. Still, if one can completely calculate the spectrum of
these Kaluza-Klein states which appear as 5d Yang-Mills instantons, one could
expect the full 6d spectrum to be calculable from the 5d theory even at $K=1$. We
assume so, and indeed confirm that we obtain the correct 6d index in
various special cases that we check.

This $\mathbb{Z}_K$ orbifold should be acting on the 6d `fields.' Since we do not know
the details of the 6d fields in the non-Abelian theory, we first consider this orbifold
for the Abelian 6d theory. We obtain the Abelian 5d action by dimensional reduction
at $K\gg 1$. The information obtained from this Abelian reduction provides us
with various constraints on the non-Abelian theory, from which one can construct the
non-Abelian SYM theory on $\mathbb{CP}^2\times\mathbb{R}$. The constraints that we
obtain are the Killing spinor equations on $\mathbb{CP}^2\times\mathbb{R}$, and also
the quadratic part of the non-Abelian SYM action which should be proportional to
the Abelian 5d action. With these constraints, supersymmetry determines
other non-Abelian curvature couplings.

Two kinds of $\mathbb{Z}_K$ orbifolds are considered in \cite{Kim:2012tr},
with the explicit construction of the 5d classical actions and SUSY transformations.
In section 2.1, we construct an infinite class of such QFT's, generalizing one theory
in \cite{Kim:2012tr}, and make the $SU(1|1)$ part of the symmetry off-shell.
We shall then study their indices, including the contribution from instantons. In
most parts of this paper, our goal is to study the 5d theory at $K=1$, or the 6d theory
on $S^5\times\mathbb{R}$ without any orbifold. (The only exception is section 4,
discussing the index at $K>1$.)

\subsection{Supersymmetric reductions to $\mathbb{CP}^2\times\mathbb{R}$}

The symmetry of the 6d $(2,0)$ theory is $OSp(8^\ast|4)$, containing
$SO(6,2)\times SO(5)_R$ as its bosonic subgroup.
Let us denote by $j_1,j_2,j_3$ the three rotation generators of $SO(6)$ acting on
$S^5$, in the convention that they rotate the three orthogonal 2-planes of $\mathbb{R}^6$.
Also, we take $R_1,R_2$ to be the two Cartans of $SO(5)$ R-symmetry, again in the
convention that they rotate 2-planes of $\mathbb{R}^5$. The five real scalars
of the 6d theory, or the reduced 5d SYM, are decomposed as follows.
$\phi$ denotes the real scalar neutral in both $R_1,R_2$, and belongs to the
5d $\mathcal{N}=1$ vector multiplet. The remaining two complex scalars, which
we call $q_1,q_2$, belong to the 5d hypermultiplet. The charges carried by
them are $(R_1,R_2)=(1,0)$ for $q_1$, and $(0,-1)$ for $q_2$. They are in a
doublet of $SU(2)_R$ with the Cartan $\frac{R_1+R_2}{2}$, which
shall be broken to the Cartan $U(1)_R$ in our curved background. The remaining
$\frac{R_1-R_2}{2}$ is called the flavor symmetry $U(1)_F$, from the viewpoint of
5d $\mathcal{N}=1$ SUSY. The preserved R-symmetry subgroup of $SO(5)$
in curved background will be minimally $U(1)^2$ (coming from $\frac{R_1\pm R_2}{2}$),
and could come with some enhancement depending on the 5d theory, which we shall
explain below.

The $\mathbb{Z}_K$ action is obtained by a discrete rotation with respect
to the following charge,
\begin{equation}\label{twisting}
  Kk\equiv j_1+j_2+j_3+\frac{3}{2}(R_1+R_2)+n(R_1-R_2)\ .
\end{equation}
The coefficient $\frac{3}{2}$ in front of $R_1+R_2$ is tuned to have $k$ to commute
with at least the following two supercharges that we want to secure in 5d:
\begin{equation}
  Q^{R_1,R_2}_{j_1,j_2,j_3}=
  Q^{+\frac{1}{2},+\frac{1}{2}}_{-\frac{1}{2},-\frac{1}{2},-\frac{1}{2}}
  \equiv Q\ ,\ \ S\equiv Q^\dag=
  Q^{-\frac{1}{2},-\frac{1}{2}}_{+\frac{1}{2},+\frac{1}{2},+\frac{1}{2}}\ .
\end{equation}
With the normalization on the left hand side of (\ref{twisting}),
$k$ has integer eigenvalues for all $OSp(8^\ast|4)$ generators and
(Abelian) 6d fields only if $n$ is half of an odd integer, which we assume.
We define a supersymmetric 6d orbifold QFT obtained by truncating
the 6d $(2,0)$ fields on $S^5\times\mathbb{R}$ to the $\mathbb{Z}_K$
invariant sector. With our current knowledge on the $(2,0)$ theory, this definition
can be made concrete only for the 6d Abelian theory. The theory after orbifold is
still a 6 dimensional theory, by having an infinite tower of KK states. The states
that survive the orbifold have integral eigenvalues of $k$.

When $K\gg 1$, the states with nonzero $k$ charge would be very heavy, whose
energies scale like $K$. It is heuristic to assume that the internal $R_1,R_2$
charges carried by the `6d fields' are of order $1$ (for the Abelian theory,
$\pm\frac{1}{2}$ for fermions, and $0$ or $\pm 1$ for bosons). On the other hand,
the angular momenta $j_1,j_2,j_3$
for the fields can be arbitrarily large. So at low energy of order $\mathcal{O}(K^0)$,
the effective description would be given by a 5d SYM theory on
$\mathbb{CP}^2\times\mathbb{R}$ obtained by a Kaluza-Klein reduction,
keeping the $j_1+j_2+j_3=\mathcal{O}(1)$ sector with $k=0$.

The sectors with nonzero $k$ charge can be seen in the non-perturbative sector
of the 5d theory, as the momentum along the circle generated by $j_1+j_2+j_3$
is identified with the Yang-Mills instanton charge. More precisely, in our setting,
the instanton charge $k$ is the momentum plus extra internal charge, as written above.
In particular, at $K=1$, we expect such instantons to recover all the modes on
$S^5$ appearing in the index.

The number of supercharges visible in 5d classical field theory depends
on $n$. Firstly, when $n=\pm\frac{1}{2}$, $8$ Hermitian supercharges are
visible at $k=0$. This can be seen by noticing that
$Kk=j_1+j_2+j_3+R_1+2R_2$ at $n=-\frac{1}{2}$. The supercharges of the 6d theory with
$k=0$ are
\begin{equation}\label{SUSY-enhanced}
  Q^{R_1,R_2}_{j_1,j_2,j_3}
  \ :\ Q^{++}_{---}\ ,\ \ Q^{+-}_{-++}\ ,\ \ Q^{+-}_{+-+}\ ,\ \ Q^{+-}_{++-}
  \ \ \ \ \big(\pm\ {\rm denoting}\ \pm\frac{1}{2}\,\big)
\end{equation}
and their conjugate $S$'s with all charge signs flipped.  The supergroup containing
the $8$ SUSY is $SU(1|1)\times SU(3|1)$. $SU(1|1)$ contains $\frac{R_1+R_2}{2}$ and
$Q\equiv Q^{++}_{---}$, $S\equiv Q^{--}_{+++}$. $SU(3|1)$ contains $SU(3)$ from
the isometry on $\mathbb{CP}^2$, $U(1)$ from $\frac{R_1-R_2}{2}$, and the latter
three $Q$'s in (\ref{SUSY-enhanced}) as well as their conjugate $S$'s.
Similar consideration can be made at $n=\frac{1}{2}$, with the roles of $R_1$, $R_2$
exchanged. At $n=-\frac{3}{2}$,
with $Kk=j_1+j_2+j_3+3R_2$, there are $4$ Hermitian supercharges with $k=0$, which are
\begin{equation}
  Q^{++}_{---}\ ,\quad Q^{-+}_{---}\ ,
\end{equation}
and their conjugate $S$'s. This QFT was first found in \cite{Kim:2012tr}. The superalgebra
containing these $4$ SUSY is $SU(1|2)$, where $SU(2)$ is a subgroup of $SO(5)_R$ and
contains $R_1$ as its Cartan. Similar $SU(1|2)$ supergroup can be found at $n=\frac{3}{2}$, again with the roles of $R_1$, $R_2$ exchanged.
At other values of $n$, one only finds $2$ Hermitian supercharges $Q^{++}_{---}$,
$S^{--}_{+++}$ at $k=0$. The superalgebra is $SU(1|1)$. This $SU(1|1)$ can be
embedded into all the enhanced supergroups discussed above.\footnote{In \cite{Kim:2012tr},
another SUSY QFT on $\mathbb{CP}^2\times\mathbb{R}$ was constructed with $SU(3|2)$ symmetry,
with supercharges $Q^{R_1-}_{-++},Q^{R_1-}_{+-+},Q^{R_1-}_{++-}$ and conjugates
for $R_1=\pm\frac{1}{2}$. In this theory, the KK momentum is $Kk=j_1+j_2+j_3+R_2$.}

These are supercharges visible in the 5d classical field theory.
At $K=1$, with no orbifolds, we expect that all $32$ supercharges of
the 6d $(2,0)$ theory exist as conserved currents, some with nonzero instanton
charges $k\neq 0$. This should be similar to the SUSY enhancement of the ABJM theory \cite{Aharony:2008ug} at the Chern-Simons level $K=1,2$, where the extra supercurrents
carry nonzero magnetic monopole charges. We designed our infinite class of 5d QFT's to
always have $Q^{++}_{---}$ and $Q^{--}_{+++}$ explicitly visible at $k=0$. These are
the pair $(Q,S)$ of Poincar\'e
and conformal supercharges that we shall use to define the 6d superconformal index.

The theories with $n=\pm\frac{1}{2}$ would be of more interest to us,
due to the presence of many SUSY visible in the classical 5d action. In particular,
for the theory with $n=-\frac{1}{2}$ which we shall focus on later, we list the
expected number of supercharges here at various values of $K$. This can be obtained
by investigating the value of $Kk=j_1+j_2+j_3+R_1+2R_2$ carried by the
$32$ supercharges.
The list is given as follows. One finds $8$ supercharges (as explained above)
at $Kk=0$; $14$ supercharges at $Kk=\pm 1$; $8$ supercharges at $Kk=\pm 2$;
$2$ supercharges at $Kk=\pm 3$. For the 6d theory on $S^5/\mathbb{Z}_K\times\mathbb{R}$,
we expect that the momentum $Kk$ is an integer multiple of $K$.
So at $K\geq 4$, one only finds the above $8$ supercharges at $Kk=0$.
At $K=3$, two supercharges at $Kk=3$ can be made gauge invariant by using
instanton operators with unit topological charge. So one expects $8+2=10$ SUSY.
At $K=2$, one expects $8+8=16$ SUSY, by including $8$ SUSY at $Kk=\pm 2$ using
unit instanton operator. At $K=1$, of course we expect maximal $32$ SUSY
by using instanton operators of topological charges $\pm 1, \pm 2, \pm 3$.
It should be interesting to confirm this by a 5d QFT calculation, similar to
the studies on the ABJM theory \cite{Kim:2009wb}. Although we do not attempt all
the studies done in \cite{Kim:2009wb} for our QFT, the fact that our index at $K=1$
agrees with large $N$ supergravity index on $AdS_7\times S^4$ (as we explain in section 3)
is a very strong signal of the enhancement of SUSY, as the spectrum of gravitons are
organized into supermultiplets of the maximal superconformal symmetry. This is
comparable to the checks made in the first reference of \cite{Kim:2009wb}, which finds
agreement of the ABJM index at $K=1$ with the supergravity index on $AdS_4\times S^7$.

Now let us actually construct the 5d QFT with the above intuitions.
The construction can be done in various ways, which have different advantages
and also provide different viewpoints. The on-shell Abelian theory can be very
easily constructed by using the Abelian 6d $(2,0)$ action (or equation of motion if
one does not like formalisms breaking covariance) in a curved background. As one can
explicitly write down the Abelian 6d tensor
theory on $S^5/\mathbb{Z}_K\times\mathbb{R}$, one reduces this to
$\mathbb{CP}^2\times\mathbb{R}$ and performs a tensor-vector dualization.
This was explicitly done in \cite{Pasti:1997gx,Aganagic:1997zk}, if we restrict their
`Born-Infeld-like' action to the quadratic one for small values of fields. Various terms
in the action can be
understood as coupling to various `type IIA background fields' during the reduction.
In particular, since the reduced circle is nontrivially fibered over $\mathbb{CP}^2$,
there appears `RR 1-form' potential $A^{RR}=\theta$, or $dA^{RR}=d\theta=2J$ where $J$
is the K\"ahler 2-form of $\mathbb{CP}^2$. So the resulting action consists of the Maxwell
term, the Wess-Zumino coupling of the form $A^{RR}\wedge F\wedge F\sim J\wedge A\wedge F$,
and all their superpartners.

The above construction seems to apply only to the Abelian theory. Perhaps a similar
logic can be developed using the non-Abelian Born-Infeld action for D4-branes, but we
will construct the non-Abelian extension by a brute-force labor with the constraints from
the Abelian theories. In particular, to the action we explain below, there could be
more terms that can be added with more parameters compatible with the symmetry. One
example is the 5d Chern-Simons term of the form $AFF$, which is killed
in our case because it does not appear in the Abelian action. More examples can
be found in appendix \ref{app:OffshellQFT}, where the off-shell supergravity analysis reveals more parameters
which we will freeze by comparing with the Abelian theory.

The classical on-shell QFT's labeled by $n$, constructed this way, are given
as follows. The 5d fields are those of 5d maximal SYM: gauge field $A_\mu$,
$5$ scalars $\phi^I$ with $I=1,2,3,4,5$, and an $SO(5)_R$ symplectic-Majorana
fermion $\lambda^i_\alpha$ with $i=1,2,3,4$ for $SO(5)_R$, all in the $U(N)$
adjoint. See appendix B for our conventions on the spinors. The (Euclidean) action
is given by
\begin{eqnarray}\label{onshell-action}
  \hspace*{-.5cm}S&=&\frac{1}{\tilde g_{YM}^2}\int d^5x\sqrt{g}\ {\rm tr}\left[\frac{1}{4}F_{\mu\nu}F^{\mu\nu}
  +\frac{1}{2}D_\mu\phi^I D^\mu\phi^I-\frac{i}{2}\lambda^\dag\gamma^\mu D_\mu\lambda
  -\frac{1}{4}[\phi^I,\phi^I]^2-\frac{i}{2}\lambda^\dag\hat\gamma^I[\lambda,\phi^I]\right.
  \nonumber\\
  \hspace*{-.5cm}&&\hspace{0cm}+\frac{2}{r^2}(\phi_I)^2-\frac{1}{2r^2}(M_n\phi^I)^2
  +\frac{1}{8r}\lambda^\dag J_{\mu\nu}\gamma^{\mu\nu}\lambda
  -\frac{i}{2r}\lambda^\dag M_n\lambda-\frac{i}{r}\left(3-2n\right)[\phi^1,\phi^2]\phi^3
  -\frac{i}{r}(3+2n)[\phi^4,\phi^5]\phi^3\nonumber\\
  \hspace*{-.5cm}&&\left.-\frac{i}{2r\sqrt{g}}\epsilon^{\mu\nu\lambda\rho\sigma}\left(
  A_\mu\partial_\nu A_\lambda-\frac{2i}{3}A_\mu A_\nu A_\lambda\right)J_{\rho\sigma}\right]\ ,
\end{eqnarray}
where $M_n\equiv\frac{3}{2}(R_1+R_2)+n(R_1-R_2)$, $I=1,2,3,4,5$ and $\tilde g_{YM}^2=  4\pi^2 r/K$.
$\phi^{1,2}$ rotate under $R_1$ while being neutral in $R_2$. $\phi^{4,5}$ rotate
under $R_2$ while being neutral in $R_1$. $\phi\equiv\phi^3$ is the 5d $\mathcal{N}=1$
vector multiplet scalar. $\hat\gamma^I$ is the gamma matrix of
the $SO(5)$ R-symmetry, which is broken to either $U(1)^2$ (for $n\neq\pm\frac{3}{2}$) or
$SU(2)\times U(1)$ (at $n=\pm\frac{3}{2}$) by the curvature coupling.
$J_{\mu\nu}$ and $\theta_\mu$ are taken to be dimensionless for coordinates which
carry dimension of length, so that $J$ becomes the canonical K\"ahler 2-form of
$\mathbb{C}^2$ in the $r\rightarrow\infty$ limit. The covariant derivative is given
by $D_\mu=\nabla_\mu-i[A_\mu,\ ]+\frac{i}{r}\theta_\mu M_n$,
which acts on fields as
\begin{eqnarray}
  &&D_\mu\phi^a=\partial_\mu\phi^a-i[A_\mu,\phi^a]-\frac{1}{r}\left(\frac{3}{2}+n\right)
  \theta_\mu\epsilon_{ab}\phi^b\ ,\ \ D_\mu\phi^i=\partial_\mu\phi^i-i[A_\mu,\phi^i]
  -\frac{1}{r}\left(\frac{3}{2}-n\right)\theta\epsilon_{ij}\phi^j\nonumber\\
  &&D_\mu\lambda=\nabla_\mu\lambda-i[A_\mu,\lambda]-\frac{1}{2r}\theta_\mu
  \left[\left(\frac{3}{2}+n\right)\hat\gamma^{12}+\left(\frac{3}{2}-n\right)
  \hat\gamma^{45}\right]\lambda
\end{eqnarray}
with $a,b=1,2$, $i,j=4,5$, and $d\theta=2J$.

The SUSY transformation is given as follows. Firstly,
the Killing spinor equation on $\mathbb{CP}^2\times\mathbb{R}$ is derived by
starting from the canonical one on $S^5\times\mathbb{R}$ and reducing the spinor to
5d, demanding $k=0$ or $\partial_y+iM_n=0$ on spinors (where $y$ is the coordinate
of the Hopf fiber with $2\pi$ periodicity). This is one of our inputs which
constrain the 5d theory. One obtains
\begin{eqnarray}
  D_m\epsilon_\pm&\equiv&\Big(\nabla_m+\frac{i}{r}\ \theta_m M_n\Big)\epsilon_\pm=\frac{i}{2r}J_{mn}
  \gamma_n\epsilon_\pm\pm\frac{1}{2r}\gamma_m\gamma_\tau\epsilon_\pm\nonumber\\
  \nabla_\tau\epsilon_\pm&=&\pm\frac{1}{2r}\epsilon_\pm\ \ ({\rm definition\ of}\
  \epsilon_\pm)\nonumber\\
  M_n\epsilon_\pm&=&\frac{i}{4}J_{mn}\gamma^{mn}\epsilon_\pm\mp\frac{1}{2}\gamma_\tau
  \epsilon_\pm\ ,
\end{eqnarray}
where $m=1,2,3,4$ is the coordinate index on $\mathbb{CP}^2$. See appendix B for the
details. The last equation
comes from the $y$ component of the 6d Killing spinor equation, and is imposing
the requirement $k=0$ for the SUSY to survive $\mathbb{Z}_K$. The number of solutions
to these equations is $8$ for $n=\pm\frac{1}{2}$, $4$ for $n=\pm\frac{3}{2}$,
and $2$ otherwise. With these Killing spinors,
the on-shell SUSY transformation rule is given by
\begin{eqnarray}
  \delta\phi^I&=&\epsilon^\dag\hat\gamma^I\lambda\ \ (=-\lambda^\dag\hat\gamma^I\epsilon)\\
  \delta A_\mu&=&i\epsilon^\dag\gamma_\mu\lambda\ \
  (=-i\lambda^\dag\gamma_\mu\epsilon)\nonumber\\
  \delta\lambda&=&\frac{1}{2}F_{\mu\nu}\gamma^{\mu\nu}\epsilon-iD_\mu\phi^I\gamma^\mu
  \hat\gamma^I\epsilon-\frac{i}{2}[\phi^I,\phi^J]\hat\gamma^{IJ}\epsilon
  +\frac{i}{r}(M_n\phi^I)\hat\gamma^I\epsilon-\frac{2i}{r}\phi^I\hat\gamma^I\eta\nonumber\ ,
\end{eqnarray}
where $\eta_\pm\equiv\pm\gamma_\tau\epsilon_\pm$, with $\epsilon=\epsilon_++\epsilon_-$
and $\eta=\eta_++\eta_-$.

The above action is consistent with the Abelian action obtained by a reduction
of 6d theory on $S^5\times\mathbb{R}$. One might worry whether there could be
extra sectors of 5d fields, from the fact that
there are $\mathbb{Z}_K$ valued discrete 1-form holonomies on $S^5/\mathbb{Z}_K$
that cannot be gauged away. This could be reducing $B_{\mu 5}=A_\mu\omega_5$ with
nontrivial holonomy $\omega_5$. We have not carefully thought about this issue at
$K\geq 2$. But since our main interest is the theory at $K=1$, we think the 5d theory
we constructed in this subsection will suffice. Even at $K>1$, in section 4, we
explain that our QFT yields the correct 6d index on $S^5/\mathbb{Z}_K\times S^1$
in the Abelian case, cross-checked with the index for the 6d free $(2,0)$ theory.
Also, since the 6d tensor field is constrained to be self-dual, the 5d vector
field can be obtained either from $B_{\mu 5}$ or by dualizing $B_{\mu\nu}$. Since
the latter does not seem to be sensitive to the discrete holonomies, it might
be that our 5d QFT is enough at general $K$.

We already explained the superalgebra of the 5d theory that we expect from
the kinematics of the 6d theory with $\mathbb{Z}_K$ orbifolds. We have not carefully
checked these on-shell algebra purely from the 5d QFT. However, an off-shell
$SU(1|1)$ subalgebra is checked and shown below.

To calculate the supersymmetric partition function, we need a QFT which makes
the $SU(1|1)$ part of the algebra off-shell. It is useful to decompose the above maximal
SYM multiplet into a 5d $\mathcal{N}=1$ vector multiplet and an adjoint hypermultiplet.
We rename the scalar and fermion fields as
\begin{equation}
  q^1=\frac{\phi^4-i\phi^5}{\sqrt{2}}\ ,\quad q^2=\frac{\phi^1+i\phi^2}{\sqrt{2}}\ ,\quad
  \phi=\phi^3
\end{equation}
and
\begin{equation}
  \lambda= \left(\begin{array}{c}\chi^1 \\ \chi^2\end{array}\right) \otimes
  			\left(\begin{array}{c} 1\\ 0 \end{array}\right) +
  			\left(\begin{array}{c}\psi^1 \\ \psi^2\end{array}\right) \otimes
  			\left(\begin{array}{c} 1\\ 0 \end{array}\right)\ .
\end{equation}
$q^A, \chi^A$ with $A=1,2$ are doublets of $SU(2)_R$ of the 5d $\mathcal{N}=1$ theory
on flat space, broken to $U(1)_R$ in our case on the curved manifold.
We decomposed the fermion $\lambda$ into $\chi$ and $\psi$ by eigenvalues of the
internal gamma matrix $\hat\gamma^3 = -{\bf 1}_2\otimes \sigma^3$.
With this decomposition, the symplectic Majorana condition on $\lambda$, which is
$\lambda^*= C\otimes \hat{C}\lambda$, reduces to $\chi^* = C\otimes i\sigma^2\chi$ where $\sigma^I$ is acting on $SU(2)$ internal indices $A$. See appendix B for the definition
of the charge conjugation matrices $C$ and $\hat{C}$.
The vector multiplet consists of with $A_\mu,\phi,\chi^A$ and the adjoint hypermultiplet
consists of $q^A,\psi\equiv\psi^2$. The `off-shell' theory we construct has
the SUSY transformation rule and action which has the $\delta^2=\cdots$ part of the
algebra off-shell, with any given SUSY transformation $\delta$. The SUSY we shall
use to calculate the path integral is $\delta\sim Q+S$. More specifically, the charge
content of $Q=Q^{++}_{---}$ demands that
\begin{equation}\label{Killing-spinor-eq2}
  D_m\epsilon_\pm=\frac{i}{2r}J_{mn}\gamma_n\epsilon_\pm
  \pm\frac{1}{2r}\gamma_m\gamma_\tau\epsilon_\pm=0\ ,\ \
  M_n\epsilon_\pm=\mp\frac{3}{2}\epsilon_\pm\ ,\ \
  i\gamma^{12}\epsilon_\pm=-i\gamma^{34}\epsilon_\pm=\mp\epsilon_\pm\ ,
\end{equation}
where $\epsilon_\pm$ parameterize $Q,S$ SUSY. For the vector multiplet part,
it is useful to employ the off-shell supergravity formalism of \cite{Festuccia:2011ws}.
We use the same supergravity convention as \cite{Kim:2012qf}, which is essentially
that of \cite{Kugo:2000af,Fujita:2001kv,Hanaki:2006pj}. The details of this
construction are explained in appendix C. For the off-shell SUSY and
action for the hypermultiplet, we used the strategy of \cite{Hosomichi:2012ek},
which is inspired by the methods of \cite{Pestun:2007rz,Hama:2012bg}.

For the vector multiplet, Euclidean off-shell action is (with $I=1,2,3$ for $SU(2)_R$)
\begin{eqnarray}\label{vector-action}
  \tilde g_{YM}^2 \mathcal{L}_V&=&{\rm tr}\left[\frac{1}{4}F_{\mu\nu}F^{\mu\nu}+
  \frac{1}{2}D_\mu\phi D^\mu\phi-\frac{i}{2}\chi^\dag \gamma^\mu D_\mu\chi
  -\frac{i}{2}\chi^\dag[\phi,\chi]
  -\frac{1}{2}\left(D^I+\frac{3}{r}\phi\delta^I_3\right)^2\right.\\
  &&\hspace{.7cm}\left.+\frac{2}{r^2}\phi^2+\frac{1}{8r}\chi^\dag
  J_{\mu\nu}\gamma^{\mu\nu}\chi-\frac{i}{2r}\chi^\dag M_n\chi-\frac{i}{2r\sqrt{g}}
  \epsilon^{\mu\nu\chi\rho\sigma}\left(A_\mu\partial_\nu A_\chi
  -\frac{2i}{3}A_\mu A_\nu A_\chi\right)J_{\rho\sigma}\right]\ .\nonumber
\end{eqnarray}
Throughout the paper,
our convention for the auxiliary fields $D^I$ is taking their path integral contours
to be along the imaginary direction, with possible shifts towards the real axis as we
shall explain below. (So the minus sign of the last term on the first line is natural.)
The SUSY variations are given by
\begin{eqnarray}\label{offshell-SUSY-vector}
  \delta\chi&=&\frac{1}{2}F_{\mu\nu}\gamma^{\mu\nu}\epsilon+iD_\mu\phi\gamma^\mu\epsilon
  -iD^I\sigma^I\epsilon-\frac{i}{r}\phi\sigma^3\epsilon\\
  \delta\phi&=&-\epsilon^\dag\chi\ \ ,\ \ \ \quad
  \delta A_\mu=i\epsilon^\dag\gamma_\mu\chi\nonumber\\
  \delta D^I&=&\epsilon^\dag\sigma^I\gamma^\mu D_\mu\chi
  -\frac{i}{4r}\epsilon^\dag\gamma^{\mu\nu}J_{\mu\nu}\sigma^I\chi
  +\epsilon^\dag\sigma^I[\phi,\chi]
  -\epsilon^\dag\frac{1}{2r}\sigma^3\sigma^I\chi\ .\nonumber
\end{eqnarray}
Comparing with the previous notation using $\epsilon_\pm$, $\sigma^I$
are defined by $\sigma^3\epsilon_\pm=\pm\epsilon_\pm$. Appendix C explains
the supergravity derivation of these action and SUSY.
The SUSY algebra closes off-shell and, with a commuting parameter $\epsilon=\epsilon_++\epsilon_-$, it is given by
\begin{eqnarray}\label{vector-off-algebra}
	\delta^2\phi &=&\xi^\mu\partial_\mu\phi+i[\Lambda,\phi] \nonumber\\
	\delta^2A_\mu &=& \xi^\nu\partial_\nu A_\mu
	+ \partial_\mu\xi^\nu A_\nu+D_\mu\Lambda \nonumber\\
	\delta^2\chi &=& \xi^\mu\partial_\mu\chi
	+\frac{1}{4}\Theta_{\mu\nu}\gamma^{\mu\nu}\chi
	+i[\Lambda,\chi] +\frac{i}{2r}\epsilon^\dagger\epsilon(\sigma^3\chi) \nonumber \\
	\delta^2 D^I &=& \xi^\mu\partial_\mu D^I+i[\Lambda,D^I]
	-\frac{i}{r}\epsilon^\dagger\sigma^3\sigma^{IJ}\epsilon D^J\ ,
\end{eqnarray}
where
\begin{eqnarray}
	\xi^\mu &= -i\epsilon^\dagger\gamma^\mu\epsilon \,, \quad
    \Theta^{\mu\nu} ~=~ \xi^\lambda\omega_\lambda^{\mu\nu} \,, \quad
    \Lambda ~=~ i\epsilon^\dagger\gamma^\mu\epsilon A_\mu -\epsilon^\dagger\epsilon\phi\ .
\end{eqnarray}

The off-shell hypermultiplet action and SUSY are obtained by introducing
two complex auxiliary fields $F^{A^\prime}$, as in \cite{Hosomichi:2012ek}.
The action is given by
\begin{eqnarray}
  \hspace*{-1cm}\tilde{g}_{YM}^2 \mathcal{L}_H&\!=\!&{\rm tr}\left[|D_\mu q^A|^2+\frac{4}{r^2}|q^A|^2
  +\frac{1}{r^2}|M_nq^A|^2+\left|[\phi,q^A]\right|^2+D^I(\sigma^I)^A_{\ B}
  [q^B,\bar{q}_A]-\frac{2n}{r}\phi[q^A,\bar{q}_A]-\bar{F}_{A^\prime}
  F^{A^\prime}\nonumber\right.\\
  \hspace*{-1cm}&\!&\!\left.-i\psi^\dag \gamma^\mu D_\mu\psi\!+\!i\psi^\dag[\phi,\psi]
  \!+\!\sqrt{2}i\psi^\dag[\chi_A,q^A]\!-\!\sqrt{2}i[\bar{q}_A,\chi^{\dag A}]\psi
  \!+\!\frac{1}{4r}\psi^\dag J_{\mu\nu}\gamma^{\mu\nu}\psi\!-\!\frac{i}{r}
  \psi^\dag M_n\psi\right]\ .
\end{eqnarray}
The SUSY transformation is given by
\begin{eqnarray}
  \delta q^A&=&\sqrt{2}\epsilon^{\dag A}\psi\ \ ,\ \ \
  \delta\bar{q}_A=\sqrt{2}\psi^\dag\epsilon_A\\
  \delta\psi&=&\sqrt{2}\left[-iD_\mu q_A\gamma^\mu\epsilon^A-i[\phi,q_A]\epsilon^A
  +\frac{i}{r}(M_nq_A)\epsilon^A-\frac{2i}{r}q_A\eta^A-i\hat\epsilon^{A'}F_{A'}\right]\nonumber\\
  \delta\psi^\dagger &=& \sqrt{2}\left[i\epsilon^\dagger_A\gamma^\mu D_\mu\bar{q}^A
  -i[\bar{q}^A,\phi]\epsilon^\dagger_A-\frac{i}{r}\epsilon^\dagger_AM_n\bar{q}^A
  -\frac{2i}{r}\eta^\dagger_A\bar{q}^A-i\hat\epsilon_{A'}^\dagger\bar{F}^{A'}\right]\nonumber \\
  \delta F^{A'} &=& \sqrt{2}(\hat\epsilon^\dagger)^{A'}\left[
  \gamma^\mu D_\mu\psi-[\phi,\psi]-\sqrt{2}[\chi_A,q^A]+\frac{i}{4r}J_{\mu\nu}\gamma^{\mu\nu}\psi  +\frac{1}{r}M_n\psi\right]\nonumber \\
  \delta\bar{F}_{A^\prime}&=&-\sqrt{2}\left[D_\mu\psi^\dagger\gamma^\mu+[\psi^\dagger,\phi]
  -\sqrt{2}[\bar{q}_A,\chi^{\dagger A}]-\frac{i}{4r}\psi^\dagger J_{\mu\nu}\gamma^{\mu\nu}
  +\frac{1}{r}M_n\psi^\dagger\right]\hat\epsilon_{A'} \nonumber \ .
\end{eqnarray}
Following \cite{Hosomichi:2012ek}, we introduced a new spinor parameter
$\hat\epsilon^{A^\prime}$ satisfying
\begin{equation}
	\epsilon^\dagger\epsilon =\hat\epsilon^\dagger\hat\epsilon \,, \ \ \
	(\epsilon^A)^TC\hat\epsilon^{B'}=0 \,, \ \ \
	\epsilon^\dagger\gamma_\mu\epsilon+\hat\epsilon^\dagger\gamma_\mu\hat\epsilon=0 \ .
\end{equation}
As before, $\eta$ is given by $\eta_\pm=\pm\gamma_\tau\epsilon_\pm$.
$F^{A'}, \hat\epsilon^{A'}$ are doublet under the auxiliary internal symmetry $SU(2)'$.
The off-shell SUSY algebra for the hypermultiplet with a commuting spinor $\epsilon$ is
\begin{eqnarray}\label{hyper-off-algebra}
    \delta^2q^A &=& \xi^\mu \partial_\mu q^A + i[\Lambda,q^A] +\frac{i}{r}\epsilon^\dagger\epsilon (M_nq)^A
    +\frac{2i}{r}\epsilon^\dagger\epsilon(\sigma^3q)^A \nonumber \\
    \delta^2 \psi &=& \xi^\mu\partial_\mu\psi+\frac{1}{4}\Theta_{\mu\nu}\gamma^{\mu\nu}\psi +i[\Lambda,\psi]+\frac{i}{r}\epsilon^\dagger\epsilon (M_n\psi) \nonumber \\
    \delta^2 F^{A'} &=& \xi^\mu\partial_\mu F^{A'} +i[\Lambda, F^{A'}] + \frac{1}{2} (\hat{R}^{IJ}\hat\sigma^{IJ} F)^{A'} \ ,
\end{eqnarray}
where
\begin{eqnarray}
    \hat{R}^{IJ} &=& -i\hat\epsilon^\dagger\hat\sigma^{IJ}\gamma^\mu D_\mu\hat\epsilon+\frac{i}{4r}\hat\epsilon^\dagger\hat\sigma^{IJ}\gamma^{\mu\nu}J_{\mu\nu}\hat\epsilon \ .
\end{eqnarray}

We shall study the 6d superconformal index
\begin{equation}\label{index-definition}
  {\rm tr}\left[(-1)^Fe^{-\beta(E-\frac{R_1+R_2}{2}-m(R_1-R_2)+aj_1+bj_2+cj_3)}\right]
  \ \ \ \ ({\rm with}\ a+b+c=0)
\end{equation}
from the 5d theory. The index acquires nonzero
contributions from the BPS states saturating the following bound for the energy
($S^5$ radius $r$ is absorbed into $E$)
\begin{equation}
  E\geq 2(R_1+R_2)+j_1+j_2+j_3\ ,
\end{equation}
where the inequality comes from the right hand side of $\{Q,S\}$ being non-negative.
In the 5d QFT language, the relevant superalgebra is
\begin{equation}
  \{Q,S\}\sim E-\left(\frac{1}{2}-n\right)R_2+\left(\frac{1}{2}+n\right)R_2-Kk\ ,
\end{equation}
where the last term $-Kk$ is the central extension of the 5d algebra, which contains
the subalgebra of $OSp(8^*|4)$ commuting with $k$. One may also take $\beta_i=\beta(1+a_i)$
(where $a_i=(a,b,c)$) and $\mu=\beta m$ to be $4$ independent chemical potentials
and write the index as
\begin{equation}
  {\rm tr}\left[(-1)^Fe^{-\beta_i(j_i+\frac{R_1+R_2}{2})+\mu(R_1-R_2)}\right]\ .
\end{equation}
The effective `energy,' namely the charge conjugate to $\beta$ in the exponent, is
\begin{equation}
 \mathcal{E}\equiv E-\frac{R_1+R_2}{2}-m(R_1-R_2)+aj_1+bj_2+cj_3\ \rightarrow\
  \frac{3}{2}(R_1+R_2)-m(R_1-R_2)+(1+a_i)j_i\ .
\end{equation}
To get to the last expression, we used the BPS relation $E=2(R_1+R_2)+j_1+j_2+j_3$
for the BPS excited states.

We shall evaluate the index using supersymmetric localization in the next
subsection, which amounts to identifying the SUSY saddle points and then calculating
the determinant over quadratic fluctuations. The last determinant will be calculated
using the equivariant localization of indices for certain differential operators.
By this we mean, the determinant to be calculated will take a product form, where
each factor acquires contribution from a 5d QFT on $\mathbb{R}^4\times S^1$ living
near certain fixed points of $\mathbb{CP}^2$. The resulting QFT on $\mathbb{R}^4\times S^1$
will be deformed by the Omega deformation
\cite{Nekrasov:2002qd,Nekrasov:2003rj,Nekrasov:2004vw,Shadchin:2005mx},
whose parameters $\epsilon_1,\epsilon_2$ depend suitably on $a,b,c$. The identification
of the parameters $\epsilon_1,\epsilon_2$, etc. can be easily made by studying the behavior
of the charge $\mathcal{E}$ near the fixed points. So let us do it here.

The fixed points of $\mathbb{CP}^2$ that are relevant for the determinant calculation
are with respect to the $U(1)^2\subset SU(3)$ rotation on $\mathbb{CP}^2$, conjugate to
$a,b,c$. One can use the following coordinates on $\mathbb{C}^3$
\begin{equation}
  z_i=rn_ie^{i\phi_i}\ \ \ (i=1,2,3)
\end{equation}
with $n_1^2+n_2^2+n_3^2=1$ to address these fixed points. The $U(1)^2$ rotation
acts on the $\phi_1-\phi_3$, $\phi_2-\phi_3$ angles by constant shifts. As the
$\mathbb{CP}^2$ coordinates are defined by the ratios like $z_1/z_3$, $z_2/z_3$,
the overall phase rotation on all $z_i$'s are invisible on $\mathbb{CP}^2$. There
are three fixed points of this $U(1)^2$, which are given by $(n_1,n_2,n_3)=(1,0,0)$,
$(0,1,0)$ and $(0,0,1)$. The index of the transversally elliptic operators that we
shall discuss in the next subsection, to calculate the quadratic fluctuations'
determinant, factorizes into three sums. Each part can be obtained from appropriate
equivariant indices used to calculate Nekrasov's partition function on
$\mathbb{R}^4\times S^1$.

We investigate the behavior of the charge $\mathcal{E}$ near these fixed points
to find the relation between Nekrasov's parameters and ours. For instance, at the third
fixed point with $n_3=1$, $j_1+j_2+j_3$ plays the role of extra circle
momentum visible through the instanton charge, while other two
$j_1$ and $j_2$ rotate $\mathbb{R}^4$. So one can write
\begin{equation}
  \mathcal{E}=(1+c)Kk-(m+n(1+c))(R_1-R_2)+(a-c)j_1+(b-c)j_2
  -\frac{3c}{2}(R_1+R_2)\ ,
\end{equation}
up to a possible addition of $\{Q,S\}$.
Introducing the Omega deformation parameters $\epsilon_1=a-c$,
$\epsilon_2=b-c$ with the self-dual part given by $\epsilon_+=
\frac{\epsilon_1+\epsilon_2}{2}=-\frac{3c}{2}$ (using $a+b+c=0$),
the above effective energy yields a factor
\begin{equation}
  e^{-Kk\beta(1+c)}e^{-\beta(\epsilon_1j_1+\epsilon_2j_2+\epsilon_+(R_1+R_2))}
  e^{\beta(m+n(1+c))(R_1-R_2)}\ .
\end{equation}
The first factor implies that one has to provide the factor $e^{-K\beta(1+c)}$
per instanton at this fixed point. We shall indeed derive this result
from the 5d QFT perspective later. The second factor is the standard measure
inserted for the states in the Omega-deformed $\mathbb{R}^4$. The last
$\epsilon_+(R_1+R_2)$ combines with the self-dual part of the angular momentum
part to yield $\epsilon_+(R_1+R_2+j_R)$, making it possible for a
right-chiral (namely, dotted) Poincar\'e SUSY generator on $\mathbb{R}^4$ to
commute with it. The last factor demands that the hypermultiplet mass $m$
on Omega-deformed $\mathbb{R}^4\times S^1$ has to be replaced by
\begin{equation}
  m\ \rightarrow\ m+n(1+c)\ ,
\end{equation}
before we use it as a building block of our determinant.
At the other two fixed points,
the parameter relations are similar, with the cyclic permutation taken
to the $a,b,c$ variables.

Apart from the formal analysis of the index in the rest of this section,
we shall later concentrate on the QFT's with $n=\pm\frac{1}{2}$ for detailed studies.
These theories have the advantage of exhibiting more supersymmetries manifestly
visible in 5d. However, the Abelian 6d index in section 3.1 will be discussed from
the 5d QFT's with other values of $n$.

\subsection{The index on $\mathbb{CP}^2\times S^1$}

In this section we calculate the supersymmetric path integral for the 5d QFT,
for the index. The 6d index (\ref{index-definition}) can be directly defined
as the 5d index, as all charges $E,R_1,R_2$ and $a_ij_i$ appearing in the measure
acquires precise meaning in 5d. In particular, since $a+b+c=0$, the charge
$aj_1+bj_2+cj_3$ is a combination of $U(1)^2$ rotations on $\mathbb{CP}^2$.
The factor $(-1)^Fe^{-\beta E}$ in (\ref{index-definition}) makes all fields
to be periodic in the Euclidean time direction with $\tau\sim\tau+\beta$.
Other measures twist the boundary conditions of all 5d fields. Equivalently,
one can untwist the boundary condition. All fields are periodic, while all
time derivatives appearing in the action and SUSY transformation are changed as
\begin{equation}
  \partial_\tau\ \rightarrow\ \partial_\tau-\frac{R_1+R_2}{2}-m(R_1-R_2)+a_ij_i\ .
\end{equation}
$R_1,R_2$, $a_ij_i$ are suitable rotations acting on the fields. Such untwisting
deformations are often phrased as putting the system in an Omega background.
As $Q$, $S$ are designed to commute with all the twistings, they survive the
deformed QFT (in the deformed version with the above substitutions). In particular,
around the three fixed points of $\mathbb{CP}^2$ under $U(1)^2$ generated by $a_ij_i$,
the deformed QFT should be exactly the same as that of
\cite{Nekrasov:2002qd,Nekrasov:2003rj}. This is because our curved space QFT reduces
locally to the 5d SUSY QFT on $\mathbb{R}^4\times S^1$, and also because our Omega
deformation exactly reduces to that on $\mathbb{R}^4\times S^1$ around the fixed points.
The parameter identifications for the local QFT's and that of
\cite{Nekrasov:2002qd,Nekrasov:2003rj} are already made. Near the third fixed point,
one finds
\begin{equation}
  (\epsilon_1,\epsilon_2,m_0)=(a-c,b-c,m+n(1+c))\ ,
\end{equation}
where the left hand side is the Omega background and mass of \cite{Nekrasov:2002qd,Nekrasov:2003rj}. Around the other two fixed points,
similar relation holds with cyclic permutations on $a,b,c$.

The index is given by the supersymmetric path integral of the deformed QFT
on $\mathbb{CP}^2\times S^1$. To be absolutely rigorous, one would have to start
from a gauge fixed path integral. One should introduce the ghost multiplet to
account for the gauge fixing term and the Faddeev-Popov determinant, and then
identify the gauge-fixed version of the off-shell SUSY transformation
$\hat\delta=\delta+\delta_B$, where $\delta_B$ is a BRST transformation and
$\hat\delta^2$ is given by a combination of bosonic symmetry similar to $\delta^2$.
All quantum calculations, especially the calculation of the determinant from Gaussian
fluctuations, should be done using $\hat\delta$. On $S^4$, this has been carefully done
in \cite{Pestun:2007rz,Hama:2012bg}. Also, on the round and squashed $S^5$, similar
procedure was discussed in \cite{Kallen:2012va} and \cite{Kim:2012qf} (v2), respectively.
In this paper, on $\mathbb{CP}^2\times S^1$, we work in a less
rigorous manner. The calculation consists of the study on the `classical' saddle points,
and then the Gaussian integration in the previous saddle point background using
the gauge-fixed theory using certain background gauge for the fluctuations. As the
latter determinant calculation can be easily obtained by combining the already known
results on $\mathbb{R}^4\times S^1$, we assume that a suitable gauge-fixed supersymmetric
path integral exists, which around all the fixed points reduce to those known on
$\mathbb{R}^4\times S^1$.

Let us start by localizing the vector multiplet part of the path integral.
The SUSY transformation of the gaugino
\begin{equation}
  \delta\chi=\frac{1}{2}F_{\mu\nu}\gamma^{\mu\nu}\epsilon
  +iD_\mu\phi\gamma^\mu\epsilon-iD^I\sigma^I\epsilon-\frac{i}{r}\phi\sigma^3\epsilon
\end{equation}
can be used to provide the $Q$-exact deformation $tQV$ with the following $V$
for the vector multiplet:
\begin{equation}\label{deformation}
  V=(\delta\chi)^\dag\chi\ .
\end{equation}
Such a $Q$-exact deformation does not change the path integral, and in particular
does not depend on $t$. So one can take $t$ to be large and use a formal saddle
point approximation to obtain the exact partition function.
The complex conjugation $\dag$ has to be understood with care, by suitably choosing
the path integral contours. Firstly, we keep the fields
$A_\mu,\phi$ to be real, apart from a minor shift of $A_\mu$'s contour to imaginary
directions coming from nonzero chemical potentials $a_i$, as explained below.
In the Euclidean
theory, the contours for $D^I$ fields are chosen to be along the imaginary
axis. However, for the convergence of the Euclidean off-shell action, the contour for
the field $D\equiv D^3$ has to be shifted towards the real direction, so that
$D+\frac{3}{r}\phi$ appearing in the classical action is imaginary for a given value
of real $\phi$. It seems safest to use the above $V$ with $(\delta\chi)^\dag$
obtained with these conjugation rules, as then both classical action and the
$Q$-exact deformation will be positive semi-definite by themselves, with a universal
contour which does not depend on the value of $t$. Imposing a more general reality
condition (meaning, contour choice) could also be acceptable if the $Q$-exact
deformation is made together with contour deformation.

We found it a bit more illuminating to work with a slightly more general one-parameter
family of possible path integral contours. By doing so, we can have a better
understanding on some crucial signs that will appear in our final result which
are important to get the correct result, correlated with the choice of
legitimate/illegal path integral contours. We use a $Q$-exact
deformation (\ref{deformation}) where $\dag$ is taken in $V$ `\textit{as if}'
$D+\frac{\xi}{r}\phi$ is imaginary with a real parameter $\xi$, for a given real
value of $\phi$. If one is uninterested in this slight generalization, one could
just plug in $\xi=3$ in all formulas below. So with
\begin{equation}
  (\delta\chi)^\dag\equiv-\frac{1}{2}\epsilon^\dag\gamma^{\mu\nu}F_{\mu\nu}
  -i\epsilon^\dag\gamma^\mu D_\mu\phi-\epsilon^\dag\sigma^a(iD^a)
  +i\epsilon^\dag\sigma^3\left(\frac{1-\xi}{r}\phi-(D+\frac{\xi}{r}\phi)\right)\ ,
\end{equation}
with $a=1,2$ and $D\equiv D^3$, the bosonic part of $QV$ is given by
\begin{equation}
\left(F_{mn}^--\frac{1-\xi}{2r}\phi J_{mn}\right)^2
-(D^a)^2-\left(D+\frac{\xi}{r}\phi\right)^2+(D_\mu\phi)^2+(F_{m\tau})^2
\end{equation}
with $m,n=1,2,3,4$, before introducing $a,b,c$ parameters. The effect coming from
nonzero $a_i$'s will be discussed below.
The saddle point conditions, which make all the complete squares to vanish,
can be written as
\begin{equation}\label{saddle-point-vector}
  D^1=D^2=0\ ,\ \ F^-=\frac{2s}{r^2}J\ ,\ \ \frac{\phi}{r}+D=\frac{4s}{r^2}
  \ ,\ \ D+\frac{\xi}{r}\phi=0\ .
\end{equation}
The second and third equations are universal SUSY requirements
(obtained from $\delta\chi=0$) which is valid with any $\xi$.
The matrix $s$ that we defined by the third equation is suitably quantized as follows.
$s$ can be taken to be an $N\times N$ diagonal matrix with
integer entries $s_1,s_2,\cdots s_N$. The equation $r^2F^-=2sJ$ means that, nontrivial
self-dual instantons can be superposed with anti-self-dual part $2sJ$, if possible.
$s_i$'s are integer parameters which partly label the saddle point space.
Solving the third and fourth equations, one obtains
\begin{equation}\label{scalar-saddle}
  r\phi=\frac{4s}{1-\xi}\equiv\sigma\ ,\ \ r^2D=-\frac{4s\xi}{1-\xi}
\end{equation}
for given $s$. As there is an extra circle factor, one can turn on the holonomy of $A_\tau$,
supposing that it satisfies the last condition $F_{m\tau}=0$.

The allowed value of $\xi$ can be determined as follows.
One should see if the net classical plus $Q$-exact deformation maintains
positive quadratic terms for all modes. For simplicity, let us consider the zero modes
of $D$ and $\phi$ on $\mathbb{CP}^2$. Denoting by $t=\varepsilon^{-1}>0$, the net action
times $\varepsilon$ is
\begin{equation}
  \varepsilon\left[(D+\frac{3i}{r}\phi)^2+\frac{4}{r^2}\phi^2\right]
  +\left(D+\frac{i\xi\phi}{r}\right)^2+
  \frac{(1-\xi)^2}{r^4}\left(r\phi-\frac{4s}{1-\xi}\right)^2\ .
\end{equation}
For clarity, only in this paragraph we use the convention of real $D$ by
inserting $-iD$ to the
$D$ used above. The last two terms come from the $Q$-exact deformation.
Let us investigate the sector with $s=0$, or the region with large enough $\phi$.
Then the above quadratic terms can be written as
\begin{equation}
  (1+\varepsilon)\left(D+\frac{i}{r}\frac{\xi+3\varepsilon}{1+\varepsilon}\phi\right)^2
  +\frac{(1-2\varepsilon-\xi)^2}{(1+\varepsilon)r^2}\phi^2\ .
\end{equation}
Thus, to guarantee the convergence of path integral in these zero mode parts,
one should continuously deform the contour as one takes $\varepsilon$ from
$\infty$ to $0$ by demanding $D+i\frac{\xi+3\varepsilon}{1+\varepsilon}\phi$
and $\phi$ be real. However, two clearly distinct cases arise from the second
complete-square term proportional to $\phi^2$. For $\xi>1$, one finds
$1-\xi-2\varepsilon<0$ for all positive $\varepsilon$, so $\phi$ maintains to have
a positive quadratic term throughout the deformation process. On the other hand,
for $\xi<1$, the $\phi$ quadratic term becomes degenerate at $\varepsilon=\frac{1-\xi}{2}>0$. So in this sense, $Q$-exact deformations with $\xi>1$ are safe while those with $\xi<1$
are dangerous by developing a flat direction during deformation. Note that the case
with $\xi=3$ has $D+\frac{3i}{r}\phi$ to be real throughout the deformation
process, which we already found to be a safe choice. So we restrict our discussion to
the cases with $\xi>1$, for which the final result will not depend on $\xi$. The only
crucial point is that, for $\xi>1$, the ratio of $\phi$ and $s$ at
the saddle point (\ref{scalar-saddle}) is negative. This fact will be crucial
for getting the correct index.

The hypermultiplet fields are all trivial at the saddle points. To see this,
we introduce the following $Q$-exact deformation
\begin{equation}
  \delta((\delta\psi)^\dag\psi+\psi^\dag(\delta\psi^\dag)^\dag)\ ,
\end{equation}
where
\begin{eqnarray}
  (\delta\psi)^\dagger &=& -i\epsilon^\dagger_A\gamma^\mu D_\mu\bar{q}^A +i\epsilon^\dagger_A[\phi,\bar{q}^A]
  -\frac{i}{r}\epsilon^\dagger_A M_n\bar{q}^A -\frac{2i}{r}\eta^\dagger_A\bar{q}^A +i\hat\epsilon_{A'}^\dagger\bar{F}^{A'} \nonumber \\
  (\delta\psi^\dagger)^\dagger &=& iD_\mu q_A\gamma^\mu\epsilon^A -i[\phi,q_A]\epsilon^A +\frac{i}{r}(M_nq_A)\epsilon^A
  -\frac{2i}{r}q_A\eta^A +i F_{A'}\hat\epsilon^{A'}
\end{eqnarray}
Note that we take the matter scalar $q^A$ to be real and the auxiliary field $F^{A'}$ to be imaginary. The bosonic part becomes
\begin{equation}
  (\delta\psi)^\dag(\delta\psi) +(\delta\psi^\dag)(\delta\psi^\dag)^\dag= 2|D_\mu q_A\gamma^\mu\epsilon^A|^2+\frac{1}{r^2}\left(|M_nq^A|^2+|2q^A|^2\right) +|[\phi,q^A]|^2 - \bar{F}_{A'}F^{A'}
\end{equation}
This ensures that the saddle point values for the hypermultiplet scalars are zero, $q^A=F^{A'}=0$.

There appear further changes and restrictions on the saddle points with nonzero $a,b,c$.
The introduction of the chemical potential $m$ for $R_1-R_2$ will only affect
the determinant calculation, as the hypermultiplet is trivial at the saddle points.
So let us consider the effect of $a,b,c$ here. As chemical potentials, they
simply twist all time derivatives appearing in the action and SUSY transformation.
This effect is captured by considering the system in the rotating frame
\begin{equation}\label{rotating}
  \phi_i\rightarrow \phi_i^\prime=\phi_i-i\frac{a_i}{r}\tau\ ,\ \ \tau^\prime=\tau\ ,
\end{equation}
or equivalently
\begin{equation}
  E=E^\prime-\frac{a_i}{r}j_i^\prime\ ,\ \ j_i=j_i^\prime\ ,
\end{equation}
where $E\sim-\partial_\tau$ and $j_i\sim-i\partial_{\phi_i}\,$.
This will replace the factor $e^{-\beta(E+a_ij_i)}$ that we have in the index
by $e^{-\beta E^\prime}$, making the fields to be periodic
in $S^1$ in this frame. This procedure is often called `untwisting,' as the
twisted boundary condition is made periodic. With imaginary chemical potential,
this is indeed going to the rotating frame.
With real $a_i$'s, we still regard it formally as a complex rotation.
The action and SUSY transformation are complexified in this way.
We shall drop the prime
superscript for the new coordinates from now on. Most of the analysis goes
the same as the case with $a,b,c=0$, apart from the two differences. Firstly,
$D^a=0$, $D+\frac{\xi}{r}\phi=0$, $D_\mu\phi=0$ remain the same.
$F_{mn}^-=\frac{1-\xi}{2r}\phi J_{mn}$ and $F_{m\tau}=0$
change into
\begin{equation}\label{new-saddle}
  F_{mn}^-=\frac{1-\xi}{2r}\phi J_{mn}\ ,\ \
  F_{m\tau}=\frac{i}{2}(1-\xi)\phi(a_in_idn_i\wedge d\tau)_{m\tau}\ .
\end{equation}
Here we used the fact $J=r^2\sum_{i=1}^3n_idn_i\wedge d\phi_i$ on $\mathbb{CP}^2$,
picking up $dn_i\wedge d\tau$ components in the new frame (\ref{rotating}).

The first change with nonzero $a_i$ is that, the moduli of self-dual Yang-Mills
instantons is lifted by the $U(1)^2$ rotations. To see this, let us study the second
equation of (\ref{new-saddle}). After the shift on $\partial_\tau$
which acts on $A_m$ in $F_{m\tau}$, this equation can be written as
\begin{equation}\label{twisted-saddle}
  D_m\left(A_\tau-i\frac{1-\xi}{4}\phi a_in_i^2\right)
  =\left(\partial_\tau+a_ij_i\right)A_m\ ,
\end{equation}
where we used the fact $D_\mu\phi=0$. The configuration of nontrivial $A_m$
yielding self-dual instantons, generically rotate under the $U(1)^2$ rotations
$a_ij_i$. So the stationary self-dual instanton configuration would cause
nonzero right hand side, unless its nonzero field profile is localized on
the fixed points of $a_ij_i$ rotations. We already identified three of them
before: $(n_1,n_2,n_3)=(1,0,0)$, $(0,1,0)$, $(0,0,1)$. So the part of $A_m$ which
yields self-dual part of $F_{mn}$ should be completely localized on these
fixed points in a singular way.\footnote{We are not sure how to rigorously argue
that these are the most general saddle point configurations solving the differential
conditions. At the very least, we can say that we identified set of saddle point solutions
with which the partition function works perfectly well as the 6d index, in all sectors we
studied. This could be a circumstantial evidence of the completeness of our findings. Also,
note that near the three fixed points, all of our equations should be the same as the saddle point equations leading to the results of \cite{Nekrasov:2002qd,Nekrasov:2003rj},
so at least the local analysis should not be missing anything.}
This is exactly the same phenomenon as the self-dual instantons on $\mathbb{R}^4$ with
lifted position moduli in the Omega background \cite{Nekrasov:2002qd}.
On the other hand, the anti-self-dual part of $F_{mn}$, proportional to the K\"ahler
2-form $J_{mn}$ of $\mathbb{CP}^2$, remains unchanged as $J$ is invariant under
the $SU(3)$ isometry, of which $a_ij_i$ is a subset. As the equations are nonlinear
due to the non-Abelian nature of the gauge fields, one might wonder whether one can
separate the potential $A_m$ into that contributing to the self-dual and anti-self-dual
field strengths. Of course generally this would be impossible. However, since $F_{mn}^+$
is singularly supported at the fixed points (with finite $\int_{\mathbb{CP}^2}F\wedge F$)
while $F_{mn}^-\sim J$ is delocalized,
one can think that $A_m^-$ leading to $F_{mn}^-\neq 0$ is dominant in one region (most
regions in $\mathbb{CP}^2$ away from the fixed points), while $A_m^+$ leading to
$F_{mn}^+\neq 0$ is dominant in very small regions near the fixed points. Also, it will
turn out that the gauge orientations of the self-dual part of the field strength
have to be in $U(1)^N\subset U(N)$, in the basis $F^-\sim sJ$ is diagonal.
So the above considerations are fine.

So we made the right hand side of (\ref{twisted-saddle}) to be zero
by taking $A_m$ to be stationary and invariant under $a_ij_i$ as above.
To have the left hand side to vanish, one should take
\begin{equation}
  A_\tau-i\frac{1-\xi}{4}\phi a_in_i^2
\end{equation}
to be covariantly constant. Away from the fixed points, the covariant derivative
$D_m$ only contains $A_m\sim s\theta_m$ in the $U(1)^N$ part, where $2J=d\theta$.
So a constant holonomy inside the same $U(1)^N$ can be given to solve this equation,
leading to ($\beta=\frac{2\pi r_1}{r}$ with the circle radius $r_1$)
\begin{equation}
  A_\tau=\frac{\lambda}{r\beta}+\frac{i(1-\xi)}{4}\phi(a_in_i^2)
  =\frac{1}{r}\left[\frac{\lambda}{\beta}+is(a_in_i^2)\right]\ .
\end{equation}
$\lambda$ is a diagonal matrix with $N$ eigenvalues being $2\pi$ periodic variables $\lambda_i\sim\lambda_i+2\pi$ in the above normalization. Thus,
\begin{equation}
  r\beta A_\tau=\lambda+is\beta(a_in_i^2)
\end{equation}
at the saddle points will play the role of the `scalar expectation value,'
effectively breaking the $U(N)$ gauge symmetry into $U(1)^N$ in all the quadratic
fluctuation calculus of the path integral.
Later, for the calculation of determinant, the saddle point value of $A_\tau+i\phi$
at the fixed point is identified as the (complex) scalar expectation value of scalar
at the Coulomb phase of QFT on $\mathbb{R}^4\times S^1$, since this combination appears
on the right hand sides of $\delta^2$ in (\ref{vector-off-algebra}),
(\ref{hyper-off-algebra}). At the $i$'th fixed point
(with $n_i=1$), the value of $r\beta(A_\tau+i\phi)$ appearing in the determinant is
\begin{equation}
  \lambda+ia_is\beta+i\beta\sigma\ ,
\end{equation}
where $\sigma=\frac{4s}{1-\xi}$. This will play the role of chemical potential for
the $U(N)$ charged modes in the quadratic fluctuations. The replacement $\lambda\rightarrow\lambda+ia_is\beta$
means that charged states will pick up nonzero $U(1)^2$ angular momentum in the $s$
flux background. Also, for the self-dual field strength $F^+_{mn}$ discussed in the
previous paragraph to be compatible with the $U(1)^N$ expectation value of $\phi$
satisfying $D_\mu\phi=0$, the self-dual instantons should be in the $U(1)^N$ part.
This is also the same as what happens in the instanton calculus of \cite{Nekrasov:2002qd}
on $\mathbb{R}^4\times S^1$ in the Coulomb branch.

Now before evaluating the determinant of quadratic fluctuations in the above
background, one should obtain the value of the classical action at the saddle points.
As the nonzero configuration of self-dual instantons are supported only at the fixed
points while the anti-self-dual part $\sim J$ is spread out in $\mathbb{CP}^2$,
the contributions from the two parts of field strength can be separately calculated.
The self-dual instantons with instanton numbers $k_1,k_2,k_3$ localized at the three
fixed points provide a factor of
\begin{equation}\label{self-dual-classical}
  e^{-S_{SD}}=e^{-\sum_{i=1}^3k_i(1+a_i)\beta}\ .
\end{equation}
The factor $k_i\beta$ naturally comes from the Yang-Mills action. The correction
$k_ia_i\beta$ comes from the Chern-Simons term, the last term in (\ref{vector-action}).
One should again remember that the twisting (\ref{rotating}) changes $J$ appearing in
(\ref{vector-action}), in the same way as (\ref{new-saddle}). To see this in more
detail, consider
\begin{equation}\label{JAF}
  S\leftarrow-i\int (J+ia_in_idn_i\wedge d\tau)\wedge
  \left(AdA-\frac{2i}{3}A^3\right)\ .
\end{equation}
The first term proportional to $J$ does not yield nonzero action in the
self-dual instanton background. (It will contribute from $F^-$ part, as we shall
see below.) The second term proportional to $a_i$ can be integrated by
parts to yield
\begin{equation}
S_{SD}\leftarrow\int\frac{1}{2}d(a_in_i^2)\wedge d\tau\wedge\left(AdA-\frac{2i}{3}A^3\right)
=\int\frac{a_in_i^2}{2}d\tau\wedge F\wedge F\ .
\end{equation}
This yields the desired contribution $a_ik_i\beta$ for $k_i$ instantons localized
at the $i$'th fixed point, with $n_i=1$.

Apart from the contribution from $k_{SD}=k_1+k_2+k_3$ self-dual instantons,
the anti-self-dual field, scalar $\phi$, holonomy $\lambda$ and $D$ also
contribute to the classical action. It comes from
$\frac{1}{4}F_{\mu\nu}^2+\frac{2}{r^2}\phi^2-\frac{1}{2}(D+\frac{3}{r}\phi)^2$ and the first $JAF$
term in (\ref{JAF}).\footnote{From (\ref{JAF}), one should calculate
$\int J\wedge (\lambda+is\beta(a_in_i^2))d\tau\wedge sJ$,
as the value of $A_\tau$ is shifted by $a_i$. The shifted term proportional to $a_i$,
however, is zero for the following reason. This term boils down to
$\int_{\mathbb{CP}^2}(a_in_i^2)J\wedge J$. Due to the symmetry of $\mathbb{CP}^2$,
$\int n_i^2 J\wedge J$ are same for all $i=1,2,3$. So this integral is zero from
$a+b+c=0$.} The result is
\begin{equation}\label{saddle-action-asd-scalar}
 e^{-S_0}=\exp\left[\sum_{i=1}^N\left(\frac{\beta s_i^2}{2}
 -is_i(\lambda_i+i\beta\sigma_i)\right)\right]\ ,
\end{equation}
where $\sigma$ is the saddle point value of $r\phi$ defined in (\ref{saddle-point-vector}),
$\sigma=\frac{4s}{1-\xi}$ with $\xi>1$. Note that the
appearance of $\lambda+i\beta\sigma$ is `holomorphic.' This is the same combination
$A_\tau+i\phi$ appearing on the right hand side of the $\delta^2$ algebra
(\ref{vector-off-algebra}), (\ref{hyper-off-algebra}).
So the same combination will appear in the 1-loop determinant, as already commented
above. The factor $e^{-i\sum_{i=1}^Ns_i\lambda_i}$ coming from the Chern-Simons term
induces electric charges to the anti-self-dual instantons proportional to $J$.
The factor $e^{\frac{\beta}{2}\sum_{i=1}^Ns_i^2}$ induces negative zero point `energy'
(or charge) to the ground states with nonzero anti-self-dual $s$ fluxes.

The determinant for small quadratic fluctuations in a given saddle point
background can be calculated using suitable index theorems of differential
operators appearing in the $Q$-exact deformation. To rigorously derive this part,
one has to carefully redevelop the localization problem in the gauge-fixed version,
also including the ghost multiplet. The basic strategy is similar to that of
\cite{Pestun:2007rz}, and in the context of $S^5$, similar story was developed in \cite{Kallen:2012va}, \cite{Kim:2012qf} (v2), without or with squashing, respectively.
To calculate the determinant on $\mathbb{CP}^2\times S^1$,
we shall proceed without filling all the technical details comparable to
the above works on $S^5$, however with clear comments on what we have not
shown. All properties that
we shall assume for our QFT on $\mathbb{CP}^2\times S^1$ are locally
completely well-developed around three fixed points of $\mathbb{CP}^2$
under $U(1)^2\subset SU(3)$, for the Omega-deformed 5d $\mathcal{N}=1$
QFT on $\mathbb{R}^4\times S^1$. The last fixed point QFT's are basically
all we need to calculate our determinant.

In the gauge-fixed action, the determinant for quadratic fluctuations
is that of differential operators appearing in the $Q$-exact deformations.
The determinant takes the form of \cite{Pestun:2007rz,Hama:2012bg}
\begin{equation}\label{det-general}
  \frac{\det(\mathcal{R})_{\rm fermion}}{\det(\mathcal{R})_{\rm boson}}\ ,
\end{equation}
where $\mathcal{R}$ denotes the differential operator for the symmetry $\mathcal{R}$
appearing as $\delta^2=\mathcal{R}$ in the superalgebra. This determinant ratio
is calculated by using the index for an auxiliary differential operator,
which is called $D_{10}$ in \cite{Pestun:2007rz}. This operator has the property
of commuting with $\mathcal{R}$, and can be canonically extracted by investigating
the quadratic part of the $Q$-exact deformation \cite{Pestun:2007rz}. The equivariant
index of $D_{10}$ is used to capture the modes which contribute to (\ref{det-general})
after a boson-fermion cancelation. More concretely, the equivariant index takes
the form of
\begin{equation}\label{equiv-index}
  \sum_i n_ie^{-w_i}\ ,
\end{equation}
where $i$ labels the eigenvalue of $\mathcal{R}$ on the BPS modes,
$w_i$ is eigenvalue of $\mathcal{R}$, and $n_i$ is the
index which is the number of bosonic modes minus the number of fermionic modes
with that eigenvalue. Once this index is known, the determinant boils down to
\cite{Pestun:2007rz,Hama:2012bg}
\begin{equation}\label{det-index}
  \frac{\det(\mathcal{R})_{\rm fermion}}{\det(\mathcal{R})_{\rm boson}}
  \rightarrow\frac{\det(\mathcal{R})_{{\rm coker}(D_{10})}}{\det(\mathcal{R}
  )_{{\rm ker}(D_{10})}}=\prod_iw_i^{-n_i}\ ,
\end{equation}
so that the knowledge of the index (\ref{equiv-index}) is
sufficient to calculate the determinant.

We assume that the operator $D_{10}$ in our case is transversally elliptic
\cite{Pestun:2007rz}, with the symmetry given by a linear combination of the $S^1$
translation and the rotation $a_ij_i$ on $\mathbb{CP}^2$, in the combination appearing
in $\mathcal{R}$. This allows us to use suitable index theorems to conveniently calculate
the index (\ref{equiv-index}). In our case of $\mathbb{CP}^2\times S^1$, we follow a
procedure similar to \cite{Drukker:2012sr} which studied the partition function
on $S^3$ or $S^2\times S^1$, or \cite{Kallen:2012cs,Kim:2012qf} on $S^5$.
We fix a momentum $t$ for the quadratic modes along the $S^1$
factor of $\mathbb{CP}^2\times S^1$. Then, one obtains infinitely many 4 dimensional
differential operators $\mathcal{R}_t$ labeled by $t$, by
inserting $t$ to $-i\partial_\tau$ in the differential operator.
One has to compute
\begin{equation}
  \prod_{t=-\infty}^\infty\frac{\det(\mathcal{R}_t)_{\rm fermion}}
  {\det(\mathcal{R}_t)_{\rm boson}}\ .
\end{equation}
The auxiliary differential operator $D_{10}$ also can be reduced to a 4 dimensional
one on $\mathbb{CP}^2$, labeled by $t$, which is transversally
elliptic on $\mathbb{CP}^2$. (On $\mathbb{CP}^2$, we actually expect the operator
$D_{10}$ to be elliptic, which is a physically important property when one considers
the $a_i\rightarrow 0$ limit: but our calculation in this subsection will not depend
on this fact.) Similar indices on $\mathbb{CP}^2$ were discussed in \cite{Kallen:2012cs,Kallen:2012va}. The index with given $t$ eigenvalue can be
calculated using equivariant index theorems \cite{atiyah} for transversally elliptic
operators, which states that the index (\ref{equiv-index}) is given by the sum of
contributions from the fixed points of the $a_ij_i$ symmetry. We already discussed the
three fixed points of the $a_ij_i$ rotation. Around these fixed points, the full analysis
of the gauge-fixed action and thus the determinant calculus (including the relevant index
theorems for transversally elliptic operators on $\mathbb{R}^4\times S^1$) has been
done in \cite{Nekrasov:2004vw,Nekrasov:2002qd}. \cite{Shadchin:2005mx} discusses
the relevant index theorems on $\mathbb{R}^4\times S^1$ in detail, for the so-called
self-dual complex (for the vector multiplet) and the Dirac complex (for the hypermultiplet),
which is reviewed in \cite{Kim:2012qf} in the form we shall use here. This is basically
the reason why we could manage to get our determinant without a full rigorous calculation.
We emphasize that, all the properties that we assume about the $D_{10}$ operator and
the gauge-fixed supersymmetric action are known to be true near the three fixed points,
from these studies.

Since the contributions from the three fixed points to the index (\ref{equiv-index})
add as
\begin{equation}
  \sum_i n_ie^{-w_i}=\sum_i(n_i^{(1)}+n_i^{(2)}+n_i^{(3)})e^{-w_i}\ ,
\end{equation}
the contributions to the determinant multiply as
\begin{equation}
  \prod_i w_i^{-n_i}=\prod_iw_i^{-n_i^{(1)}}\cdot\prod_iw_i^{-n_i^{(2)}}
  \cdot\prod_iw_i^{-n_i^{(3)}}\ .
\end{equation}
This determinant factor is given for a fixed number of self-dual instantons
$k_1$, $k_2$, $k_3$ at the three fixed points. Since the classical measure
for self-dual instantons also factorizes as (\ref{self-dual-classical}),
one can combine
\begin{equation}
  \left(\sum_{k_1=0}^\infty e^{-k_1\beta(1+a)}\prod_iw_i^{-n_i^{(1)}}\right)
  \left(\frac{}{}\!a,b,c\rightarrow b,c,a\right)
  \left(\frac{}{}\!a,b,c\rightarrow c,a,b\right)\ .
\end{equation}
Thus, the determinant combined with the self-dual
instantons' classical factor can be factorized into three parts after
$k_1,k_2,k_3$ are summed over. Each part is given by the gauge theory
partition function on Omega deformed $\mathbb{R}^4\times S^1$
of \cite{Nekrasov:2002qd,Nekrasov:2003rj,Nekrasov:2004vw,Shadchin:2005mx}.

So without a further ado, we present the form of our index and then explain the
well-known partition function on $\mathbb{R}^4\times S^1$ which is the building
block of our index. The index takes the form of
\begin{equation}\label{final}
  \frac{1}{N!}\sum_{s_1,s_2,\cdots s_N=-\infty}^\infty
  \oint\left[\frac{d\lambda_i}{2\pi}\right]
  e^{\frac{\beta}{2}\sum_{i=1}^Ns_i^2-i\sum_is_i\lambda_i}
  Z_{\rm pert}^{(1)}Z_{\rm inst}^{(1)}\cdot Z_{\rm pert}^{(2)}Z_{\rm inst}^{(2)}
  \cdot Z_{\rm pert}^{(3)}Z_{\rm inst}^{(3)}\ .
\end{equation}
Each factor $Z^{(1,2,3)}_{\rm pert}$, $Z_{\rm inst}^{(1,2,3)}$ depends on
$\lambda$, $a,b,c,m$ and also on the flux $s$ in the way explained below.
We defined the perturbative indices so that $Z^{(1)}_{\rm pert}Z^{(2)}_{\rm pert}
Z^{(3)}_{\rm pert}$ will also include the Haar measure for the gauge group
$G_s=\prod_iU(n_i)\subset U(N)$ unbroken by the flux $s$, which in turn will contain
the Vandermonde measure (or the Faddeev-Popov determinant) for $\lambda_i$ coming from
some gauge fixings. (This gauge-fixing part is implicitly built-in by borrowing the
results from the partition function on $\mathbb{R}^4\times S^1$, obtained from
equivariant indices which include the ghost multiplet.) $\frac{1}{N!}$ factor comes
from the division by Weyl group of $U(N)$. $\lambda_i$ includes the possible shift
$i\sigma$ by the scalar expectation
value, which re-scales the integration contour for $\lambda$ by the flux.
Naively, this contour integral appears to be along the $0\leq\lambda_i\leq 2\pi$
in the real direction, up to possible constant shifts from $i\sigma$.
Equivalently, the naive expectation is that the integration contour is the
unit circle on the complex plane for $e^{-i\lambda_i}$. This will turn out to be
wrong, after a careful inspection on the localization calculus we did.
As the correct choice of contour is very important for getting a sensible index
from (\ref{final}), we explain in detail the subtle aspects and
the final contour choice in the next subsection. In the rest of
this subsection, we explain the details of $Z_{\rm pert}^{(i)}$ and
$Z_{\rm inst}^{(i)}$ appearing in the integrand.

At the $i$'th fixed point, the factor $Z^{(i)}_{\rm pert}Z^{(i)}_{\rm inst}$
is the index of the 5d $\mathcal{N}=1^\ast$ gauge theory on Omega-deformed
$\mathbb{R}^4\times S^1$. Let us again summarize the rules for
parameter identification, explained in the previous subsection, to obtain our
factors from the $\mathbb{R}^4\times S^1$ partition function in the literature.
Without losing generality, we restrict the discussion to
the third fixed point $i=3$. Results for the other fixed points can be obtained
by cyclically permuting the roles of $a_i=(a,b,c)$. At the third fixed point,
Omega deformation parameters $\epsilon_1,\epsilon_2$ and the mass parameter $m_0$ on
$\mathbb{R}^4\times S^1$ are identified with our parameters $a,b,c,m$ as
\begin{equation}
  \epsilon_1=a-c\ ,\ \ \epsilon_2=b-c\ ,\ \ m_0=m+n(1+c)
\end{equation}
as explained before.
Moreover, as explained above in the classical action, $k$ self-dual instantons
localized at this fixed point yields the instanton factor $e^{-k\beta(1+c)}$.
This replaces the normal
$\exp\left(-\frac{4\pi^2k r_1}{g_{YM}^2}\right)$
factor which controls the instanton expansion. Finally, the partition function
on $\mathbb{R}^4\times S^1$ depends on the VEV of $i(A_\tau+i\phi)$. We
replace it by $i(\lambda+isc\beta+i\beta\sigma)$, as explained.

Let us first explain the perturbative part. On $\mathbb{R}^4\times S^1$,
the index is given by\footnote{For the term with $i=j$, namely for the Cartans,
$I_+$ has to be replaced by $I_++1$ for two of the three fixed point contributions.
This extra $+1$ is a contribution from a mode in the ghost multiplet in the gauge-fixed
action \cite{Pestun:2007rz}. We also refer to
\cite{Kim:2012qf} (v2), which explains exactly the same issue in the context
of $S^5$ partition function. In particular, this will be the reason why the product
in (\ref{haar}) does not acquire contribution from the Cartan modes with $i=j$, because
the $-2$ contribution from (\ref{pert-sum-0}) is canceled out.}
\begin{equation}
  Z_{\rm pert}=PE\left[\frac{1}{2}I_+(\epsilon_1,\epsilon_2,m_0)
  \sum_{i,j=1}^Ne^{-i\lambda_{ij}}\right]\ ,
\end{equation}
where $PE$ is normally defined in the literature as
\begin{equation}\label{PE-literature}
  PE[f(x)]=\exp\left[\sum_{n=1}^\infty\frac{1}{n}f(nx)\right]\ ,
\end{equation}
with all the `chemical potential like' variables $x=(\epsilon_1,\epsilon_2,m_0,\lambda_i)$.
$I_+$ is the index coming from a single superparticle on $\mathbb{R}^{4,1}$, preserving
the SUSY of a half-BPS W-boson (perturbative particle):
\begin{equation}
  I_+(\epsilon_1,\epsilon_2,m_0)=
  \frac{\sinh\frac{\beta(m_0+\epsilon_+)}{2}\sinh\frac{\beta(m_0-\epsilon_+)}{2}}
  {\sinh\frac{\beta\epsilon_1}{2}\sinh\frac{\beta\epsilon_2}{2}}\ .
\end{equation}
The definition of $PE$ above has to be slightly refined, as the above is only
taking into account the information on the `excited states.' To account for the
`vacuum energy factor,' $PE$ is given as follows. Suppose the `single particle index'
or the `letter index' appearing inside $PE$ admits a series expansion of the form
\begin{equation}\label{letter-expand}
  f(x)=\sum_i n_ie^{-\mu_i x}\ ,
\end{equation}
where $x$ again denotes all chemical potentials. Then $PE[f]$ is given by
\begin{equation}\label{PE-precise}
  PE[f]=\prod_i\left(2\sinh\frac{\mu_i x}{2}\right)^{-n_i}=
  \prod_i\frac{e^{-\frac{n_i\mu_i x}{2}}}{(1-e^{-\mu_i x})^{n_i}}\ .
\end{equation}
The prefactor $e^{-\frac{1}{2}\sum_in_i\mu_i x}$, calculated with suitable
regularization/renormalization, contributes to the vacuum `energy' or more
precisely the charges carried by the vacuum. The factor in the denominator
is the index of the Fock space made of the $i$'th single particle state,
and is essentially (\ref{PE-literature}). Another way (\ref{PE-precise}) generalizes
(\ref{PE-literature}) is that, $\mu_i x$ appearing in (\ref{letter-expand})
may be negative, which comes from a complex conjugate mode to a mode with
opposite value $-\mu_i x$ of charge. Up to an overall sign of $PE[f]$ on which
we do not try to be careful in this paper, one thus finds
\begin{equation}\label{PE-final}
  PE[f]=\prod_i\frac{e^{-\frac{1}{2}n_i|\mu_i x|}}{(1-e^{-|\mu_i x|})^{n_i}}\ ,
\end{equation}
which will be the precise form of $PE$ that we shall use.

The three factors of perturbative indices, with $\epsilon_1,\epsilon_2,m_0$
suitably replaced, combine to yield a simple expression. To be definite, let us explain
the combined factor for the 5d QFT with $n=-\frac{1}{2}$, namely a QFT with $8$
manifest SUSY. When we analyze the Abelian index, we shall explain similar structures
for all QFT$_n$ (here, QFT$_n$ denotes QFT associated with given $n$). As the flux $s$
factor multiplies $e^{i\lambda}$ by $e^{-\beta a_is}$,
which is different for three fixed points, the combined perturbative determinant depends
on the effective flux $s_i-s_j$ felt by the $ij$-th component of the adjoint mode.
We describe the determinant as a function of this effective flux $s=s_i-s_j\equiv s_{ij}$.
At $n=-\frac{1}{2}$, inside $PE$, three perturbative `letter indices' for a mode
with $s=0$ add to be
\begin{equation}\label{pert-sum-0}
  I_+\left(b\!-\!a,c\!-\!a,m\!-\!\frac{1+a}{2}\right)+I_+\left(c\!-\!b,a\!-\!b,m\!-\!\frac{1+b}{2}\right)
  +I_+\left(a\!-\!c,b\!-\!c,m\!-\!\frac{1+c}{2}\right)=-2\ .
\end{equation}
As the perturbative letter index is multiplied by the adjoint character of $U(N)$,
this part of the perturbative index at zero flux sector simply yields the Haar
measure for the unbroken gauge group in $s$ background:
\begin{equation}\label{haar}
  Z_{\rm pert}^{(1)}Z_{\rm pert}^{(2)}Z_{\rm pert}^{(3)}\leftarrow
  \prod_{i<j}\left(2\sin\frac{\lambda_{ij}}{2}\right)^2\ .
\end{equation}
The product is taken over the positive roots of unbroken subgroup $G_s$.
More generally, for nonzero $s$, the sum of three perturbative indices $I_+$ are
given as follows. Apart from the overall factor $e^{-i(\lambda_{ij}+i\beta\sigma_{ij})}$
common for all three fixed points, one has to compute
\begin{equation}\label{pert-sum}
  e^{sa\beta}I_+\left(b\!-\!a,c\!-\!a,m\!-\!\frac{1+a}{2}\right)+
  e^{sb\beta}I_+\left(c\!-\!b,a\!-\!b,m\!-\!\frac{1+b}{2}\right)
  +e^{sc\beta}I_+\left(a\!-\!c,b\!-\!c,m\!-\!\frac{1+c}{2}\right)
\end{equation}
inside $PE$, where $s=s_{ij}$. For some low values of $s$, this sum becomes
\begin{eqnarray}\label{pert-sum-low}
  s=\pm 1&:&-\left(e^{\pm \beta a}+e^{\pm\beta b}+e^{\pm\beta c}
  -e^{\mp\beta\left(m-\frac{1}{2}\right)}\right)\\
  s=\pm 2&:&-\left(\sum_{p+q+r=2}e^{\pm\beta(pa+qb+rc)}
  -e^{\mp\beta(m-\frac{1}{2})}(e^{\pm \beta a}+e^{\pm\beta b}+e^{\pm\beta c})
  -e^{\pm\beta(m-\frac{1}{2})}\right)\nonumber\\
  s=\pm 3&:&-\left(\sum_{p+q+r=3}e^{\pm\beta p_ia_i}
  -e^{\mp\beta(m-\frac{1}{2})}\sum_{p+q+r=2}e^{\pm\beta p_ia_i}-e^{\pm\beta(m-\frac{1}{2})}
  (e^{\pm\beta a}+e^{\pm\beta b}+e^{\pm\beta c})+1\right)\ .\nonumber
\end{eqnarray}
$p,q,r$ inside the summations are nonnegative integers satisfying the constraint.
The sum for general $|s|\geq 3$ follows the patten for $s=\pm 3$ above.
To write them down neatly, let us define the rank $n$ homogeneous polynomial $P_n$ by
\begin{equation}
  P_n(x,y,z)\equiv\sum_{p+q+r=n}x^py^qz^r\ .
\end{equation}
Then, the sum (\ref{pert-sum}) for $|s|\geq 3$ is given by
\begin{equation}
  -\left[P_{|s|}-e^{\mp\beta(m-\frac{1}{2})}P_{|s|-1}
  -e^{\pm\beta(m-\frac{1}{2})}P_{|s|-2}+P_{|s|-3}\right]\ ,
\end{equation}
where the arguments of $P_n$ are all
$(x,y,z)=(e^{\pm\beta a}, e^{\pm\beta b}, e^{\pm\beta c})$,
and upper/lower signs are for $s\gtrless 0$, respectively.
The corresponding determinants are easily obtained by
multiplying the fugacity $e^{-i(\lambda+i\beta\sigma)}$ for the gauge charges, and
taking $PE$. As the effective fluxes $s_{ij}$ and $s_{ji}$ only take
relative minus signs, it is convenient to combine the two contributions from
these mutually conjugate modes. For instance, we can order the $s_i$ fluxes in a
non-increasing way, $s_1\geq s_2\geq\cdots\geq s_N$. Then the mode with $i<j$ has $s_{ij}>0$,
and the conjugate mode with $i>j$ has $s_{ji}=-s_{ij}<0$.
Let us take $\alpha_{ij}=e_i-e_j$ to be positive roots when $i<j$.
Then, the total perturbative determinant in the background of anti-self-dual
flux $s$ is given by
\begin{equation}
  \hspace{-.3cm}Z_{\rm pert}^{(1)}Z_{\rm pert}^{(2)}Z_{\rm pert}^{(3)}
  =\prod_{\alpha\in\Delta_+}\frac{\prod_{\sum_{i=1}^3p_i=\alpha(s)}
  2\sin\frac{\alpha(\lambda+i\beta\sigma)+i\beta p_ia_i}{2}\cdot
  \prod_{\sum_{i=1}^3p_i=\alpha(s)-3}
  2\sin\frac{\alpha(\lambda+i\beta\sigma)+i\beta p_ia_i}{2}}
  {\prod_{\sum_{i=1}^3p_i=\alpha(s)-1}
  2\sin\frac{\alpha(\lambda+i\beta\sigma)+i\beta p_ia_i-i\beta\hat{m}}{2}
  \cdot\prod_{\sum_{i=1}^3p_i=\alpha(s)-2}
  2\sin\frac{\alpha(\lambda+i\beta\sigma)+i\beta p_ia_i+i\beta\hat{m}}{2}}
\end{equation}
where we defined $\hat{m}\equiv m+n=m-\frac{1}{2}$. If there are modes (positive roots)
for which $\alpha(s)=0,1,2$, the product with $\sum_{i=1}^3p_i=\alpha(s)-n$ is absent
when $\alpha(s)-n<0$, as the consequence of (\ref{pert-sum-low}). Note that, for
any given $\alpha\in\Delta_+$, the number of sine functions in the numerator minus
the number of sine functions in the denominator is always $2$.

Recall that $\sigma=\frac{4s}{1-\xi}$. In our organization of the $s$ fluxes,  one
finds that $\alpha(\sigma)\leq 0$ since $\xi>1$ in our localization. Below,
we shall find that the value of $\xi$ will not affect the final result of the
$\lambda$ integral as long as $\xi>1$, which is the condition of
the safe path integral contour.

The instanton part on $\mathbb{R}^4\times S^1$ that we take to construct
our $Z^{(1,2,3)}_{\rm inst}$ is given as follows. The $U(N)$ result on
$\mathbb{R}^4\times S^1$ takes the following form,
\begin{equation}
  Z_{\rm inst}(\epsilon_{1,2},m,\lambda,q)=\sum_{k=1}^\infty q^k
  Z_k(\epsilon_{1,2},m,\lambda)\ ,
\end{equation}
with $Z_0=1$ by definition. Apart from the parameters $\epsilon_{1,2}$, $m$, $\lambda$
explained above, $q=e^{-\beta(1+c)}$ is understood at the third fixed point.
$Z_k$ is given by a sum over $N$-colored Young diagrams ($Y_1,Y_2,\cdots,Y_N$) with $k$
boxes \cite{Nekrasov:2002qd,Bruzzo:2002xf}. The expression is given by
\begin{equation}\label{Zk}
  Z_k=\sum_{Y;\ \sum_i|Y_i|=k}\
  \prod_{i,j=1}^N\prod_{s\in Y_i}
  \frac{\sinh\frac{\beta(E_{ij}+m_0-\epsilon_+)}{2}
  \sinh\frac{\beta(E_{ij}-m_0-\epsilon_+)}{2}}
  {\sinh\frac{\beta E_{ij}}{2}\sinh\frac{\beta(E_{ij}-2\epsilon_+)}{2}}
\end{equation}
with
\begin{equation}
  E_{ij}=i(\lambda_i-\lambda_j)-\epsilon_1h_i(s)+\epsilon_2(v_j(s)+1)\ .
\end{equation}
Here, $s$ labels the boxes in the $i$'th Young diagram $Y_i$.
$h_i(s)$ is the distance from the box $s$ to the edge on the right side of
$Y_i$ that one reaches by moving horizontally to the right. $v_j(s)$ is the distance from
$s$ to the edge on the bottom side of $Y_j$ that one reaches by moving down vertically
(which may be negative if one has to move up). See \cite{Bruzzo:2002xf,Kim:2011mv} for
more detailed explanations. From this, to obtain $Z^{(3)}_{\rm inst}$, one replaces
\begin{equation}
  (\epsilon_1,\epsilon_2,m_0,\lambda,q)\rightarrow\left(a-c,b-c,m+n(1+c),
  \lambda+is\beta c+i\beta\sigma,e^{-\beta(1+c)}\right)\ .
\end{equation}
$Z^{(1)}_{\rm inst}$, $Z^{(2)}_{\rm inst}$ are obtained by similar replacements,
changing the roles of $a,b,c$ by cyclic permutations.

Let us explain one curious aspect of the index (\ref{final}). The factor
$e^{\frac{\beta}{2}\sum_{i=1}^Ns_i^2}$ that appears in the summation of
$s_i$'s might look very dangerous and physically unacceptable, as it seems to
say that large $s$ fluxes lead to indefinitely negative `energies,' in particular
making the sum over $s_i$ to be divergent. As we shall explain in section 3,
with the correct contour integral rules for $z_i=e^{-i\lambda_i}$, the integral
over $z_i$ is nonzero only for a finitely many $s$ flux configurations.
In particular, the state appearing in the index with lowest energy (the vacuum)
will come with nonzero $s$ flux. The `energy' $-\frac{1}{2}\sum_{i=1}^Ns_i^2$
with this nonzero flux will be a contribution to the negative vacuum `energy.'

Let us finish this subsection by explaining the convenient basis and scalings
of the chemical potentials, that we shall use in the remaining part of this paper.
For a conceptual purpose in the next subsection, and also for a practical purpose in
the later sections to easily expand the index with fugacities, let us focus
on the special scaling regime of the four fugacity variables $\beta$, $\beta m$,
$\beta a$, $\beta b$ (with $\beta c=-\beta a-\beta b$). Namely, we take $\beta\gg 1$,
$\beta a_i\ll\beta$ and $\beta\hat{m}\equiv\beta(m+n)\ll \beta$. This is taking
three combinations of chemical potentials to be much smaller than $\beta$,
but it should not be confused with turning them off. We keep all four of them,
without losing any information in the index, but just prescribe the order of
expansion parameters for convenience. A technical
benefit (or convenience) of this ordering is as follows. In a single factor of
$Z^{(i)}_{\rm inst}Z^{(i)}_{\rm inst}$, if one takes the basis of $4$ fugacities
to be
\begin{equation}
  q=e^{-\beta}\ll 1\ ,\ \ y=e^{\beta\hat{m}}=e^{\beta(m+n)}\ ,\ \
  y_i=e^{-\beta a_i}\ ,
\end{equation}
then the first fugacity $e^{-\beta}$ appears only through the instanton
number expansion in $Z_{\rm inst}^{(i)}=\sum_{k=0}^\infty Z_ke^{-k\beta(1+a_i)}$,
namely, as $e^{-k\beta}$ in $e^{-k\beta(1+a_i)}$. Inside $Z_k$, or in $Z_{\rm pert}^{(i)}$,
$e^{-\beta}$ appears only through the relation between the mass $m_0$ on
$\mathbb{R}^4\times S^1$ on $m$ given by
$e^{\beta m_0}=e^{\beta m}e^{\beta n}e^{\beta na_i}$ in the $\sinh$ factors on
the numerators. Having redefined $e^{\beta\hat{m}}=e^{\beta(m+n)}\gg e^{-\beta}$
to be one of the four expansion parameters, $e^{-\beta}\ll 1$ appears only through
$e^{-k\beta(1+a_i)}$, as claimed. Thus, in this setting, one can identify the
expansion with $e^{-\beta}$ as being made with the total number of self-dual
instantons localized at the three fixed points. This is as expected,
as the measure $\mathcal{E}$ in the index can be written as
\begin{equation}
  \mathcal{E}=E-\frac{R_1+R_2}{2}-m(R_1-R_2)+a_ij_i=k-(m+n)(R_1-R_2)+a_ij_i+\{Q,S\}\ ,
\end{equation}
where $k=j_1+j_2+j_3+\frac{3}{2}(R_1+R_2)+n(R_1-R_2)$ and $\{Q,S\}=E-2(R_1+R_2)-(j_1+j_2+j_3)$.
So changing the basis of chemical potentials from $\beta m$ to $\beta\hat{m}$ makes
$\beta$ to be conjugate to the instanton number $k$ for the BPS excitations with
$\{Q,S\}=0$. Including the contribution from anti-self-dual fluxes, extra factor
of $e^{-\beta}$ appears in the classical factor $e^{\frac{\beta}{2}\sum_{i=1}^Ns_i^2}$,
whose minimum corresponds to the vacuum `energy.' If one includes the vacuum energy
contribution, we cannot say $E=2(R_1+R_2)+j_1+j_2+j_3$. However, we can still say
that $e^{-\beta}$ is conjugate to the total instanton number including the above
classical factor, coming from both $F^+$ at the fixed points and also from
the uniform $F^-=2sJ$ fluxes. This can be shown by
\begin{eqnarray}
  k&\equiv&\frac{1}{8\pi^2}\int_{\mathbb{CP}^2}{\rm tr}F\wedge F=
  \frac{1}{8\pi^2}\int_{\mathbb{CP}^2}{\rm tr}F^+\wedge F^+ +\frac{1}{8\pi^2}\int_{\mathbb{CP}^2}{\rm tr}F^-\wedge F^-\\
  &=&k_{SD}+\frac{1}{2\pi^2}\sum_{i=1}^Ns_i^2\ \int_{\mathbb{CP}^2}J\wedge J
  =k_{SD}-\frac{1}{\pi^2}\sum_{i=1}^Ns_i^2\ {\rm vol}(\mathbb{CP}^2)=
  k_{SD}-\frac{1}{2}\sum_{i=1}^Ns_i^2\ .\nonumber
\end{eqnarray}
Here, at the second step on the first line, we used the fact that the self-dual
fluxes are singularly localized on the fixed points, and on the second line
we used the volume ${\rm vol}(\mathbb{CP}^2)=\frac{\pi^2}{2}$, when it appears
as the base of the Hopf fibration of $S^5$ of unit radius. The last expression
$-\frac{1}{2}\sum_{i=1}^Ns_i^2$, which is the contribution of $F^-$ to the instanton
number, is exactly what we had from the classical factor in our index. So the
`instanton charge' defined in the above sense (as a topological charge in
5d QFT) can be interpreted as $k=j_1+j_2+j_3+\frac{3}{2}(R_1+R_2)+n(R_1-R_2)$ defined
in the 6d perspective for excited states plus the vacuum `energy' contribution as
captured by the index.

So to summarize, in the above parametrization of chemical potentials,
$e^{-\beta}$ is the fugacity conjugate to the instanton charge $k$. As our
calculus of the index
is naturally decomposed into the instanton sums, this is a very convenient basis
for concrete studies. Setting $q\ll e^{\beta\hat{m}},e^{\beta a_i}$, one can even
make the instanton number to take the highest charge cost in the index than other
charges, which happens in all semi-classical instanton calculus. In many ways,
$k$ will be playing the role of `energy level' of states in the index. Also,
setting $q\ll e^{\beta\hat{m}},e^{\beta a_i}$ will let us to understand the contour
integral prescriptions in the next subsection more clearly.

\subsection{The contour integral}

Let us finally explain the integral over the holonomies $\lambda_i$.
It might very naturally look that the integral over $z_i\equiv e^{-i\lambda_i}$
has to be along the unit circle on the complex plane. However, there is
a subtlety that one has to consider carefully, which we explain now.

At given `energy,' or the instanton number $k$ in our setting, we expect the index to be
a finite polynomial in all fugacities but $e^{-\beta}$ (the main `temperature-like'
fugacity, conjugate to $k$). This will be concretely illustrated below, using the 6d
unitarity bounds from $OSp(8^\ast|4)$ superconformal algebra. One subtlety
in our calculation is that, although
each part $Z_{\rm pert}^{(i)}Z_{\rm inst}^{(i)}$ for $i=1,2,3$ takes a manifest index
form, it appears in an infinite series in the rest of the fugacity variables at
given $k$, most importantly in $e^{\beta a_i}$. The
way it can eventually provide the correct $S^5\times S^1$ index in a finite polynomial
at given $k$ is as follows. Firstly, we combine three such factors after which the
three infinite sums are partially canceled. Then the integral over $\lambda_i$ also
projects the states
into a `gauge invariant' subset, eventually yielding a finite polynomial. So with the
infinitely many
`fictitious' BPS states that one observes in the $\mathbb{R}^4\times S^1$ partition function
$Z_{\rm pert}^{(i)}Z_{\rm inst}^{(i)}$ (with $\lambda$ kept), one should first decide how
to correctly expand the factors in the denominators as geometric series in the fugacities.
It might appear that, with real
$a_i$'s, expanding in $e^{\pm \beta a_i}$ with one of the two signs which is smaller than
$1$ will do. This is not the case. Even if we expand `formally' with a fugacity which
is larger than $1$, the gauge-invariant projection and combination of all three fixed
point contributions can make the infinite series canceled. If one wishes, one could
have taken all the chemical potentials (except $\beta$) to be purely imaginary, to
highlight the ambiguity on how to expand an $\mathbb{R}^4\times S^1$ partition function.
The fugacities $e^{\beta a_i}$ are then phases. We assume so, at least as an
intermediate step during calculation.
We can of course continue back to real $a_i$ after all calculations are done.

To summarize, we encounter an expression (\ref{final}) in which
one should carefully expand each $Z_{\rm pert}^{(i)}Z_{\rm inst}^{(i)}$ for
$i=1,2,3$. All factors in the denominator to be expanded take the form of
\begin{equation}\label{denom-expand}
  \frac{1}{1-t}\ ,
\end{equation}
where $t$ is a combination $e^{-\beta(p\epsilon_++q\epsilon_-)}z$ with $p>0$.
This can be expanded in two ways:
\begin{equation}
  \left[\frac{1}{1-t}\right]_+=1+t+t^2+\cdots\ \ ,\ \ \
  \left[\frac{1}{1-t}\right]_-=\left[-\frac{t^{-1}}{1-t^{-1}}\right]_-
  =-(t^{-1}+t^{-2}+\cdots)\ .
\end{equation}
One finds that the difference between the two expansions is
\begin{equation}\label{delta-expand}
  \left[\frac{1}{1-t}\right]_+-\left[\frac{1}{1-t}\right]_-=
  \sum_{n=-\infty}^\infty t^n=2\pi\delta(\theta)\ ,
\end{equation}
where $t\equiv e^{i\theta}$. Thus, as long as $t$ does not include the integral
variable $\lambda$, or $z$, the difference is zero for generic nonzero choice of
$\epsilon_\pm$, for which one can set $\delta(\theta)=0$. However, for non-Abelian
self-dual instantons, some factors of the form (\ref{denom-expand}) contain the holonomy
variable to be integrated, and taking different expansions could yield different results
due to the delta function in the integrand. This nature is also obvious when we naturally
take imaginary chemical potentials, as $z_i$ integrals on the unit circle hit poles.
Thus, the question of expanding the denominator is equivalent to that of going
around the poles one hits on the unit circle contour.

This ambiguity in expanding $\mathbb{R}^4\times S^1$ determinant comes from the fact
that the local determinant calculation around a fixed point, using Nekrasov's result,
forgets some details on our initial problem. We explain below why, also providing
a way to restore this information and making the expression unambiguous.

In the determinant calculation around a fixed point, say the third fixed point,
one decomposes the effective energy $\mathcal{E}$ appearing in the index as
\begin{equation}
  \mathcal{E}=(1+c)k+\epsilon_1j_1+\epsilon_2j_2+\epsilon_+(R_1+R_2)
  -(m+n(1+c))(R_1-R_2)+\{Q,S\}\ .
\end{equation}
The first term $(1+c)k$ is accounted for by the classical instanton action, and
the last term $\{Q,S\}$ can be ignored for BPS excitations. So in the measure,
$e^{-\beta\mathcal{E}}$, the rest of the terms
\begin{equation}\label{measure-local}
  e^{-\beta\epsilon_-(j_1-j_2)}e^{-\beta\epsilon_+(R_1+R_2+j_1+j_2)}
  e^{\beta(m+n(1+c))(R_1-R_2)}
\end{equation}
provide weights to the BPS quadratic fluctuation modes which contribute to the
$\mathbb{R}^4\times S^1$ determinant. In the original 6d index, one finds various
non-negativity conditions for the charges from the unitarity bound and the BPS
energy relation $E_{BPS}=2(R_1+R_2)+j_1+j_2+j_3$. (We can use the latter relation
because all modes appearing in the determinant are BPS). Consider
the following unitarity bounds from the 6d algebra
\begin{equation}\label{unitarity}
  \{Q^{\pm\pm}_{\mp\mp\mp},S^{\mp\mp}_{\pm\pm\pm}\}\geq 0\ \rightarrow\ \ E_{BPS}\geq
  2(\pm R_1\pm R_2)\pm j_1\pm j_2\pm j_3
\end{equation}
with various signs, acting on BPS modes. The two signs in front of $R_1,R_2$ are independent,
while the signs in front of $j_1,j_2,j_3$ are constrained to have their product
to be $+1$ from the chirality of supercharges. From these, one can derive
\begin{equation}
  R_1\geq 0\ ,\ \ R_2\geq 0\ ,\ \ j_1+j_2\geq 0\ ,\ \ j_2+j_3\geq 0\ ,\ \ j_3+j_1\geq 0\ ,
\end{equation}
for the BPS modes. This requires that the sign of $R_1+R_2+j_1+j_2$ charge
conjugate to $\epsilon_+$ in (\ref{measure-local}) be non-negative, providing
a definite way to
expand the denominators for each $\mathbb{R}^4\times S^1$ determinant. So
one has to expand each factor in the denominator in a positive series in
$e^{-\beta\epsilon_+}$ at each fixed point. Note that this is where the local
$\mathbb{R}^4\times S^1$ calculation forgets the original problem of our index
and exhibits expansion ambiguities. Our sign constraints on the charges
come from the superconformal algebra, while the Nekrasov's results is obtained
from a flat space QFT with 5d $\mathcal{N}=1$ Poincar\'e SUSY and is thus ignorant
on the unitarity bounds for charges in our problem. Of course, if one had done the
localization calculus much more carefully than we did here, all prescriptions that
we obtain here could have been derived rigorously, without an ambiguity at any step.
Here we simply restore this information by remembering our original problem, which
we think is sufficiently convincing.

Perhaps one might think that the positivity of $R_1+R_2+j_1+j_2$ may not be
requiring positive expansions in $e^{-\beta\epsilon_+}$ at each fixed point,
for two possible reasons. Firstly, it might appear that, in the decomposition of
$\mathcal{E}$ into $(1+c)k$ and $\epsilon_+(R_1+R_2+j_2+j_2)$, $R_1+R_2+j_1+j_2$
appearing in the latter term does not have to be positive by itself because
$R_1+R_2$ and $j_1+j_2$ also appear in $(1+c)k$. So one might think that requiring
the net sum to be positive may suffice. Now let us consider $k$ in more detail,
here for the QFT with $n=\pm\frac{1}{2}$, which we shall mostly focus on in the rest of
this paper. One can easily show using the above charge unitarity bounds that
\begin{equation}\label{inst-charge-bound}
  k=j_1+j_2+j_3+\frac{3}{2}(R_1+R_2)+n(R_1-R_2)\geq R_1+R_2+\frac{j_1+j_2}{2}
  \geq\frac{R_1+R_2+j_1+j_2}{2}
\end{equation}
at $n=\pm\frac{1}{2}$. (Other values of $n$ will be briefly commented on below.)
So with given $k\geq 0$, the values of $R_1+R_2+j_1+j_2$ carried by the classical
factor $e^{-k\beta(1+c)}$ is bounded from above. One the other hand, if we expand the
$\mathbb{R}^4\times S^1$ determinant with $e^{-\beta\epsilon_+}$ in the wrong way,
the expansion can pick an arbitrarily large value of this charge with the negative sign,
making the net value of $R_1+R_2+j_1+j_2$ to be negative. So the expansion has to be made
properly. Secondly, one might think that the (partially) wrong expansion at a given fixed
point
with negative $R_1+R_2+j_1+j_2$ could be canceled against other wrong terms from different
fixed points in a subtle way. With examples discussed in section 3, we find that
such a thing cannot happen. Namely, one finds that the above charge unitarity bounds
are satisfied only when one expands all fixed point determinants in the positive powers
of their own $e^{-\beta\epsilon_+}$.

Thus, the restoration of the forgotten information on our original problem into the
$\mathbb{R}^4\times S^1$ determinants provides a unique way to expand the infinite
series in the integrand. One can rephrase the rule in a
more convenient way. As the integrand is a holomorphic function of $z_i=e^{-i\lambda_i}$,
the way one expands a factor like (\ref{denom-expand}) can be phrased as constraining
the integral contours on the complex $z_i$ planes. Namely, first regard all three
$\epsilon_+$ variables at the fixed points as `formally' being positive. Then, keeping
all the residues at poles which are inside the unit circles of all $z_i$'s yields the
correct integral as determined in the previous paragraph. Note that, due to the constraint
$a+b+c=0$, it is impossible to have all three $e^{-\beta\epsilon_+}$ to be smaller than $1$.
Thus, the above prescription for the integral is nontrivially deforming the contour
away from the unit circle, to include the desired poles only.

A similar expansion ambiguity has been observed in the non-relativistic superconformal
indices for the instantons' superconformal quantum mechanics \cite{Kim:2011mv}, describing
a DLCQ M5-brane theory \cite{Aharony:1997th,Aharony:1997an}. There, two possible expansions
with $\epsilon_+>0$ and $\epsilon_+<0$ yield different indices counting different BPS
sectors.

Once the $\lambda$ integrations are done, we could still have infinite series in
$e^{\beta a_i}$ expansions before combining contributions from all the fixed points.
As explained in (\ref{delta-expand}) and below, expansion prescription does not change
the answer after $\lambda$ integral is finished, since the right hand side of
(\ref{delta-expand}) is zero without $\lambda_i$ variables being involved.
So one can unambiguously combine the remaining infinite series to finally obtain
finite polynomials in $e^{\beta a_i}$ at a fixed order in $e^{-\beta}$, free of
expansion ambiguities. It does not look obvious at all, from the expressions we have,
that the final index will indeed become a finite polynomial in $e^{\beta a_i}$ at
given $k$ (although it should be, if we correctly calculated the 6d index).
The detailed illustration of the last statement will be given with some concrete
examples in the next section.

Minor subtleties in the above discussions, which we do not attempt to clarify in
full detail, are as follows. Firstly, in deriving the unitarity bounds (\ref{unitarity}),
we used the supercharges which are expected to exist only at $K=1$ with SUSY enhancement.
As our main interest is the theory and index at $K=1$, we do not consider this question
further. Also, in (\ref{inst-charge-bound}),
$n=\pm\frac{1}{2}$ was used to show that $k\geq 0$ provides an upper bound for
$R_1+R_2+j_1+j_2$. At $|n|=\frac{3}{2}$, similar analysis provides an insufficient
bound $j_1+j_2\leq 2k$. The situation is worse for $|n|>\frac{3}{2}$. So perhaps
the arguments for the integration contour may be subtler for $|n|\geq\frac{3}{2}$.
Again, as our main interest is the simplest theories at $n=\pm\frac{1}{2}$, which
anyway provides the 6d index at $K=1$, we do not further consider this issue either.
We consider the 5d index with general value of $n$ only in the Abelian case in the
next section, in which the issue of contour choice is absent.

\section{Tests }

\subsection{The Abelian index}

In this subsection, we compare the index we got from $\mathbb{CP}^2\times S^1$ with
the known index for the Abelian 6d $(2,0)$ theory. Only in this subsection, we discuss
the family of QFT$_n$ labeled by half-integral $n$'s other than $n=\pm\frac{1}{2}$.

The 6d index is given by \cite{Bhattacharya:2008zy,Kim:2012qf}
\begin{equation}
  PE\left[\frac{e^{-\frac{3}{2}\beta}(e^{\beta m}+e^{-\beta m})-e^{-2\beta}
  (e^{\beta a}+e^{\beta b}+e^{\beta c})+e^{-3\beta}}
  {(1-e^{-\beta(1+a)})(1-e^{-\beta(1+b)})(1-e^{-\beta(1+c)})}\right]
\end{equation}
in our convention. The allowed ranges of the parameters are
\begin{equation}
  -\frac{3}{2}<m<\frac{3}{2}\ ,\ \ -1<a,b,c<1\ \ \ ({\rm subject\ to}\ a+b+c=0)\ .
\end{equation}
We shall reproduce this index from our 5d indices.

Let us reconsider the perturbative and instanton corrections with general $n$. As for
the perturbative part, one should combine the three factors
\begin{equation}
  PE\left[\frac{1}{2}\Big(
  I_+(b\!-\!a,c\!-\!a,m\!+\!n(1\!+\!a))+I_+(c\!-\!b,a\!-\!b,m\!+\!n(1\!+\!b))
  +I_+(a\!-\!c,b\!-\!c,m\!+\!n(1\!+\!c))\Big)+1\right]\ ,
\end{equation}
where the last $+1$ is the contribution from a bosonic zero mode in the ghost multiplet
\cite{Pestun:2007rz,Kim:2012qf}. For $n=-\frac{1}{2}$, we already showed that
the sum inside $PE$ is $0$, making the net perturbative part to be $1$.
More generally, for other half-integral $n$ with $|n|\geq\frac{3}{2}$, one obtains
\begin{equation}
  Z_{\rm pert}=PE\left[\frac{1}{2}\left(e^{\mp\beta(m+n)}P_{|n|-\frac{3}{2}}
  (e^{\beta a_i})+e^{\pm\beta(m+n)}P_{|n|-\frac{3}{2}}(e^{-\beta a_i})\right)\right]\ ,
\end{equation}
where upper/lower signs are for $n\lessgtr0$, respectively. Here, the two terms
appearing in the $PE$ are conjugate to each other, as one part changes into another
by sign flips of all charges. Since $|m|<\frac{3}{2}$ and $|n|\geq\frac{3}{2}$,
one of the two terms contains a fugacity which is greater than $1$. Thus from
the precise definition of $PE$ that we explained around (\ref{PE-final}), one obtains
\begin{equation}\label{abelian-pert}
  Z_{\rm pert}=PE\left[e^{-\beta(|n|\mp m)}P_{|n|-\frac{3}{2}}(e^{-\beta a_i})\right]\ .
\end{equation}

As for the instanton part, the $U(1)$ $Z_{\rm inst}$ on $\mathbb{R}^4\times S^1$
is simply rearranged as \cite{Iqbal:2008ra}
\begin{equation}
  PE\left[I_-(\epsilon_{1,2},m_0)\frac{q}{1-q}\right]\ ,
\end{equation}
where
\begin{equation}
  I_-=\frac{\sinh\frac{\beta(m_0+\epsilon_-)}{2}\sinh\frac{\beta(m_0-\epsilon_-)}{2}}
  {\sinh\frac{\beta\epsilon_1}{2}\sinh\frac{\beta\epsilon_2}{2}}
\end{equation}
is the index for a single superparticle preserving the SUSY of a half-BPS instanton
particle on $\mathbb{R}^{4,1}$ \cite{Kim:2011mv}. This index satisfies a useful property
$I_-=I_++1$. Inserting $q=e^{-\beta(1+a_i)}$ and appropriate values of $\epsilon_{1,2},m_0$
at the three fixed points, one finds the following properties.
Firstly, at $n=\mp\frac{1}{2}$, one finds
\begin{equation}
  Z_{\rm inst}^{(1)}Z_{\rm inst}^{(2)}Z_{\rm inst}^{(3)}=
  PE\left[\frac{e^{-\frac{3}{2}\beta}(e^{\beta m}+e^{-\beta m})-e^{-2\beta}
  (e^{\beta a}+e^{\beta b}+e^{\beta c})+e^{-3\beta}}
  {(1-e^{-\beta(1+a)})(1-e^{-\beta(1+b)})(1-e^{-\beta(1+c)})}\right]\ .
\end{equation}
The right hand side is the 6d index. Since $Z_{\rm pert}=1$ at $n=\mp\frac{1}{2}$,
this proves that the correct Abelian index is obtained for these values of $n$.
As our $PE$ given by (\ref{PE-precise}) includes the factor of Casimir `energy,'
one should also discuss it which is implicit in the above expression.
This factor is discussed in detail in section 5.1.
Then, for other $n$'s satisfying $|n|\geq\frac{3}{2}$, we found
\begin{eqnarray*}
  \hspace*{-0.5cm}Z_{\rm inst}^{(1)}Z_{\rm inst}^{(2)}Z_{\rm inst}^{(3)}=
  PE\left[-e^{-\beta(|n|\mp m)}P_{|n|-\frac{3}{2}}(e^{-\beta a_i})\right]
  PE\left[\frac{e^{-\frac{3}{2}\beta}(e^{\beta m}+e^{-\beta m})-e^{-2\beta}
  (e^{\beta a}+e^{\beta b}+e^{\beta c})+e^{-3\beta}}
  {(1-e^{-\beta(1+a)})(1-e^{-\beta(1+b)})(1-e^{-\beta(1+c)})}\right]
\end{eqnarray*}
which we checked for several values of $n$'s with a computer. Presumably one
should be able to prove this equation in full generality by a more closer look
at the expressions, which we did not attempt. (It can be easily proven with a
computer at various values of $n$, which was enough to convince us.) Since the first
factor on the right hand side is the inverse of the perturbative index (\ref{abelian-pert}),
this proves that the full index on $\mathbb{CP}^2\times S^1$ agrees with the
known 6d index.

Thus, at least in the Abelian case where several subtle aspects of section 2.3
are absent, we proved that the 5d QFT's on $\mathbb{CP}^2\times\mathbb{R}$ with
various values of $n$ capture the correct index of the 6d theory.

One general comment about the subtlety of this 5d QFT approach should be made
here, which also applies to the non-Abelian theories and their indices.
Namely, this calculation acquires crucial contributions from Abelian self-dual
instantons. Self-dual instantons do not exist in 5d QFT as a regular field theory
configuration. However, without instantons, it would be hopeless to reproduce the 6d KK
modes' contribution to the index. So the instanton contribution in this subsection should
all come from singular field configurations. The implicit way we
ended up including their contributions to our index is as follows.

Apparently, after reducing our path integral to the determinant calculus of a QFT
on $\mathbb{R}^4\times S^1$, we borrowed the results of \cite{Nekrasov:2002qd} which
contain the singular instantons' contributions. More generally, in the non-Abelian
case, the moduli space of smooth instanton configurations has regions in
which the instantons' sizes are small, and exhibits singularities.
So an unambiguous calculation can be done only after giving certain `UV prescriptions'
near these singular field theory configurations. For instantons on the flat space,
one such prescription for the $U(N)$ gauge theory was introducing an anti-self-dual
noncommutative deformation to the QFT in the spatial $\mathbb{R}^4$ part
\cite{Seiberg:1999vs}. This makes the instanton moduli space smooth by giving nonzero
minimal sizes to instantons \cite{Nekrasov:1998ss}. In particular, this deformation
makes the $U(1)$ instantons into regular field theory solitons. The $\mathbb{R}^4\times S^1$
partition function of \cite{Nekrasov:2002qd} may be understood as acquiring contributions
from these configurations, although the value of non-commutative deformation parameter
actually does not appear in the final partition function. By (somewhat blindly) borrowing
these results into our determinant calculations, we could be implicitly providing certain
UV prescriptions beyond the naive 5d QFT. Similar issues exist for the partition function
on $S^5$. It is not clear to us at the moment what the physical meaning of this
implicit prescription could be. (We do not even have an a priori justification
on this either.) Apart from the fact that this is a `technically natural' resolution
of the moduli space singularities for instantons in flat space, the results in this
subsection (for $U(1)$ theory) and the rest of this section (for non-Abelian theory) could
be regarded as more practical supports for our implicit UV prescription. A more physical
understanding on this issue would definitely be desirable.

\subsection{Some unrefined indices}

In \cite{Kim:2012av,Kim:2012qf}, it was shown that the non-Abelian index
for the 6d theory simplifies when one tunes the chemical potentials to
$m=\pm\frac{1}{2}$, $a=b=c=0$. From the viewpoint of the $S^5$ partition function,
the path integral uses the maximal SYM action as the measure. This can be easily
understood, as the unrefined indices for $a=b=c=0$ and $m=\pm\frac{1}{2}$ are given by
\begin{equation}
  {\rm tr}\left[(-1)^Fe^{-\beta(E-R_1)}\right]\ ,\ \   {\rm tr}\left[(-1)^Fe^{-\beta(E-R_2)}\right]\ ,
\end{equation}
respectively, and each measure inside the trace commutes with $16$ of the
$32$ SUSY of the 6d $(2,0)$ theory. The two indices count different BPS subsectors,
with different boson-fermion cancelation structures, but the functional form of the
index is the same. The result known from the $S^5$ partition function is
\begin{equation}
  e^{\beta\left(\frac{N(N^2-1)}{6}+\frac{N}{24}\right)}
  \prod_{n=0}^\infty\prod_{s=1}^N\frac{1}{1-e^{-\beta(n+s)}}
\end{equation}
for the $U(N)$ gauge group. The simplification which admits one to get
this explicit result is the extra cancelation between the 5d vector multiplet and
hypermultiplet. In fact, the enhanced SUSY and simplification of the $S^5$
partition function happens more generally. For instance, let us just impose
$m=\frac{1}{2}-c$, and keep the other three chemical potentials unconstrained.
(Two more unrefinement limits, replacing $c$ by $a$ or $b$, are similar.)
In this case, one expects a SUSY enhancement from $2$ to $4$:
\begin{equation}
  Q^{++}_{---}\ \rightarrow\ \ Q^{++}_{---}\ ,\ Q^{+-}_{--+}
\end{equation}
and similar enhancement of the conjugate $S$ supercharge.
The extra SUSY $Q^{+-}_{--+}$, $S^{-+}_{++-}$ mix the fields in the vector
multiplet and the hypermultiplets, significantly simplifying the $S^5$ partition
function. The result is
\begin{equation}\label{unrefined-S5}
  e^{\beta\frac{N(N^2-1)}{24}\frac{1+c}{(1+a)(1+b)}
  \left(2+a+b\right)^2+\frac{N}{24}\beta(1+c)}
  \prod_{n=0}^\infty\prod_{s=1}^N\frac{1}{1-e^{-\beta(1+c)(n+s)}}\ .
\end{equation}
Namely, the index part takes the same form, simply with a replacement $\beta\rightarrow\beta(1+c)$. The overall `Casimir energy' factor, which is
rather complicated, will be briefly discussed in section 5.1.

In this subsection, we show that our $\mathbb{CP}^2\times S^1$ index
reproduces this known result. At $m=\frac{1}{2}-c$, $Z_{\rm pert}$ and $Z_{\rm inst}$
in the $\lambda$ integrand simplify as follows. Firstly, on the third fixed point,
one finds that
\begin{equation}
  \epsilon_+-m_0=-\frac{3c}{2}-m+\frac{1+c}{2}=0\ .
\end{equation}
At $\epsilon_+=\pm m_0$, one finds that $I_+=0$ and $I_-=I_++1=1$.
The former condition implies that $Z^{(3)}_{\rm pert}=1$. The latter condition,
together with similar simplifications of various $\frac{\sinh\sinh}{\sinh\sinh}=1$
factors in (\ref{Zk}) at $\epsilon_+=\pm m_0$ , one finds that
\begin{equation}
  Z^{(3)}_{\rm inst}=\prod_{n=1}^\infty\frac{1}{(1-e^{-n\beta(1+c)})^N}\ .
\end{equation}
This is simply $\eta(e^{-\beta(1+c)})^{-N}$ upon multiplying a factor
$e^{\frac{N\beta(1+c)}{24}}$.
On the other hand, at the first and second fixed points, one finds that
\begin{eqnarray}
  {\rm 1st}&:&\epsilon_--m_0=\frac{b-c}{2}-m+\frac{1+a}{2}=0\nonumber\\
  {\rm 2nd}&:&\epsilon_-+m_0=\frac{c-a}{2}+m-\frac{1+b}{2}=0
\end{eqnarray}
at $m=\frac{1}{2}-c$. At $\epsilon_-=\pm m_0$, one finds that $I_-=0$ and
$I_+=I_--1=-1$. From the former condition, one obtains
$Z_{\rm inst}^{(1)}=Z_{\rm inst}^{(2)}=1$. The latter condition implies that
the perturbative contributions at the two fixed points are
\begin{equation}
  Z_{\rm pert}^{(1)}=\prod_{i<j}
  2\sin\frac{\lambda_{ij}+i\beta\sigma_{ij}+is_{ij}\beta a}{2}\ ,\ \
  Z_{\rm pert}^{(2)}=\prod_{i<j}
  2\sin\frac{\lambda_{ij}+i\beta\sigma_{ij}+is_{ij}\beta b}{2}\ ,
\end{equation}
where $\sigma=\frac{4s}{1-\xi}$ is the saddle point value of the scalar,
with $\xi>1$. To obtain the final results, we divided the $+1$ contribution
inside the $PE$ coming from the zero mode in the ghost multiplet into
$+\frac{1}{2}+\frac{1}{2}$, and combined each
$\frac{1}{2}$ with $Z_{\rm pert}^{(1)}$ or $Z_{\rm pert}^{(2)}$.
From the point of view of the physics on $\mathbb{R}^4\times S^1$, the
extra $2$ SUSY which appear at $m=\frac{1}{2}-c$ are manifested differently
in three fixed points. At the third fixed point, $I_+=0$ implies that
the enhanced SUSY become part of the half-BPS subset preserved by the
perturbative superparticle (W-boson). At the other two fixed points, they
are part of the half-BPS subset preserved by the instanton superparticle.

So we consider the following integral:
\begin{equation}
  \frac{Z_{\rm inst}^{(3)}}{N!}\oint[d\lambda_i]\sum_{\{s_i\}}
  e^{\frac{\beta}{2}\sum_is_i^2}e^{-i\sum_is_i\tilde\lambda_i}\prod_{i<j}
  \left(2\sinh\frac{i\tilde\lambda_{ij} -\beta s_{ij}a}{2}\cdot
  2\sinh\frac{i\tilde\lambda_{ij} -\beta s_{ij}b}{2}
  \right)\  ,
\end{equation}
where $\tilde\lambda_i \equiv \lambda_i + i\beta\sigma_i$. At the unrefinement point $m=\frac{1}{2}-c$, all possible poles in the
$z_i=e^{-i\lambda_i}$ planes disappear after cancelations, except for the
poles at the origin $z_i=0$ coming from $d\lambda_i=\frac{idz_i}{z_i}$ and the perturbative contribution.
So one can change the contours of $z_i$'s as long as they include the origins $z_i=0$.
We take the contours of $\tilde{z}_i=e^{-i\tilde\lambda_i} $
to be along the unit circles $|\tilde{z}_i|=1$.
Dropping the tilde of the new variables, we evaluate the integral, setting aside
the overall factor of $Z_{\rm inst}^{(3)}$ for a while. Using
\begin{equation}
  \prod_{i<j}2\sinh\frac{x_i-x_j}{2}=\sum_\sigma(-1)^\sigma\prod_j
  e^{\left(\frac{N+1}{2}-\sigma(j)\right)x_j}\ ,
\end{equation}
where $\sigma$ is the permutation of $N$ objects, one finds
\begin{eqnarray}\label{unrefined-calculation}
  (Z_{\rm inst}^{(3)})^{-1}I&=&\sum_{\{s_i\}}\sum_\sigma(-1)^\sigma
  \int\prod_{j=1}^N\left[\frac{d\lambda_j}{2\pi}
  e^{\frac{\beta}{2}s_j^2+i(N+1-j-\sigma(j)-s_j)\lambda_j
  +\beta(c\frac{N+1}{2}+ja+b\sigma(j))s_j}\right]\\
  &=&\sum_\sigma(-1)^\sigma\prod_j e^{\frac{\beta}{2}(N+1-j-\sigma(j))^2
  +\beta(c\frac{N+1}{2}+ja+\sigma(j)b)(N+1-j-\sigma(j))}\nonumber\\
  &=&e^{-\frac{\beta N(N+1)(N+2)}{6}+\frac{\beta cN(N+1)^2}{2}
  +c\frac{\beta N(N+1)(2N+1)}{6}-c\beta N(N+1)^2}
  \sum_\sigma(-1)^\sigma\prod_j e^{\beta(1+c)j\sigma(j)}\nonumber\\
  &=&e^{-\frac{\beta(1+c)N(N+1)(N+2)}{6}+\frac{\beta(1+c)N(N+1)^2}{4}}
  \sum_\sigma(-1)^\sigma\prod_j e^{-\beta(1+c)(\frac{N+1}{2}-\sigma(j))j}\nonumber\\
  &=&e^{\frac{\beta(1+c)N(N^2-1)}{12}}\prod_{m>n}2\sinh\frac{\beta(1+c)(m-n)}{2}
  =e^{\frac{\beta(1+c)N(N^2-1)}{6}}\prod_{n=1}^{N-1}(1-e^{-n\beta(1+c)})^{N-n}\ .
  \nonumber
\end{eqnarray}
So the full partition function is given by
\begin{equation}
  I=e^{\beta(1+c)\frac{N(N^2-1)}{6}}Z_{\rm inst}^{(3)}
  \prod_{n=1}^{N-1}(1-e^{-n\beta(1+c)})^{N-n}=
  e^{\beta(1+c)\frac{N^3-N}{6}}
  \prod_{n=0}^\infty\prod_{s=1}^N\frac{1}{1-e^{-\beta(1+c)(n+s)}}\ .
\end{equation}
The double product part, which contains the information on the excitations,
is the same as (\ref{unrefined-S5}) obtained from the $S^5$ partition function.
 The subtle discrepancy of the zero point charges is
discussed in section 5.1.

Note that the negative Casimir energy scaling like $N^3$ is also obtained. It is helpful
to trace where this $\frac{N(N^2-1)}{6}$ comes from. From (\ref{unrefined-calculation}),
the values of the anti-self-dual fluxes $s_i$ for the vacuum are given by
\begin{equation}
  (s_1,s_2,\cdots,s_N)=(N-1,N-3,\cdots,-(N-1))\ \ \ \ \ ({\rm namely}\ s_i=N+1-2i) \ ,
\end{equation}
or its permutations. So one obtains $-\sum_i\frac{s_i^2}{2}=-\frac{N(N^2-1)}{6}$.
So somehow the anti-self-dual instantons at the saddle point are contributing to
this vacuum energy, which scales like $N^3$ in the large $N$ limit.~\footnote{With roots $e_i-e_j$ with $i,j =1,...,N$ for $A_{N-1}$, the Weyl vector $\boldsymbol{\rho}$ is one half of the sum of all positive roots and so $2\boldsymbol{\rho} = (N-1)e_1+ (N-3) e_2+\cdots -(N-1)e_N$. The Weyl vector and its square have also appeared in the counting of 1/4 BPS junctions~\cite{Lee:2006gqa}.}
See section 5.1 for more discussions.

As explained in \cite{Kim:2012qf} (v2), the BPS states captured by this
index are rather simple. Writing $q=e^{-\beta(1+c)}$,
the index can be written as
\begin{equation}
  \prod_{s=1}^N\prod_{n=1}^\infty\frac{1}{1-q^{n+s}}=
  PE\left[\frac{q+q^2+\cdots+q^N}{1-q}\right]\ .
\end{equation}
Note that $q+q^2+\cdots+q^N$ is the partition function for the generator of
half-BPS states. Had it been, say, the 4d $\mathcal{N}=4$ Yang-Mills theory, these
$N$ generators could have been chosen as ${\rm tr}(Z^n)$ for $n=1,2,\cdots,N$
with a complex scalar. The factor $\frac{1}{1-q}$ comes from one of the three holomorphic
derivatives of $\mathbb{C}^3$ acting on these generators many times.
Taking the multi-particle excitations by doing $PE$, one obtains the full
partition function made out of one complex scalar and one kind of derivatives.
See \cite{Kim:2012qf} for the similar discussion for the $SO(2N)$ and $E_n$
gauge groups. Of course this gauge theory account would not be a precise
one in the case of 6d $(2,0)$ theory. But the presence of a similar sector in
the 4d $\mathcal{N}=4$ theory makes our findings rather natural. Also, in the
large $N$ limit, the above explanation makes a precise sense, replacing the
single trace operators above as single graviton particles. The gravity indeed
has such particle states in its Hilbert space. So not surprisingly,
the large $N$ limit of this index agrees with the gravity index \cite{Bhattacharya:2008zy},
as shown in \cite{Kim:2012av}.

\subsection{General finite $N$ indices and the $AdS_7$ gravity dual}

In this subsection, we study the low energy spectrum of the index at $n=-\frac{1}{2}$,
by expanding the quantity up to several orders in the fugacities. The instanton charge
$k=E-R_1$ in this case plays the role of `energy' with the scaling of chemical potentials
explained in section 2.2. By this, we mean that its conjugate fugacity $e^{-\beta}$ plays
the role of the main expansion parameter, meaning that other fugacities may be turned
off to $1$ but $e^{-\beta}$ should be kept  smaller than $1$. Of course, we shall keep
all $4$ of them in the low energy expansion.

In this subsection, let us denote by $k$ the instanton number of the excited state
minus that of the vacuum: so `$k=0$' for the vacuum, and so on.
It turns out that the form of the general index that we obtain will differ
depending on whether $k\leq N$ or $k>N$. At $k\leq N$, we will find
that our finite $N$ indices completely agree with the large $N$ supergravity
index on $AdS_7\times S^4$. This is natural as the gravity approximation of
M-theory is valid only at low energy compared to $N$. Strictly speaking,
the general supergravity spectrum is reliable only when $E\ll N$, or
in our case $k\ll N$. However, although not rigorously proven, the BPS
supergravity spectrum often turns out to acquire corrections at energy
$E\sim N$, and exact in the regime $E\lesssim N$.\footnote{It often has to do
with the `stringy exclusion' from the giant graviton physics at the
energy of order $N$ \cite{McGreevy:2000cw}.} We shall indeed find that our finite
$N$ indices completely
agree with the supergravity index for $k\leq N$ with some low values of $N$.
At $k>N$, our findings are new predictions.

It should also be of interest to take the large $N$ limit of our index in the
proper sense, and compare with the gravity index on $AdS_7\times S^4$. Note that
taking large $N$ limit is not as straightforward as, say, the index of 4d
superconformal theories \cite{Kinney:2005ej}. In the latter case, standard large $N$
technique of introducing `eigenvalue distribution' (for the $N$ integral variables)
was used to obtain the exact large $N$ index. We find it is more difficult to do
similar manipulations for a couple of reasons. Firstly, the integral is not along
the unit circle but along a more complicated contour, as explained in section 2.3.
Although it might be possible to apply more refined matrix model techniques to find
a complicated large $N$ eigenvalue distribution on this contour, we thought it will
not be easily doable by ourselves. Also, the index comes with an extra summation over
anti-self-dual fluxes $s_i$, partly breaking the $U(N)$ symmetry. In particular,
since nonzero fluxes appear in the low energy spectrum, starting immediately from
the vacuum, we are not sure if a systematic large $N$ analysis could be easily done.
However, we do study the large $N$ limit of our index for $k\leq 2$ and successfully
compare with the gravity dual, using a rather tedious analysis. We shall first obtain
the general finite $N$ index at $k\leq 2$, and then take the large $N$ limit of this
exact result.

Let us start by summarizing the supergravity index on $AdS_7\times S^4$,
which will be compared to our field theory index at various points.
The full gravity spectrum is obtained by first studying the single particle
spectrum. The single particle index is given by \cite{Bhattacharya:2008zy}
($\hat{m}\equiv m-\frac{1}{2}$)
\begin{eqnarray}\label{sugra-single}
  I_{\rm sp}&=&\frac{e^{-3\beta/2}(1-e^{-3\beta})(e^{\beta m}+e^{-\beta m})
  -e^{-2\beta}(e^{\beta a}+e^{\beta b}+e^{\beta c})+
  e^{-4\beta}(e^{-\beta a}+e^{-\beta b}+e^{-\beta c})}
  {(1-e^{-\beta(1+a)})(1-e^{-\beta(1+b)})(1-e^{-\beta(1+c)})(1-e^{-3\beta/2}e^{\beta m})
  (1-e^{-3\beta/2}e^{-\beta m})}\nonumber\\
  &=&\frac{(1-e^{-3\beta})(e^{-\beta}e^{\beta\hat{m}}+e^{-2\beta}e^{-\beta\hat{m}})
  -e^{-2\beta}(e^{\beta a}+e^{\beta b}+e^{\beta c})+
  e^{-4\beta}(e^{-\beta a}+e^{-\beta b}+e^{-\beta c})}
  {(1-e^{-\beta(1+a)})(1-e^{-\beta(1+b)})(1-e^{-\beta(1+c)})(1-e^{-\beta}e^{\beta\hat{m}})
  (1-e^{-2\beta}e^{-\beta\hat{m}})}\nonumber\\
  &=&e^{-\beta}e^{\beta\hat{m}}+e^{-2\beta}\left[e^{-\beta\hat{m}}
  -(e^{\beta a}+e^{\beta b}+e^{\beta c})+e^{\beta\hat{m}}
  (e^{-\beta a}+e^{-\beta b}+e^{-\beta c}+e^{\beta\hat{m}})\right]+\cdots\ .\qquad\quad
\end{eqnarray}
To obtain the multi-particle index, one takes
\begin{equation}\label{sugra-multiple}
  I_{\rm mp}=PE[I_{\rm sp}]=\exp\left[\sum_{n=1}^\infty\frac{1}{n}
  I_{\rm sp}(\beta\rightarrow n\beta)\right]\ .
\end{equation}
Then, keeping all $a,b,c,\hat{m}=m-\frac{1}{2}\ll 1$ and $e^{-\beta}\ll 1$,
one can expand $I_{\rm mp}$ in the fugacity $e^{-\beta}$ conjugate to the instanton
number in the 5d sense. The coefficient of $e^{-k\beta}$ with given $k$ is a finite
polynomial of other fugacities $e^{\beta a}, e^{\beta b}, e^{\beta c}, e^{\beta\hat{m}}$.

Now let us consider our finite $N$ index, with all $4$ chemical potentials
$\beta,\beta\hat{m},\beta a,\beta b$ turned on.
As we discussed the $U(1)$ index already, we mostly
discuss the $U(N)$ index with $N\geq 2$ in this subsection. We first discuss
the index for general $N\geq 2$ up to $k\leq 2$, which yields a universal
result for $N\geq 2$ and agrees with $AdS_7\times S^4$ supergravity. So this
finding implies that the large $N$ index agrees with the gravity dual up to $k\leq 2$.

We start by explaining some general setting. Using the Weyl symmetry, we first assume
\begin{equation}
  s_1\geq s_2\geq\cdots\geq s_N\ ,
\end{equation}
and then replace $\frac{1}{N!}$ by the inverse of the Weyl group of the
subgroup unbroken by the $s_i$ fluxes. Then, let us view the integrand
as a holomorphic function of $z_i\equiv e^{\sigma_i-i\lambda_i}$. The integral
measure from change of variables is
\begin{equation}
  \prod_{i=1}^N\frac{d\lambda_i}{2\pi}=(-1)^N\prod_{i=1}^N\frac{dz_i}{2\pi iz_i}\ .
\end{equation}
As we have not been careful about the overall sign of the path integral during our
derivation, we simply tune the overall sign to yield the positive degeneracy (i.e.
$+1$ rather than $-1$) for the vacuum. So in foresight, we replace $(-1)^N$ by $-1$
here. The perturbative and instanton parts of the determinant are given in terms of
positive roots value in $\lambda$, or for the case of $U(N)$ come with the variables $z_i/z_j=e^{\sigma_{ij}-i\lambda_{ij}}$ with $i<j$. The classical measure also
depends on the ratios of $z_i$ variables since
$\sum_is_i=0$. So one can decompose the $N$ contour integrals into one trivial integral
(for the overall $U(1)$) and other $N-1$ integrals. Let us define the other $N-1$
variables to be $\zeta_i=\frac{z_i}{z_{i+1}}$ with $i=1,2,\cdots,N-1$ for later use.
Since $\sigma_i=\frac{4s_i }{1-\xi}\leq 0$ with $s_{ij}\geq 0$,
one finds that $|z_i|\ll|z_j|$ for $i<j$ in our setting $\beta\gg 1$ when
$s_{ij}>0$. Thus, the contour for $\zeta_i$ is a very small circle on
the complex plane when $s_i>s_{i+1}$. On the other hand, if $s_i=s_{i+1}$,
the contour is a small deformation away from the unit circle dictated by the
rules explained in section 2.3. (Some non-Abelian symmetry is unbroken by $s$
in this case.) With these understandings, let us investigate the index by
starting from the vacuum and then increasing the instanton number ($\sim$ `energy level') or
the order of $e^{-\beta}$ expansion.

\hspace*{-.6cm}{\bf Vacuum:} Firstly, in the self-dual instanton part
$Z_{\rm inst}^{(1)}Z_{\rm inst}^{(2)}Z_{\rm inst}^{(3)}$ of the measure,
the `energy cost' for $e^{-\beta}$ is always positive when $k_{SD}>0$ and is $1$
when $k_{SD}=0$. So we replace this factor by $1$ for the vacuum. Then the vacuum
is coming simply with lowest value of $-\frac{1}{2}\sum_{i=1}^Ns_i^2$ from the
classical measure. We should decide what is the anti-self-dual flux configuration
$s_i$ which survives the contour integral at lowest energy.
From the structure of the perturbative measure, the poles for $\zeta_i$ appear only
from the modes which feel nonzero relative flux $s_{ij}$. However, we have seen
in the above paragraph that these poles are always outside the integration contour,
since the integration contour for $\zeta_i$ is a very small circle around $\zeta_i=0$
when the relative flux $s_i-s_{i+1}$ is nonzero.
So all the poles one has to consider are $\zeta_i=0$.
This implies that the allowed nonzero fluxes $s_i$ are exactly
the same as those fluxes allowed in the previous subsection for the unrefined index.
(The last fact applies not just to the vacuum, but to all excited states as well.)
So the ground state comes with the flux
\begin{equation}
  s=(N-1,N-3,\cdots,-(N-1))\equiv s_0\ ,
\end{equation}
as in the case with $m=\frac{1}{2}-c$ discussed in the last subsection.
Although we essentially discussed it in the last subsection, for concreteness
we repeat the discussion in full generality again.
After replacing $Z_{\rm inst}^{(1)}Z_{\rm inst}^{(2)}Z_{\rm inst}^{(3)}\rightarrow 1$,
one reorganizes the remaining integrand $Z_{\rm pert}$ as a
rational function of $\zeta_i\equiv z_i/z_{i+1}$ variables by expanding various $\sinh$ functions,
taking the form of
\begin{equation}
  \prod_{i<j}\frac{(z_i/z_j-B_1)(z_i/z_j-B_2)\cdots}{(z_i/z_j-A_1)(z_i/z_j-A_2)\cdots}\ .
\end{equation}
In $\zeta_i$ variables, the above rational function takes the form of
\begin{equation}\label{rational}
  \prod_{i<j}\frac{(\zeta_i\zeta_{i+1}\cdots\zeta_{j-1}-B_1)
  (\zeta_i\zeta_{i+1}\cdots\zeta_{j-1}-B_2)}{(\zeta_i\zeta_{i+1}\cdots\zeta_{j-1}-A_1)
  (\zeta_i\zeta_{i+1}\cdots\zeta_{j-1}-A_2)}\ .
\end{equation}
The polynomial in the denominator will not have any pole inside the integration
contour with the flux $s_0$, as just explained. The $z_i$ dependent part of the prefactor
from $Z_{\rm pert}$, multiplying (\ref{rational}), is
\begin{equation}\label{phase-ground}
  z_1^{-(N-1)}z_2^{-(N-3)}\cdots z_N^{N-1}=
  \zeta_1^{-(N-1)}\zeta_2^{-(N-1)-(N-3)}\zeta_3^{-(N-1)-(N-3)-(N-5)}
  \cdots\zeta_{N-1}^{-(N-1)}\ ,
\end{equation}
which is exactly the same prefactor that one obtains by rewriting the Haar measure
used in the previous subsection into the form of (\ref{rational}). So from
the contour integral perspective, the classical phase factor
$e^{-i\sum_is_i\lambda_i}=z_1^{s_1}z_2^{s_2}\cdots z_N^{s_N}$ with too large
$\sum_is_i^2$ would kill the pole at some $\zeta_i=0$, making the integral to be zero.
The values of $s_i$'s which survive the contour integral is exactly the same as those
in the unrefined case of previous subsection, as the phase (\ref{phase-ground})
is the same as that coming from the unrefined perturbative measure.

So coming back to the flux $s_0$, the classical measure is given by
\begin{equation}
  e^{\frac{\beta N(N^2-1)}{6}}z_1^{N-1}z_2^{N-3}\cdots z_N^{-(N-1)}\ ,
\end{equation}
where the first factor comes from $\exp(\frac{\beta}{2}\sum s_i^2)$.
The residue obtained from the perturbative part is
$\prod_{i<j}e^{-\frac{\beta\hat{m}s_{ij}}{2}}$. Using $\frac{1}{2}\sum_{i<j}s_{ij}=\frac{N(N^2-1)}{6}$,
the vacuum index becomes
\begin{equation}\label{vacuum-index}
  I_{k=0}=e^{\beta(1-\hat{m})\frac{N(N^2-1)}{6}}\ .
\end{equation}
The vacuum degeneracy is of course $1$.

\hspace*{-.6cm}{\bf First excited states:}
Now, the `first excited states' at $k=1$ order come in many different
ways. Firstly, one can try to put a self-dual instanton localized at one of the three
fixed points, with the anti-self-dual flux $s_0$. Secondly, one may
choose a different $s$ flux with higher energy, keeping $k_{SD}=0$. The first
contribution takes the form of
\begin{equation}\label{k=1-first}
  e^{\beta(1+\hat{m})\frac{N(N^2-1)}{6}}\oint\prod_{i=1}^{N-1}
  \frac{d\zeta_i}{2\pi i\zeta_i}\left[\prod_{\alpha\in\Delta_+}
  \frac{(\zeta-\cdots)(\zeta-\cdots)\cdots}{(\zeta-\cdots)(\zeta-\cdots)\cdots}\right]
  \left[Z_{k_{SD}=1}^{(1)}+Z_{k_{SD}=1}^{(2)}
  +Z_{k_{SD}=1}^{(3)}\right]\ .
\end{equation}
The schematically written part inside the square bracket, with $\cdots$,
is a rational function obtained from the perturbative measure, taking the
form of (\ref{rational}). Now with the flux $s_0$ breaking
the $U(N)$ symmetry into $U(1)^N$, this contribution simplifies as follows.
Firstly, as explained in (\ref{Zk}), the instanton partition function on
$\mathbb{R}^4\times S^1$ takes the form of the sum over colored Young diagrams.
The contribution of each Young diagram is weighted by a
product of many factors of the following form:
\begin{equation}
  F(x)\equiv
  \frac{\sinh\frac{\beta(x+m_0-\epsilon_+)}{2}\sinh\frac{\beta(x-m_0-\epsilon_+)}{2}}
  {\sinh\frac{\beta x}{2}\sinh\frac{\beta(x-2\epsilon_+)}{2}}\ .
\end{equation}
The variable $x$ is a suitable linear combination of $\lambda_{ij}$ and
$\epsilon_{1,2}$. Now one can rewrite this expression in the following two ways:
\begin{equation}\label{charged-decompose}
  F(x)=1+\frac{e^{\beta x}(1-e^{-\beta(\epsilon_++m_0)})(1-e^{-\beta(\epsilon_+-m_0)})}
  {(1-e^{\beta x})(1-e^{\beta x}e^{-2\beta\epsilon_+})}=
  1+\frac{e^{-\beta x}(1-e^{\beta(\epsilon_++m_0)})(1-e^{\beta(\epsilon_+-m_0)})}
  {(1-e^{-\beta x})(1-e^{-\beta x}e^{2\beta\epsilon_+})}\ .
\end{equation}
In the product of $F(x)$ functions at a given Young diagram, the variable $x$
may either contain $\lambda_{ij}$, or only depend on $\epsilon_{1,2}$ without
$\lambda_{ij}$ dependence. When it is the former case, using one of the two
expressions above, the overall factor of $e^{\pm\beta x}$ in the numerator of
the second term can be made into a positive product of $\zeta_i$ variables.
Since the rest of the integrand only contains a simple pole from
$\frac{d\zeta_i}{2\pi i\zeta_i}$, the second terms in $F(x)$ containing
$\zeta_i$ dependence kill one of these $N-1$ simples poles for $\zeta_i$ at
the origin, yielding zero. Thus, the only nonzero contribution (not killing the
simple poles at the origins) is obtained by effectively replacing all $F(x)$'s
that contain $\lambda_{ij}$ by $1$, the first term in (\ref{charged-decompose}).
This means that the contributions of all the `charged modes' to
the $U(N)$ instanton partition disappears, and one is effectively left with the
$U(1)^N$ instanton partition function, coming from multiplications of $F(x)$ functions
not containing $\lambda_{ij}$. (This phenomenon, effectively replacing non-Abelian
self-dual instantons by instantons in the subgroup unbroken by the $s$ flux, will
turn out to appear repeatedly for other excited states, unless the rest of the
integrand from classical and perturbative contributions exhibits multiple poles at
the origin.) So one adds the contributions of $U(1)^N$ self-dual instantons from three
fixed points at the $k_{SD}=1$ order. The result is the vacuum factor (\ref{vacuum-index})
multiplied by
\begin{equation}\label{general-k=1-first}
  Ne^{-\beta}e^{\beta(m-\frac{1}{2})}\ ,
\end{equation}
which is $N$ times the $U(1)$ index at $k=1$ ($N$ coming from $N$ copies, $U(1)^N$).

The second contribution at $k=1$ is not exciting self-dual instantons, while changing
the flux $s$ to have its energy $-\frac{1}{2}\sum_is_i^2$ to be higher than the vacuum
by $1$ unit. The fluxes with the energy cost $e^{-\beta}$ are
\begin{equation}
  s=s_{1j}\equiv s_0+(0,0,\cdots,0,-1_{j},+1_{j\!+\!1},0,\cdots,0)\ ,
\end{equation}
with $j=1,2,\cdots,N-1$. For instance, at $N=4$, there are $3$ possible fluxes
at this order: $(2,2,-1,-3)$, $(3,0,0,-3)$, $(3,1,-2,-2)$. The symmetry unbroken by
this flux is $U(2)\times U(1)^{N-2}$. By repeating the analysis of classical and
perturbative measure in $\zeta_i$ variables, one obtains
\begin{equation}\label{k=1-second}
  e^{\beta(1+\hat{m})\frac{N(N^2-1)}{6}}\cdot e^{-\beta(1+\hat{m})}
  \oint\frac{d\zeta_j}{2\pi\zeta_j}\cdot
  \frac{1}{2!}\zeta_{j}^{-1}(\zeta_j-1)^2
  \oint\prod_{i(\neq j)=1}^{N-1}
  \frac{d\zeta_i}{2\pi i\zeta_i}\left[\prod_{\alpha(\neq (j,j\!+\!1))\in\Delta_+}
  \frac{(\zeta-\cdots)(\zeta-\cdots)\cdots}{(\zeta-\cdots)(\zeta-\cdots)\cdots}\right]
  \ .
\end{equation}
Again the part with $\cdots$ in the last square bracket is arranged into
the form (\ref{rational}). The factor $e^{-\beta}$ in front is the energy cost
for the $s_j$ flux relative to $s_0$, and the prefactors containing $\hat{m}$ come from
$e^{\frac{\beta\hat{m}}{2}\sum_{i<j}s_{ij}}$. . $2!$ is the order of Weyl group
for $U(2)$ unbroken by the flux. The integrand again does not contain
any pole inside the small circular contour for $\zeta_i$ for $i\neq j$.
The contour integral for $\zeta_j$ yields $-1$,
while the rest yields $e^{-2\beta\hat{m}\frac{N(N^2-1)}{6}}e^{2\beta\hat{m}}$. So
the contribution from the sector with flux $s_{1j}$ is
$-e^{\beta(1-\hat{m})\frac{N(N^2-1)}{6}}e^{-\beta}e^{\beta\hat{m}}$. As there
are $N-1$ possible $s_{1j}$ fluxes ($j=1,2,\cdots,N-1$) at the same energy, the
net contribution is the vacuum factor (\ref{vacuum-index}) multiplied by
\begin{equation}\label{s1j-integral}
  -(N-1)e^{-\beta}e^{\beta\hat{m}}\ .
\end{equation}
Adding this with the first contribution (\ref{general-k=1-first}) from
$s_0$ flux and $k_{SD}=1$ self-dual instantons, one obtains
\begin{equation}
  I_{k=1}=I_{k=0}\cdot e^{-\beta}e^{\beta\hat{m}}\ .
\end{equation}
for any $N$. (The derivation above sometimes used $N\geq 2$, but the final result
is correct also for $N=1$.) In particular,
the result is independent of $N$ and so remains the same in the large $N$ limit.
This agrees with the supergravity result (\ref{sugra-single}), (\ref{sugra-multiple})
at $\mathcal{O}(e^{-\beta})$.

\hspace*{-.6cm}{\bf Second excited states:}
We consider the index for $N\geq 2$ at $k=2$. The contributions come from $3$
different kinds of $s$ fluxes. Firstly, the flux $s_0$ with $k_{SD}=2$ self-dual
instantons can contribute to $k=2$. Secondly, the flux $s_{1j}$ with $k_{SD}=1$
self-dual instanton instanton can contribute. Finally, depending on the value of $N$,
there may also be fluxes $s$ which contribute to this sector at $k_{SD}=0$. As the
fluxes of the former two cases are clear, let us start by discussing the possible
anti-self-dual fluxes in the last case. For $U(2)$ and $U(3)$, there are no $s$ fluxes
which give energy cost $k=2$ with $k_{SD}=0$. Firstly, all allowed anti-self-dual
fluxes for $U(2)$ are $(1,-1)$ for ground state and $(0,0)$ at $k=1$. $U(3)$ fluxes
are $(2,0,-2)$ for the ground state, $(1,1,-2),(2,-1,-1)$ at $k=1$, $(1,0,-1)$ at
$k=3$ and $(0,0,0)$ at $k=4$. So there are no fluxes giving $k=2$. At $k\geq 4$,
there appear more possibilities at $k=2$. One has
\begin{equation}
  s_{2ij}=s_0+(0,0,\cdots,0,-1_i,+1_{i\!+\!1},0,\cdots,0,-1_{j},+1_{j\!+\!1},0,\cdots,0)
\end{equation}
for $1\leq i<j\leq N-1$ and $i+1<j$. The last condition is possible only for $N\geq 4$.

Now we consider the first case, with flux $s_0$ and $k_{SD}=2$. The integral is
\begin{equation}\label{k=2-first}
  e^{\beta(1+\hat{m})\frac{N(N^2-1)}{6}}
  \oint\prod_{i=1}^{N-1}\frac{d\zeta_i}{2\pi i\zeta_i}
  \left[\prod_{\alpha\in\Delta_+}
  \frac{(\zeta-\cdots)(\zeta-\cdots)\cdots}{(\zeta-\cdots)(\zeta-\cdots)\cdots}\right]
  \left[Z_{\rm inst}^{(1)}Z_{\rm inst}^{(2)}Z_{\rm inst}^{(3)}\right]_{k_{SD}=2}\ .
\end{equation}
Again, since the combination of classical and perturbative measures only show
simple poles in $\zeta_i$, one can use the arguments around (\ref{charged-decompose})
to effectively replace the self-dual instanton part by the $U(1)^N$ instanton
partition function at $k_{SD}=2$. This can be obtained by taking the
$U(1)^N$ index (already known from section 2.1),
\begin{equation}\label{multi-abelian}
  PE\left[N\frac{qy+q^2y^{-1}-q^2(y_1^{-1}+y_2^{-1}+y_3^{-1})+q^3}
  {(1-qy_1)(1-qy_2)(1-qy_3)}\right]
\end{equation}
and taking the term at $q^2$ order.
Here and below, we shall define and use $q=e^{-\beta}$, $y_i=e^{-\beta a_i}$ for
$i=1,2,3$ (satisfying $y_1y_2y_3=1$), and $y=e^{\beta\hat{m}}=e^{\beta(m-\frac{1}{2})}$.
The result is
\begin{equation}\label{k=2-first-final}
  q^2\left[\frac{N(N+1)}{2}y^2+Ny(y_1+y_2+y_3)-N\left(y_1^{-1}+y_2^{-1}+y_3^{-1}\right)
  +Ny^{-1}\right]\ .
\end{equation}
After the contour integration, this result multiplied by
$I_{k=0}=e^{\beta(1-\hat{m})\frac{N(N^2-1)}{6}}$ is the net
contribution from the sector with $s=s_0$, $k_{SD}=2$.

We move on to the second cases, at $s=s_{1j}$, $k_{SD}=1$. As the gauge symmetry
is broken by $s$ to $U(2)\times U(1)^{N-2}$, the integral takes the form of
\begin{eqnarray}\label{k=2-second}
  &&e^{\beta(1+\hat{m})\frac{N(N^2-1)}{6}}\cdot e^{-\beta(1+\hat{m})}
  \oint\frac{d\zeta_j}{2\pi\zeta_j^2}\cdot
  \frac{1}{2!}(\zeta_j-1)^2
  \oint\prod_{i(\neq j)=1}^{N-1}
  \frac{d\zeta_i}{2\pi i\zeta_i}\\
  &&\hspace{3cm}\times\left[\prod_{\alpha(\neq (j,j\!+\!1))\in\Delta_+}
  \frac{(\zeta-\cdots)(\zeta-\cdots)\cdots}{(\zeta-\cdots)(\zeta-\cdots)\cdots}\right]
  [Z_{\rm inst}^{(1)}Z_{\rm inst}^{(2)}Z_{\rm inst}^{(3)}]_{k_{SD}=1}\ .\nonumber
\end{eqnarray}
The self-dual instantons' contribution at $k_{SD}=1$ can be divided into two parts.
As $\zeta_i$ with $i\neq j$ all have simple poles at $\zeta_i=0$ apart from $Z_{\rm inst}$,
the contributions to $Z_{\rm inst}$ from the modes charged in $U(1)^{N-2}$ effectively
reduce to $1$ for a given colored Young diagram, as explained around
(\ref{charged-decompose}). The modes charged
within the unbroken $U(2)$ carrying the charge factor $\zeta_j$ remain nontrivial,
as $\zeta_j$ has a double pole at the origin. So
$[Z_{\rm inst}^{(1)}Z_{\rm inst}^{(2)}Z_{\rm inst}^{(3)}]_{k_{SD}=1}$
is given by the contribution from $U(1)^{N-2}$ instanton at $k_{SD}=1$, plus the
contribution from $U(2)$ instantons at $k_{SD}=1$. The former can be obtained
from (\ref{multi-abelian}), which is $(N-2)qy$.
The remaining integral for this part is evaluated around (\ref{s1j-integral}), which is $I_{k=0}$ times $-qy$
per given $s_{1j}$. So the contribution from $s=s_{1j}$, $k_{SD}=1$ $U(1)^{N-2}$
instantons is $I_{k=0}$ times
\begin{equation}\label{k=2-second-1-final}
  -(N-1)(N-2)q^2y^2\ ,
\end{equation}
where we summed over the $N-1$ possible $s_{1j}$ fluxes. The contribution from the
$k_{SD}=1$ $U(2)$ instanton is obtained as follows. There are $3$ possible saddle
point configurations of this sort, with the single $U(2)$ instanton localized
at one of the three fixed points. At each fixed point, the $U(2)$ single instanton
measure is given from (\ref{Zk}) by
\begin{equation}
  \frac{\sinh\frac{\beta(m_0+\epsilon_-)}{2}\sinh\frac{\beta(m_0-\epsilon_-)}{2}}
  {\sinh\frac{\beta\epsilon_1}{2}\sinh\frac{\beta\epsilon_2}{2}}
  \left[2+\frac{\zeta_j(e^{\beta\epsilon_+}+e^{-\beta\epsilon_+})
  (e^{\beta\epsilon_+}+e^{-\beta\epsilon_+}-e^{\beta m_0}-e^{-\beta m_0})}
  {(\zeta_j-e^{-2\beta\epsilon_+})(\zeta_j-e^{2\beta\epsilon_+})}\right]
\end{equation}
at each fixed point (substituting suitable fixed point values of $\epsilon_{1,2},m_0$),
multiplied by the classical $e^{-\beta(1+a_i)}$ factor at the $i$'th fixed point.
The addition of $3$ such contributions becomes the contour integral measure for $\zeta_j$,
together with the $\zeta_j$ dependent part on the first line of (\ref{k=2-second}).
The poles for $\zeta_j$ inside the contour are $\zeta_j=0,e^{-2\beta\epsilon_+}$
for all these contributions. Taking the two residues, and adding the contributions
from three fixed points, one obtains $I_{k=0}$ times
\begin{equation}\label{k=2-second-2-final}
  -(N-1)q^2\left[y^2+y(y_1+y_2+y_3)-(y_1^{-1}+y_2^{-1}+y_3^{-1})+y^{-1}\right]\ ,
\end{equation}
where the factor $N-1$ comes from possible $s_{1j}$ fluxes. The finite polynomial
in $y,y_i$ is obtained only after adding the three contributions.

The third case with $s=s_{2ij}$, $k_{SD}=0$ comes with $U(2)^2\times U(1)^{N-4}$ unbroken
gauge symmetry. The number of possible fluxes satisfying $i+2\leq j$ is
\begin{equation}
  \sum_{i=1}^{N-3}(N-2-i)=\frac{(N-2)(N-3)}{2}\ .
\end{equation}
At each flux, the integral is given by $I_{k=0}\cdot q^3y^2$.  So
the net contribution from this sector is $I_{k=0}$ times
\begin{equation}\label{k=2-third-final}
  \frac{(N-2)(N-3)}{2}q^2y^2\ .
\end{equation}

Collecting all the contributions (\ref{k=2-first-final}), (\ref{k=2-second-1-final}),
(\ref{k=2-second-2-final}), (\ref{k=2-third-final}), $I_{k=2}$
is given by $I_{k=0}$ times
\begin{eqnarray}\label{k=2-general}
  &&q^2\left[\frac{N(N+1)}{2}y^2+Ny(y_1+y_2+y_3)-N\left(y_1^{-1}+y_2^{-1}+y_3^{-1}\right)
  +Ny^{-1}\right]\nonumber\\
  &&-(N-1)(N-2)q^2y^2%\nonumber\\
  %&&
  -(N-1)q^2\left[y^2+y(y_1+y_2+y_3)-(y_1^{-1}+y_2^{-1}+y_3^{-1})+y^{-1}\right]\nonumber\\
  &&+\frac{(N-2)(N-3)}{2}q^2y^2\nonumber\\
  &&=q^2\left[2y^2+y(y_1+y_2+y_3)-(y_1^{-1}+y_2^{-1}+y_3^{-1})+y^{-1}\right]\ .
\end{eqnarray}
The first line comes from $s=s_0$, $k_{SD}=2$, the first term on the second line
from $s=s_{1j}$ and $k_{SD}=1$ $U(1)^{N-2}$ instantons,
the second term on the second line from $s=s_{1j}$ and $k_{SD}=1$
$U(2)$ instantons, the third line from $s=s_{2ij}$ and $k_{SD}=0$.
This formula is valid for $N\geq 2$: although the sectors yielding the first term
on the second line and the third line do not exist for $N=2$ and $N=2,3$, respectively,
inserting these values of $N$'s make the irrelevant terms to be zero
so that the final formula on the last line is still valid.
The formula is invalid for $N=1$ because of the third line.
(Erasing the third line, the remainders yield the correct $U(1)$ index at $k=2$.)
The above index at $k=2$ is also independent of $N$ at $N\geq 2$. So again,
the large $N$ index at $k=2$ is the last line of (\ref{k=2-general}).
This precisely agrees with the supergravity index on $AdS_7\times S^4$
given by (\ref{sugra-single}), (\ref{sugra-multiple}) at $\mathcal{O}(e^{-2\beta})$.

The general analysis becomes more cumbersome for larger $k$. Below, for $k\geq 3$,
we restrict our studies to some lower gauge group ranks. (Actually, we made a concrete
study only up to $k=3$ so far.) Still, at $k\leq N$, we shall continue to
find a complete agreement of the finite $N$ index with the gravity dual,
which we regard as a kind of test of the AdS$_7$/CFT$_6$ duality
(although we did not take the strict large $N$ limit).

\subsubsection{$U(2)$ index}

Let us study the $U(2)$ index. The allowed anti-self-dual
fluxes are only $s=s_0=(1,-1)$ and $s=(0,0)$. At $s=s_0=(1,-1)$,
the index becomes
\begin{equation}
  e^{\beta}\oint\frac{d\lambda}{2\pi}e^{-i\lambda}
  \frac{4\sin\frac{\lambda+2i\beta a}{2}\sin\frac{\lambda+2i\beta b}{2}
  \sin\frac{\lambda+2i\beta c}{2}\sin\frac{\lambda-i\beta a}{2}\sin\frac{\lambda-i\beta b}{2}
  \sin\frac{\lambda-i\beta c}{2}}{\sin\frac{\lambda+i\beta(a-\hat{m})}{2}
  \sin\frac{\lambda+i\beta(b-\hat{m})}{2}\sin\frac{\lambda+i\beta(c-\hat{m})}{2}
  \sin\frac{\lambda+i\beta\hat{m}}{2}}Z_{\rm inst}^{(1)}Z_{\rm inst}^{(2)}Z_{\rm inst}^{(3)}
\end{equation}
where $\lambda\equiv\lambda_{12}+i\beta\sigma$ with
$\sigma=\frac{4s_{12}}{1-\xi}=\frac{8}{1-\xi}<0$. The prefactor
$e^{\beta}$ comes from $e^{\beta\frac{N(N^2-1)}{6}}$ at $N=2$ for $s=s_0$.
The contour for the $z\equiv e^{-i\lambda}$ integral
includes the pole at $z=0$ only, since $|z|=e^{\beta\sigma}\ll 1$.
The integral can be rewritten as
\begin{equation}
  I_{(1,-1)}=e^\beta e^{\beta\hat{m}}\oint\frac{dz}{2\pi i}\frac{(z-e^{-2\beta a})
  (z-e^{-2\beta b})(z-e^{-2\beta c})
  (z-e^{\beta a})(z-e^{\beta b})(z-e^{\beta c})}{z(z-e^{-\beta (a-\hat{m})})
  (z-e^{-\beta(b-\hat{m})})(z-e^{-\beta(c-\hat{m})})(z-e^{-\beta\hat{m}})}
  Z_{\rm inst}^{(1)}Z_{\rm inst}^{(2)}Z_{\rm inst}^{(3)}\ .
\end{equation}
At $s=(0,0)$, the perturbative measure is simply the Haar measure. The index
in this sector is given by
\begin{equation}
  I_{(0,0)}=\frac{1}{2!}\oint\frac{d\lambda}{2\pi}\ 4\sin^2\frac{\lambda}{2}
  Z_{\rm inst}^{(1)}Z_{\rm inst}^{(2)}Z_{\rm inst}^{(3)}=\oint\frac{dz}{2\pi i z^2}
  (z-1)^2\ Z_{\rm inst}^{(1)}Z_{\rm inst}^{(2)}Z_{\rm inst}^{(3)}\ ,
\end{equation}
where $\lambda=\lambda_{12}$ and $z=e^{-i\lambda}$. The integral contour for
$z$ is very close to the unit circle (at $\beta\gg 1$, $\beta\hat{m},\beta a_i\ll\beta$).
The precise shape of the contour was explained in section
2.3: the contour includes all poles of the form $z=e^{-\beta(p\epsilon_++q\epsilon_-)}$,
with $p>0$.

Since we have already discussed the index at $k\leq 2$ in full generality,
we consider the index at $k=3$. It will turn out that
$k>N$ is the regime in which the large $N$ supergravity approximation becomes
invalid, so we should be making new predictions on the BPS spectrum here.
Still, there are many consistency requirements that one can check in this sector.
These include the unitarity bound $|R_1-R_2|\leq k$, or index at given $k$ being a
finite polynomial. There are two sectors which contribute to the $k=3$ index:
$s=(1,-1)$ with $k_{SD}=3$, and $s=(0,0)$ with $k_{SD}=2$.

In the first case, the classical and perturbative measures yield a simple pole in $z$,
so one has to consider the $U(1)^2$ self-dual instantons at $k_{SD}=3$.
The result is $I_{k=0}\cdot e^{-3\beta}$ times
\begin{eqnarray}\label{U(2)-k=3-1}
  &&2\left[1+(y^{-1}-y_1^{-1}-y_2^{-1}-y_3^{-1})(y_1+y_2+y_3)+y(y_1^2+y_2^2+y_3^2
  +y_1y_2+y_2y_3+y_3y_1)\right]\nonumber\\
  &&+4\left[1-y(y_1^{-1}+y_2^{-1}+y_3^{-1})+y^2(y_1+y_2+y_3)\right]+4y^3\ ,
\end{eqnarray}
from (\ref{multi-abelian}).

For the second case, one has to consider the $U(2)$ self-dual instantons'
partition function at $k_{SD}=2$. It first acquires contributions from both
instantons localized at one of the three fixed points, and also from two
instantons localized at two different fixed points.

Let us first consider the case with both instantons at the same fixed point. From the
$\mathbb{R}^4\times S^1$ partition function, the partition function at $k_{SD}=2$ is
\begin{equation}\label{kSD=2}
  Z_{\rm inst}(\mathbb{R}^4\times S^1)=Z_{(2)_1}+Z_{(1,1)_1}+Z_{(2)_2}+Z_{(1,1)_2}
  +Z_{(1)_1(1)_2}\ ,
\end{equation}
where $(p_1,p_2,\cdots)_i$ is a notation for the colored Young diagram,
with $p_1\geq p_2\geq\cdots$ denoting the length of the rows of a Young diagram, and
the subscript $i=1,2,\cdots, N$ labels $N$ Young diagrams. (When $i$'th Young
diagram is absent in the above formula, this diagram is void.)
The sum of all $p_1,p_2,\cdots$ is $k_{SD}=2$, the self-dual instanton number.
The five expressions can be obtained from (\ref{Zk}), and the concrete expressions
at $k_{SD}=2$ can be found in, say, \cite{Kim:2011mv}. The results are
\begin{eqnarray}
  Z_{(2)_1}&=&I_{\rm com}(\epsilon_1,\epsilon_2)
  I_{\rm com}(2\epsilon_1,\epsilon_2\!-\!\epsilon_1)\cdot
  \frac{(z-ty_0)(z-ty_0^{-1})(z-t_1ty_0)(z-t_1ty_0^{-1})}
  {(z-1)(z-t_1)(z-t_1t_2)(\zeta-t_1^2t_2)}\\
  Z_{(1,1)_1}&=&Z_{(2)_1}\ \ \ \ \ {\rm with\ a\ replacement}
  \ (\epsilon_+,\epsilon_-)\ \rightarrow\ (\epsilon_+,-\epsilon_-)\nonumber\\
  Z_{(2)_2}&=&Z_{(2)_1}\ \ \ \ \ \ {\rm with}\ \lambda_1\leftrightarrow\lambda_2
  \ \ \ ({\rm or}\ z\rightarrow z^{-1})\nonumber\\
   Z_{(1,1)_2}&=&Z_{(1,1)_1} \ \ \ \ \ {\rm with}\ \lambda_1\leftrightarrow\lambda_2
  \ \ \ ({\rm or}\ z\rightarrow z^{-1})\nonumber\\
  Z_{(1)_1(1)_2}&=&I_{\rm com}(\epsilon_1,\epsilon_2)^2\cdot
  \frac{(z-(t_1/t_2)^{\frac{1}{2}}y_0)
  (z-(t_1/t_2)^{\frac{1}{2}}y_0^{-1})(z-(t_2/t_1)^{\frac{1}{2}}y_0)
  (z-(t_2/t_1)^{\frac{1}{2}}y_0^{-1})}
  {(z-t_1)(z-t_2)(z-t_1^{-1})(z-t_2^{-1})}\ ,\nonumber
\end{eqnarray}
where $y_0=e^{\beta m_0}$, $t_{1,2}=e^{-\beta\epsilon_{1,2}}$,
$t=e^{-\beta\epsilon_+}=(t_1t_2)^{\frac{1}{2}}$,
$z=e^{-i\lambda_{12}}$, and
\begin{equation}
  I_{\rm com}(\epsilon_1,\epsilon_2)=I_-\equiv\frac{\sinh\frac{\beta(m_0+\epsilon_-)}{2}
  \sinh\frac{\beta(m_0-\epsilon_-)}{2}}
  {\sinh\frac{\beta\epsilon_1}{2}\sinh\frac{\beta\epsilon_2}{2}}\ .
\end{equation}
The contour integral in this sector (both self-dual instantons
at the same fixed point) is
\begin{equation}\label{kSD=2-1}
  \frac{1}{2}\oint\frac{dz}{2\pi iz^2}(z-1)^2\left[(Z_{\rm inst}^{(1)})_{k_{SD}=2}+
  (Z_{\rm inst}^{(2)})_{k_{SD}=2}+(Z_{\rm inst}^{(3)})_{k_{SD}=2}\right]\ ,
\end{equation}
where each $(Z_{\rm inst}^{(i)})_{k_{SD}=2}$ takes the form of (\ref{kSD=2})
with suitable identifications of $\epsilon_1,\epsilon_2,m_0$ as explained before.
The contour for $z$ includes the following poles.
In all five expressions above, the pole at $z=0$ is included. For
$Z_{(2)_1}$, extra poles at $z=t_1,t_1t_2,t_1^2t_2$ are included.
For $Z_{(1,1)_1}$, poles at $z=t_2,t_1t_2, t_1t_2^2$ are included.
For $Z_{(2)_2}$ and $Z_{(1,1)_2}$, no extra poles are included.
Finally, for $Z_{(1)_1(1)_2}$, two more poles $z=t_1,t_2$
are included. Taking one of the three terms in (\ref{kSD=2-1}),
the sum of all residues is given by
\begin{equation}\label{SD-U(2)-k=2}
  \frac{1}{2}\oint\frac{dz}{2\pi iz^2}(z-1)^2(Z_{\rm inst}^{(i)})_{k_{SD}=2}
  =-e^{-2\beta(1+a_i)}\frac{\left(y_0+y_0^{-1}-s-s^{-1}\right)
  ({\rm numerator})}{(1-t_1)(1-t_1^2)(1-t_2)(1-t_2^2)(1-t_1t_2)}
\end{equation}
with
\begin{eqnarray}\label{SD-U(2)-k=2-1}
  ({\rm numerator})&=&t+2t^2\chi_{1/2}(y_0)+t^3\left(\chi_1(y_0)-2\chi_1(s)
  -\chi_{1/2}(y_0)\chi_{1/2}(s)-1\right)\nonumber\\
  &&+t^4\left(\chi_{3/2}(y_0)+\chi_{1/2}(s)\chi_1(y_0)-2\chi_{1/2}(y_0)\chi_2(s)
  +2\chi_{1/2}(s)-2\chi_{1/2}(y_0)\right)\nonumber\\
  &&+t^5\left(4\chi_1(s)-\chi_1(y_0)-\chi_1(s)\chi_1(y_0)+6\right)\nonumber\\
  &&+t^6\left(4\chi_{1/2}(y_0)\chi_1(s)-2\chi_{1/2}(s)\chi_1(y_0)-\chi_{3/2}(s)
  -3\chi_{1/2}(s)\right)\nonumber\\
  &&+t^7\left(\chi_{3/2}(s)\chi_{1/2}(y_0)+4\chi_{1/2}(s)\chi_{1/2}(y_0)
  -3\chi_1(s)-\chi_1(y_0)-4\right)\nonumber\\
  &&+t^8\left(-\chi_{3/2}(s)-\chi_{1/2}(s)\right)
  +t^9\left(1-\chi_{1/2}(s)\chi_{1/2}(y_0)\right)+t^{10}\chi_{1/2}(s)\ ,
\end{eqnarray}
where $t\equiv(t_1t_2)^{\frac{1}{2}}$, $s\equiv(t_1/t_2)^{\frac{1}{2}}$,
$y_0\equiv e^{\beta m_0}$. $\chi_j(s)=s^{2j}+s^{2j-2}+\cdots+s^{-2j}$ etc. are
the spin $j$ character of $SU(2)$. As a small cross-check, we compare this
to the known result in the $y_0\rightarrow\infty$ limit, for the pure 5d
$\mathcal{N}=1$ SYM with $SU(2)$ gauge group \cite{Hanany:2012dm}.
The leading term in the numerator of (\ref{SD-U(2)-k=2-1}) is $t^4y_0^4$
in the $y_0\rightarrow\infty$ limit (first term on the second line).
Absorbing the $y_0^4$ factor into the instanton fugacity (changing microscopic coupling
into the dynamically generated scale in asymptotically free theories), and taking
into account that \cite{Hanany:2012dm} ignores the `ground state energy' factor $t^4$,
we reproduce our result from theirs (in our notation):
\begin{equation}
  \frac{1}{(1-ts)(1-ts^{-1})(1-t^4)}\sum_{m=0}^\infty
  \chi_{m}(s)t^{2m}=
  \frac{1}{(1-ts)(1-ts^{-1})(1-t^2)(1-t^2s^2)(1-t^2s^{-2})}\ .
\end{equation}
The left hand side is the expression in \cite{Hanany:2012dm}. The right hand side
is our (\ref{SD-U(2)-k=2}) in the $y\rightarrow\infty$ limit, apart from the
overall minus sign which arises since we use $-1$ times the $SU(2)$ Haar measure
in the integral (\ref{SD-U(2)-k=2}).
The three contributions of the form (\ref{SD-U(2)-k=2-1}), from
three fixed points, have to be summed over to obtain (\ref{SD-U(2)-k=2}).
The sum is still very complicated, being an infinite
series in fugacities.

One has to further consider the two self-dual $U(2)$ instanton contributions,
where the two instantons are localized at different fixed points. There are three
such cases. We consider the case with a pair of first and second fixed points,
as the other two cases are obtained by permuting $a,b,c$.
The contour integral measure is given by
\begin{equation}
  \hspace*{-0.5cm}\frac{1}{2}\oint\frac{dz}{2\pi iz^2}(z-1)^2
  e^{-\beta(1+a)}e^{-\beta(1+b)}I_{\rm com}^{(1)}
  \frac{2(z^2+1)-z[(t_{(1)}-t_{(1)}^{-1})^2+(t_{(1)}+t_{(1)}^{-1})
  (y_{(1)}+y_{(1)}^{-1})]}{(\zeta-t_{(1)}^2)(\zeta-t_{(1)}^{-2})}
  \cdot\left[(1)\!\rightarrow\!(2)\right]\ .
\end{equation}
The $(1)$ or $(2)$ superscripts or subscripts mean that one has to insert
the suitable values of parameters at the corresponding fixed points.
The poles inside the $z$ contour are $z=0,t_{(1)}^2,t_{(2)}^2$. One should
finally sum over the other two contributions by replacing $a,b,c$ by
$b,c,a$ and $c,a,b$.

Summing over all the above six contributions for two self-dual $U(2)$ instantons
explained above, one obtains $I_{k=0}\cdot e^{-3\beta}$ times
\begin{equation}\label{U(2)-k=3-2}
  -2y^3-2y^2(y_1+y_2+y_3)-y\left(y_1^2+y_2^2+y_3^2-\frac{1}{y_1}-\frac{1}{y_2}-\frac{1}{y_3}
  \right)+\left(\frac{y_1}{y_2}+\frac{y_2}{y_3}+\cdots\right)-y^{-1}(y_1+y_2+y_3)\ .
\end{equation}
Note that the final result is now a finite polynomial in $y,y_1,y_2,y_3$ at
the order $\mathcal{O}(e^{-3\beta})$, which should be the case from the 6d
unitarity bounds for various charges.

%Adding (\ref{U(2)-k=3-1}), (\ref{U(2)-k=3-2}),
Combining  (\ref{U(2)-k=3-1}) and  (\ref{U(2)-k=3-2}), one finds that
the final $U(2)$ index at $k=3$ is given by $I_{k=0}$ times
\begin{eqnarray}\label{U(2)-k=3}
  &&q^3\left[2y^3+2y^2(y_1+y_2+y_3)+y\left(y_1^2+y_2^2+y_3^2-\frac{1}{y_1}-\frac{1}{y_2}
  -\frac{1}{y_3}\right)\right.\nonumber\\
  &&\hspace{2cm}\left.-\left(\frac{y_1}{y_2}+\frac{y_2}{y_1}+\frac{y_2}{y_3}+\frac{y_3}{y_2}
  +\frac{y_3}{y_1}+\frac{y_1}{y_3}\right)+y^{-1}(y_1+y_2+y_3)\right]\ ,
\end{eqnarray}
where $q=e^{-\beta}$. This deviates from the supergravity index (\ref{sugra-single}),
(\ref{sugra-multiple}) only in the first term, where the latter is obtained by
just replacing $2y^3$ in (\ref{U(2)-k=3}) by $3y^3$.
As these terms come from exciting only one complex scalar in the s-wave, the reduction
by $-y^3$ should be due to the finite $N$ restrictions of the chiral operators' spectrum,
somehow similar to the constraints on the multi-trace operators in gauge theories. Also,
as a trivial check, one can take the unrefinement limit $m=\frac{1}{2}-c$,
or $y=y_3$, to find that the result is consistent with that in section 3.2.
One also finds that the absolute values of the powers of $y$ in all terms are
no greater than $3$. This is consistent with the unitarity bound $|R_1-R_2|\leq k$
that one expects from the $OSp(8^\ast|4)$ algebra.

\subsubsection{$U(3)$ index}

Let us turn to the $U(3)$ index. There are many possible anti-self-dual fluxes,
\begin{equation}
  s_0=(2,0,-2)\ ;\ \ s_{1,1}=(1,1,-2)\ ,\ \ s_{1,2}=(2,-1,-1)\ ;\ \
  s_{3}=(1,0,-1)\ ;\ \ s_4=(0,0,0)\ .
\end{equation}
The first subscripts $0,1,3,4$ denote the `energy costs' by these fluxes via
$e^{\frac{\beta}{2}\sum_is_i^2}$, or what we have been calling $k$ in this subsection.

In the sector with the flux $s_0$, the contour integral for the index is given by
\begin{eqnarray}
  I_{(2,0,-2)}&=&e^{4\beta}\oint\frac{zwdzdw}{(2\pi i)^2}\
  \frac{e^{\beta\hat{m}}\ (z-e^{-2\beta a})(z-e^{-2\beta b})(z-e^{-2\beta c})
  (z-e^{\beta a})(z-e^{\beta b})(z-e^{\beta c})}{z(z-e^{-\beta (a-\hat{m})})
  (z-e^{-\beta(b-\hat{m})})(z-e^{-\beta(c-\hat{m})})(z-e^{-\beta\hat{m}})}\nonumber\\
  &&\hspace{1cm}\times
  \frac{e^{\beta\hat{m}}(w-e^{-2\beta a})(w-e^{-2\beta b})(w-e^{-2\beta c})
  (w-e^{\beta a})(w-e^{\beta b})(w-e^{\beta c})}{w(w-e^{-\beta (a-\hat{m})})
  (w-e^{-\beta(b-\hat{m})})(w-e^{-\beta(c-\hat{m})})(w-e^{-\beta\hat{m}})}\nonumber\\
  &&\hspace{1cm}\times
  \frac{e^{2\beta\hat{m}}\prod_{p_1+p_2+p_3=4}(zw-e^{-\beta\sum_ip_ia_i})\cdot
  \prod_{p_1+p_2+p_3=1}(zw-e^{-\beta\sum_ip_ia_i})}
  {zw\prod_{p_1+p_2+p_3=3}(zw-e^{-\beta(\sum_ip_ia_i-\hat{m})})\cdot
  \prod_{p_1+p_2+p_3=2}(zw-e^{-\beta(\sum_ip_ia_i+\hat{m})})}\nonumber\\
  &&\hspace{1cm}\times Z_{\rm inst}^{(1)}Z_{\rm inst}^{(2)}Z_{\rm inst}^{(3)}\ ,
\end{eqnarray}
where $z=z_1/z_2$ and $w=z_2/z_3$. The classical phase from $e^{-is_i\lambda_i}$
is $z^2w^2$, and combining it with the $\frac{dzdw}{zw}$ factor
yields $zwdzdw$ on the first line. Rational functions of $z,w$ on the first, second
and third lines are the perturbative measures from the modes which feel the flux
$s_{12}=2$, $s_{23}=2$, $s_{13}=4$, respectively. $e^{4\beta}$ is from
$e^{\frac{\beta}{2}\sum_i s_i^2}$. Collecting the overall factors of $z,w$,
one finds that this integral has simple poles at $z=0$ and $w=0$. So the
self-dual instantons' contribution
$Z_{\rm inst}^{(1)}Z_{\rm inst}^{(2)}Z_{\rm inst}^{(3)}$ can effectively be
replaced by the $U(1)^3$ instanton partition function in this sector.

The index
in the sector with $s_{1,2}=(2,-1,-1)$ is given by
\begin{eqnarray}
  \hspace*{0cm}I_{(2,-1,-1)}&=&\frac{e^{3\beta}}{2!}\oint\frac{dzdw}{(2\pi i)^2}\
  \frac{e^{\frac{3\beta\hat{m}}{2}}\prod_{p_1+p_2+p_3=3}(z-e^{-\beta\sum_ip_ia_i})\cdot
  (z-1)}{z\prod_{p_1+p_2+p_3=2}(z-e^{-\beta(\sum_ip_ia_i-\hat{m})})\cdot
  \prod_{p_1+p_2+p_3=1}(z-e^{-\beta(\sum_ip_ia_i+\hat{m})})}\nonumber\\
  \hspace*{0cm}&&\hspace{1cm}\times
  \frac{e^{\frac{3\beta\hat{m}}{2}}\prod_{p_1+p_2+p_3=3}(w-e^{-\beta\sum_ip_ia_i})\cdot
  (w-1)}{w\prod_{p_1+p_2+p_3=2}(w-e^{-\beta(\sum_ip_ia_i-\hat{m})})\cdot
  \prod_{p_1+p_2+p_3=1}(w-e^{-\beta(\sum_ip_ia_i+\hat{m})})}\nonumber\\
  \hspace*{0cm}&&\hspace{1cm}\times\frac{(z/w-1)^2}{z/w}\cdot
  Z_{\rm inst}^{(1)}Z_{\rm inst}^{(2)}Z_{\rm inst}^{(3)}\ .
\end{eqnarray}
Here, we defined $z=z_1/z_3$, $w=z_2/z_3$.
The sector with flux $s_{1,1}=(1,1,-2)$ will yield an identical contribution,
namely $I_{(1,1,-2)}=I_{(2,-1,-1)}$. To clearly see the structure of this integral,
it is perhaps more convenient to introduce new integral variables $w$, $\zeta\equiv z/w$.
Then the above integral is given by
\begin{eqnarray}
  \hspace*{-.5cm}I_{(2,-1,-1)}&=&\frac{e^{3\beta(1+\hat{m})}}{2!}
  \oint\frac{dwd\zeta}{(2\pi i)^2w\zeta^2}\
  \frac{\prod_{p_1+p_2+p_3=3}(w\zeta-e^{-\beta\sum_ip_ia_i})\cdot
  (w\zeta-1)}{\prod_{p_1+p_2+p_3=2}(w\zeta-e^{-\beta(\sum_ip_ia_i-\hat{m})})\cdot
  \prod_{p_1+p_2+p_3=1}(w\zeta-e^{-\beta(\sum_ip_ia_i+\hat{m})})}\nonumber\\
  \hspace*{-.5cm}&&\hspace{1cm}\times
  \frac{\prod_{p_1+p_2+p_3=3}(w-e^{-\beta\sum_ip_ia_i})\cdot
  (w-1)}{\prod_{p_1+p_2+p_3=2}(w-e^{-\beta(\sum_ip_ia_i-\hat{m})})\cdot
  \prod_{p_1+p_2+p_3=1}(w-e^{-\beta(\sum_ip_ia_i+\hat{m})})}\nonumber\\
  \hspace*{-.5cm}&&\hspace{1cm}\times(\zeta-1)^2\cdot
  Z_{\rm inst}^{(1)}Z_{\rm inst}^{(2)}Z_{\rm inst}^{(3)}\ .
\end{eqnarray}
The $U(3)$ gauge symmetry of the theory is broken by the $(2,-1,-1)$ flux
to $U(1)\times U(2)$. In $Z_{\rm inst}^{(1)}Z_{\rm inst}^{(2)}Z_{\rm inst}^{(3)}$,
the contributions of charged modes in the broken part of the gauge group depend on
$w$ or $z=w\zeta$. Since the classical plus perturbative measure yields a
simple pole for $w$ on the first line, one finds that the last charged modes'
contribution can be replaced by $F(x)\rightarrow 1$, as explained around
(\ref{charged-decompose}). So $Z_{\rm inst}^{(1)}Z_{\rm inst}^{(2)}Z_{\rm inst}^{(3)}$
can be replaced by the instanton partition function for the $U(1)\times U(2)$ gauge group,
and the $w$ contour integral is a small circle surrounding the pole $w=0$.
Upon inserting $w=0$ to the perturbative measure to extract the residue,
one obtains a factor $e^{-6\beta\hat{m}}$.
The $\zeta$ contour is nontrivial, as explained in section 2.3.

The index in the sector with $s_3=(1,0,-1)$ flux is given by
\begin{eqnarray}\label{U(3)-3}
  \hspace*{-1.2cm}I_{(1,0,-1)}&=&e^{\beta}\oint\frac{dzdw}{(2\pi i)^2}
  \frac{e^{\beta\hat{m}/2}(z\!-\!e^{-\beta a})(z\!-\!e^{-\beta b})
  (z\!-\!e^{-\beta c})}{z(z\!-\!e^{\beta\hat{m}})}\cdot
  \frac{e^{\beta\hat{m}/2}(w\!-\!e^{-\beta a})(w\!-\!e^{-\beta b})
  (w\!-\!e^{-\beta c})}{w(w\!-\!e^{\beta\hat{m}})}\\
  \hspace*{-1.2cm}&&\times
  \frac{e^{\beta\hat{m}}\ (zw-e^{-2\beta a})(zw-e^{-2\beta b})(zw-e^{-2\beta c})
  (zw-e^{\beta a})(zw-e^{\beta b})(zw-e^{\beta c})}{zw(zw-e^{-\beta (a-\hat{m})})
  (zw-e^{-\beta(b-\hat{m})})(zw-e^{-\beta(c-\hat{m})})(zw-e^{-\beta\hat{m}})}
  \cdot Z_{\rm inst}^{(1)}Z_{\rm inst}^{(2)}Z_{\rm inst}^{(3)}\nonumber
\end{eqnarray}
where $z=z_1/z_2$, $w=z_2/z_3$. The contour integrals for $z,w$ are both
along small circles surrounding the poles at $z=0$ and $w=0$. Note that
both poles are double poles. So the self-dual instantons' contribution
$Z_{\rm inst}^{(1)}Z_{\rm inst}^{(2)}Z_{\rm inst}^{(3)}$ could be more than
just $U(1)^3$ instantons.

The index in the sector with $s_4=(0,0,0)$ is given by
\begin{equation}
  I_{(0,0,0)}=\frac{1}{3!}\oint\frac{dzdw}{(2\pi i)^2zw}\ \frac{(z-1)^2}{z}
  \ \frac{(w-1)^2}{w}\ \frac{(zw-1)^2}{zw}\
  Z_{\rm inst}^{(1)}Z_{\rm inst}^{(2)}Z_{\rm inst}^{(3)}\ ,
\end{equation}
where $z=z_1/z_2$, $w=z_2/z_3$.

Again we start from $k=3$.
The $k=3$ contributions come from: (i) $s=(2,0,-2)$ and $k_{SD}=3$ $U(1)^3$
self-dual instantons, (ii) $s=(1,1,-2)$ or $(2,-1,-1)$ and $k_{SD}=2$
$U(1)\times U(2)$ self-dual instantons, (iii) $s=(1,0,-1)$ and no self-dual instantons.

The first case with $s=(2,0,-2)$, $k_{SD}=3$ is simply a multiplication of the vacuum index
and $U(1)^3$ Abelian index at $k_{SD}=3$. This is $I_{k=0}\cdot e^{-3\beta}$ times
\begin{eqnarray}
  &&3\left[y^2(y_1+y_2+y_3)+y(y_1^2+y_2^2+y_3^2)+y^{-1}(y_1+y_2+y_3)
  -(1+\frac{y_1}{y_2}+\frac{y_2}{y_1}+\cdots)+y^3\right]\nonumber\\
  &&+6y\left[y(y_1+y_2+y_3)-(y_1^{-1}+y_2^{-1}+y_3^{-1})+y^{-1}+y^2\right]+y^3
\end{eqnarray}
from (\ref{multi-abelian}).
The third contribution with $s=(1,0,-1)$ and no self-dual instantons is obtained
by picking the residues at $z=0$, $w=0$ from (\ref{U(3)-3}). The result is
\begin{eqnarray}
  &&e^{-2\beta\hat{m}}
  \left[e^{-\beta\hat{m}}-(e^{\beta a}+e^{\beta b}+e^{\beta c})\right]^2\nonumber\\
  &&+e^{-2\beta\hat{m}}\left[e^{\beta\hat{m}}+e^{-\beta\hat{m}}(e^{\beta a}+
  e^{\beta b}+e^{\beta c})-(e^{2\beta a}+e^{2\beta b}+e^{2\beta c})-
  (e^{-\beta a}+e^{-\beta b}+e^{-\beta c})\right]\nonumber\\
  &&=e^{-\beta\hat{m}}+e^{-2\beta\hat{m}}(e^{-\beta a}+e^{-\beta b}+e^{-\beta c})
  -e^{-3\beta\hat{m}}(e^{\beta a}+e^{\beta b}+e^{\beta c})+e^{-4\beta\hat{m}}\nonumber\\
  &&=I_{k=0}\cdot e^{-3\beta}\left[e^{3\beta\hat{m}}
  +e^{2\beta\hat{m}}(e^{-\beta a}+e^{-\beta b}+e^{-\beta c})
  -e^{\beta\hat{m}}(e^{\beta a}+e^{\beta b}+e^{\beta c})+1\right]\ .
\end{eqnarray}

The second cases with $s=(1,1,-2)$ or $(2,-1,-1)$ come in three possible
ways: two self-dual instantons being in the Abelian sector, one being in Abelian
and another in the $U(2)$ sector, and finally both being in the $U(2)$ sector.
Let us only consider the flux $s=(2,-1,-1)$, as the sector with $s=(1,1,-2)$ will
give the same result. The first case with two Abelian self-dual instantons is simple,
\begin{equation}
  I_{k=0}\cdot(-2e^{-\beta}e^{\beta\hat{m}})\cdot
  e^{-2\beta}\left[e^{\beta\hat{m}}(e^{-\beta a}+e^{-\beta b}+e^{-\beta c})
  -(e^{\beta a}+e^{\beta b}+e^{\beta c})+e^{-\beta\hat{m}}+e^{2\beta\hat{m}}\right]\ ,
\end{equation}
where the factor $2$ comes by adding contributions from two $s$ fluxes. The
second case with one Abelian and one $U(2)$ single instantons can also be obtained
by using calculations that we already did. Firstly, the Abelian instanton just
yields an overall factor which does not affect the contour integral. The remaining
contour integral for $z,w$, or $\zeta=z/w,w$ with single $U(2)$ instanton, has been
already done, as a $U(3)$ $k=2$ contribution at the same $s$ flux. So multiplying
them all, one obtains
\begin{equation}
  -2I_{k=0}\cdot e^{-2\beta}\left[e^{2\beta\hat{m}}+e^{-\beta\hat{m}}
  -(e^{\beta a}+e^{\beta b}+e^{\beta c})+e^{\beta\hat{m}}(e^{-\beta{a}}+
  e^{-\beta b}+e^{-\beta c})\right]\cdot e^{-\beta}e^{\beta\hat{m}}\ ,
\end{equation}
where the last factor comes from the Abelian single instanton. Again, the
overall factor $2$ comes from two possible $s_{1j}$ fluxes.

Finally, we have to consider the case with two $U(2)$ self-dual instantons.
It can first get contributions from both instantons localized at one of the
three fixed points, and also from two instantons localized at different fixed points.
The calculation is basically the same as $U(2)$ $k=3$ calculation in section 3.3.1.
The $Z_{\rm inst}^{(1)}Z_{\rm inst}^{(2)}Z_{\rm inst}^{(3)}$ part of the integrand
is the same,
\begin{equation}
  \hspace*{-1.7cm}(Z_{\rm inst}^{(1)})_{k_{SD}=2}+(Z_{\rm inst}^{(2)})_{k_{SD}=2}
  +(Z_{\rm inst}^{(3)})_{k_{SD}=2}+(Z_{\rm inst}^{(1)})_{k_{SD}=1}
  (Z_{\rm inst}^{(2)})_{k_{SD}=1}+(Z_{\rm inst}^{(2)})_{k_{SD}=1}
  (Z_{\rm inst}^{(3)})_{k_{SD}=1}+(Z_{\rm inst}^{(3)})_{k_{SD}=1}
  (Z_{\rm inst}^{(1)})_{k_{SD}=1}
\end{equation}
with $z$ variable in the $U(2)$ case replaced by $\zeta$. The only difference
is that the perturbative measure picks up a factor of $e^{-3\beta\hat{m}}$ during
$w$ integration. Doing the same exercise as in section 3.3.1, one obtains
$I_{k=0}\cdot e^{-3\beta}$ times
\begin{equation}
  -4y^3-4y^2(y_1+y_2+y_3)-2y\left(y_1^2+y_2^2+y_3^2-\frac{1}{y_1}-\frac{1}{y_2}-\frac{1}{y_3}
  \right)+2\left(\frac{y_1}{y_2}+\frac{y_2}{y_3}+\cdots\right)+-2y^{-1}(y_1+y_2+y_3)\ .
\end{equation}
The extra factor of $2$ is multiplied because there is another sector with flux
$(1,1,-2)$, providing an identical contribution.

Summarizing all the contributions at $U(3)$ $k=3$, one obtains $I_{k=0}e^{-3\beta}$ times
\begin{eqnarray}
  \hspace*{-1cm}
  Z_{(2,0,-2)}&=&3\left[y^2(y_1+y_2+y_3)+y(y_1^2+y_2^2+y_3^2)+y^{-1}(y_1+y_2+y_3)
  -(1+\frac{y_1}{y_2}+\frac{y_2}{y_1}+\cdots)+y^3\right]\nonumber\\
  \hspace*{-1cm}
  &&+6y\left[y(y_1+y_2+y_3)-(y_1^{-1}+y_2^{-1}+y_3^{-1})+y^{-1}+y^2\right]+y^3\nonumber\\
  \hspace*{-1cm}
  2Z_{(2,-1,-1)}&=&-2y\left[y(y_1+y_2+y_3)-(y_1^{-1}+y_2^{-1}+y_3^{-1})
  +y^{-1}+y^2\right]\nonumber\\
  \hspace*{-1cm}&&-2y\left[y(y_1+y_2+y_3)-(y_1^{-1}+y_2^{-1}+y_3^{-1})
  +y^{-1}+y^2\right]\nonumber\\
  \hspace*{-1cm}
  &&-4y^3\!-\!4y^2(y_1\!+\!y_2\!+\!y_3)\!-\!2y\left(\!y_1^2\!+\!y_2^2\!+\!y_3^2\!
  -\!\frac{1}{y_1}\!-\!\frac{1}{y_2}\!-\!\frac{1}{y_3}\!\right)\!+\!
  2\left(\!\frac{y_1}{y_2}\!+\!\frac{y_2}{y_3}\!+\!\cdots\!\right)\!-\!
  2y^{-1}(y_1\!+\!y_2\!+\!y_3)\nonumber\\
  \hspace*{-1cm}Z_{(1,0,-1)}&=&y^3+y^2(y_1+y_2+y_3)
  -y(y_1^{-1}+y_2^{-1}+y_3^{-1})+1\ .
\end{eqnarray}
The first, second, third lines of $2Z_{(2,-1,-1)}=Z_{(2,-1,-1)}+Z_{(1,1,-2)}$
come from two Abelian self-dual instantons, one Abelian and one $U(2)$ instantons,
and two $U(2)$ instantons, respectively. From (\ref{sugra-single}), (\ref{sugra-multiple}),
the large $N$ gravity index at $k=3$ is given by $e^{-3\beta}$ times
\begin{equation}
  3y^3+2y^2(y_1+y_2+y_3)+y\left(y_1^2+y_2^2+y_3^3-\frac{1}{y_1}-\frac{1}{y_2}
  -\frac{1}{y_3}\right)-\left(\frac{y_1}{y_2}+\frac{y_2}{y_1}+\cdots\right)
  +y^{-1}(y_1+y_2+y_3)\ .
\end{equation}
From the gauge theory expressions above, one finds that  $Z_{(2,0,-2)}+2Z_{(2,-1,-2)}+Z_{(1,0,-1)}$ completely agrees with
the supergravity index at $k=3$. This supports the fact that the finite $N$
gauge theory index agrees with supergravity at $k\leq N$.

\section{Aspects of the theory with $\mathbb{Z}_K$ quotient}

In this section, we explain some structure of the 6d theory with nontrivial
$\mathbb{Z}_K$ quotient with $K\geq 2$, and its index. We first discuss the index
of the Abelian theory at general $K$, slightly generalizing our index in
section 3. As the Abelian index at general $K$ can also be calculated directly from 6d,
this will provide a further modest support for the correctness of our index.
Secondly, we explain how the `vacuum energy' from anti-self-dual fluxes behave
for various values of $N$ and $K$. One purpose of this discussion is to illustrate
that the appearance of $-\frac{N(N^2-1)}{6}\sim -\frac{N^3}{6}$ scaling is indeed
a strong coupling phenomenon from 5d QFT's viewpoint, which is absent for $K\geq N$
with small 't Hooft coupling $\frac{N}{K}\leq 1$.

Let us first discuss the index at general $K$.
The localization calculus we explained for $K=1$ can be generalized
to general $K$, and the result is as follows:
\begin{equation}
  \sum_{s_1,s_2,\cdots,s_N=-\infty}^\infty\oint\prod_{i=1}^N
  \frac{d\lambda_i}{2\pi}e^{\frac{K\beta}{2}\sum_{i=1}^Ns_i^2}
  e^{-iK\sum_{i=1}^Ns_i\lambda_i}Z_{\rm pert}^{(1)}Z_{\rm inst}^{(1)}\cdot
  Z_{\rm pert}^{(2)}Z_{\rm inst}^{(2)}\cdot Z_{\rm pert}^{(3)}Z_{\rm inst}^{(3)}\ ,
\end{equation}
where each perturbative part on $\mathbb{R}^4\times S^1$ takes the same form
as above that we explained for $K=1$, and the instanton part is given by
\begin{equation}\label{K-Zinst}
  Z_{\rm inst}^{(i)}=\sum_{k=0}^\infty e^{-Kk\beta(1+a_i)}
  Z_{k}^{(i)}(\epsilon_1,\epsilon_2,m_0,\lambda)\ .
\end{equation}
The rules for identifying $\epsilon_1,\epsilon_2,m$ at each fixed point
is the same as before, and the functional forms of $Z_{k}^{(i)}$
are again given by (\ref{Zk}). The only difference is the multiplication
of $K$ to all the contributions coming from the classical action in the localization
calculus. Presumably, at least for the theory with $n=\pm\frac{1}{2}$, the contour
prescription that we derived should hold. We expect so because the 6d spectrum
after $\mathbb{Z}_K$ is still constrained by the $OSp(8^\ast|4)$ algebra of the
mother theory before the orbifold, so all the unitarity bound arguments used in
section 2.3 will be applicable.

Our focus here is the Abelian index.  The contour integral in this case is trivial,
and the flux sum is just over $s=0$. Namely, the result from the 5d QFT is just
\begin{equation}
  Z_{\rm pert}^{(1)}Z_{\rm inst}^{(1)}\cdot
  Z_{\rm pert}^{(2)}Z_{\rm inst}^{(2)}\cdot Z_{\rm pert}^{(3)}Z_{\rm inst}^{(3)}\ ,
\end{equation}
where all factors in the $U(1)$ case do not depend on $\lambda$,
and the only appearance of $K$ is through the coupling to the self-dual instanton
number in (\ref{K-Zinst}). For simplicity, let us only consider the QFT with
$n=-\frac{1}{2}$. The perturbative contribution is completely the same as the
that for $K=1$,
\begin{equation}
  Z_{\rm pert}^{(1)}Z_{\rm pert}^{(2)}Z_{\rm pert}^{(3)}=1\ ,
\end{equation}
since $K$ does not appear in this part at all. The product of $3$ instanton parts
is given by
\begin{equation}\label{ZK-Abelian-5d}
  PE\left[I_-\left(b-a,c-a,m-\frac{1+a}{2}\right)\frac{e^{-K\beta(1+a)}}
  {1-e^{-K\beta(1+a)}}+(a,b,c\rightarrow b,c,a)+(a,b,c\rightarrow c,a,b)\right]\ .
\end{equation}
The only difference compared to the case with $K=1$ is replacing
$e^{-\beta(1+a_i)}$ by $e^{-K\beta(1+a_i)}$ in the instanton factors.

Since the $\mathbb{Z}_K$ orbifold acts on the 6d fields, the expected result
is given as follows. Firstly, the 6d `letter index' at $K=1$ is given (as explained
in section 3.1) by
\begin{equation}\label{letter-6d}
  I_{\rm letter}=
  \frac{e^{-\beta}e^{\beta\hat{m}}+e^{-2\beta}e^{-\beta\hat{m}}
  -e^{-2\beta}(e^{\beta a}+e^{\beta b}+e^{\beta c})+e^{-3\beta}}
  {(1-e^{-\beta(1+a)})(1-e^{-\beta(1+b)})(1-e^{-\beta(1+c)})}\ ,
\end{equation}
where we again introduced $\hat{m}=m-\frac{1}{2}$.
The $\mathbb{Z}_K$ orbifold on the 6d fields requires the charge $j_1+j_2+j_3+R_1+2R_2\stackrel{BPS}{=}
E-R_1$ to take eigenvalues which are integer multiples of $K$. In
the index, the measure factor includes
\begin{equation}
  e^{-\beta(E-\frac{R_1+R_2}{2})}e^{m\beta(R_1-R_2)}=
  e^{-\beta(E-R_1)}e^{\beta\hat{m}(R_1-R_2)}\ .
\end{equation}
Thus, in the chemical potential basis which uses $\beta\hat{m}, \beta a_i$
instead of $\beta m, \beta a_i$, the effect of $\mathbb{Z}_K$ on the letter index is
to expand (\ref{letter-6d}) in a power series of $e^{-\beta}$, and then
keep the terms which are integer powers of $e^{-K\beta}$ only.
But we have shown the identity
\begin{equation}
  I_{\rm letter}=I_-(b-a,c-a,\hat{m}-a/2)\frac{e^{-\beta(1+a)}}{1-e^{-\beta(1+a)}}
  +(a,b,c\rightarrow b,c,a)+(a,b,c\rightarrow c,a,b)
\end{equation}
in section 3.1, where the right hand side is the exponent in $PE$ from the 5d QFT
calculation. So we can expand this right hand side in $e^{-\beta}$ and project to
the integer powers of $e^{-K\beta}$, to obtain the $S^5/\mathbb{Z}_K\times S^1$ index
from the 6d perspective. This is nothing but the exponent of $PE$ in
(\ref{ZK-Abelian-5d}), exactly agreeing with the 5d QFT calculation on
$\mathbb{CP}^2\times S^1$ at general $K$.

We now turn to the discussion of the vacuum `energy' contribution from the
anti-self-dual flux for the non-Abelian theory at general $K$. Again,
when we say `energy,' it is understood as $E-R_1$ at $n=-\frac{1}{2}$.
Of course the true vacuum energy factor in principle could be a combination of
the anti-self-dual fluxes, 5d perturbative modes, and also extra subtle contributions
from self-dual instantons on which we comment in section 5.1. However, since
the most nontrivial and novel part is the first one, exhibiting $N^3$
scaling behavior at large $N$, we concentrate on this part only. Also, it is not
completely clear to us whether the index we derived above is the fully general one
for $K>1$. However, our index would at least account for a subsector of BPS states,
and showing that the ground state energy from the $s$ flux in there is a strong coupling
effect seems to be a valuable exercise to do.

The flux $s=(N-1,N-3,\cdots,-(N-1))$ for the ground state is allowed only at
$K=1$. It is interesting to see what is the maximal value of $K$, or minimal value
of the 't Hooft coupling $\lambda$, which admits nonzero $s$ at the vacuum.
To ease the analysis, let us work at the point $m=\frac{1}{2}$,
$a=b=c=0$. The perturbative integration measure is simply the Haar measure, and
the instanton correction factors out as it does not depend on $\lambda$.
One obtains the following expression:
\begin{equation}
  \frac{1}{N!}\oint[d\lambda_i]\sum_s e^{\frac{K\beta}{2}\sum_is_i^2}
  e^{-iK\sum_is_i\lambda_i}\prod_{i<j}\left(2\sin\frac{\lambda_{ij}}{2}\right)^2\ .
\end{equation}
We would like to check what is the value of $s$ (up to permutation of $N$ fluxes)
that yields lowest energy $-\frac{K}{2}\sum_{i=1}^Ns_i^2$. The Haar measure can
be expanded to a finite polynomial of the $N$ phases $e^{i\lambda_i}$.
For the integral to be nonzero, the classical phase $e^{-iKs_i\lambda_i}$ should
be canceled against a term in the Haar measure.

Firstly, the maximal phase that can be provided by the Haar measure
for a given $\lambda_i$ is $e^{\pm(N-1)i\lambda_i}$. So for large enough $K$,
say at $K\geq N$, clearly one cannot turn on nonzero quantized $s$ fluxes for
the ground state, as the classical phase
$e^{-iKs\lambda}$ cannot be canceled by the Haar measure. Thus, having nonzero $s$
fluxes (and thus negative energy contribution from it) for the vacuum is forbidden
for the 't Hooft coupling constant $\lambda\equiv\frac{N}{K}\leq 1$, and in particular in the weak-coupling regime
$\lambda\ll 1$. This is consistent with the expectation that the degrees of freedom in weak coupling should be order $N^2$ as it becomes a free theory.

Introducing $z_i=e^{-i\lambda_i}$, $\zeta_i=\frac{z_i}{z_{i+1}}$ and again
ordering the fluxes to satisfy $s_1\geq s_2\geq\cdots\geq s_N$,
the measure one obtains is
\begin{equation}
  \left(\zeta_1^{Ks_1}\zeta_2^{K(s_1+s_2)}\cdots\zeta_{N-1}^{K(s_1+\cdots s_{N-1})}\right)
  \cdot\frac{1}{\prod_{n=1}^{N-1}(\zeta_n)^{n(N-n)}}\left(1+\cdots\right)\ .
\end{equation}
The first factor comes from the classical phase, and all $\zeta_n$'s take
positive exponents $K(s_1+s_2+\cdots+s_n)\geq 0$ for $K\geq 1$. The second
factor comes from the Haar measure, where $\cdots$ denote terms in non-negative
powers of $\zeta_n$'s. Thus, a necessary condition for the above expression to
contain a term with canceling phases is $K\sum_{i=1}^ns_i\leq n(N-n)$
for all $n=1,2,\cdots,N-1$. Once one finds solutions for $s$ satisfying these
inequalities, one should actually check if there is precisely a term among $1$
above or the omitted $\cdots$ terms in the Haar measure, which cancels
the classical phase. If there are more than one such terms, the $s$ flux with
lower energy would yield the vacuum.

\begin{table}[t!]
$$
\begin{array}{c|ccccccccc}
  \hline K&U(2)&U(3)&U(4)&U(5)&U(6)&U(7)&U(8)&U(9)&U(N)\\
  \hline 1&-1&-4&-10&-20&-35&-56&-84&-120&-\frac{N(N^2-1)}{6}\\
  2&0&-1&-2&-5%_{2,1}
  &-8%_{2,2}
  &-14%_{3,2,1}
  &-20%_{3,3,1,1}
  &-30%_{4,3,2,1}
  &
  %\left[-\frac{n(n+1)(2n+1))}{6}\right]_{N=2n+1}
  \\
  3&&0&-1&-2&-3&-6%_{2,1,1}
  &-9%_{2,2,1}
  &-12%_{2,2,2}
  \\
  4&&&0&-1&-2&-3&-4&-7%_{2,1,1,1}
  \\
  5&&&&0&-1&-2&-3&-4\\
  6&&&&&0&-1&-2&-3\\
  7&&&&&&0&-1&-2\\
  8&&&&&&&0&-1\\
  9&&&&&&&&0\\
  \hline
\end{array}
$$
\caption{$K^{-1}$ times the vacuum energy,
$K^{-1}\epsilon_0=-\frac{1}{2}\sum_{i=1}^Ns_i^2$}\label{vacuum}
\end{table}
After a case-by-case study, the vacuum `energies' for various values of $K$ and $N$
are listed in Table \ref{vacuum}. Clearly the vacuum energy from the flux increases
to $0$ as $K$ increases towards $N$. In the regime $K>N$ which is blank in the table,
the values are all $0$ as generally argued above. This clearly illustrates that the
$N^3$ scaling of the vacuum energies from the $s$ flux is a strong coupling effect,
disappearing in the regime $K\gg N$ which admits a weakly coupled 5d description
at low energy.

Also, for $N\gg K$, the leading behavior becomes $\epsilon_0\approx-\frac{N^3}{6K}$.
We checked that the error of this approximation is very small for large values of $N$
and moderate values of $K$. The dependence on the combination $\frac{N^3}{K}$ agrees
with what one expects from the supergravity dual on $(AdS_7\times S^1)/\mathbb{Z}_K$,
where the $\frac{1}{K}$ factor comes from the $\mathbb{Z}_K$ quotient on the 11
dimensional geometry. The coefficient of $\frac{N^3}{K}$ in the gravity side
has not been properly calculated yet. We think its proper calculation should reflect
the aspects of this quantity that we explain in section 5.1 in the gravity dual version.
But if it is nonzero, the only possible $N,K$ dependence is $\frac{N^3}{K}$.

\section{Remarks}

We derived an expression for the 6d $(2,0)$ superconformal index, with a number of
nontrivial supports for the correctness of the result by comparing it to other known
results. Still, there are many subtle aspects to be clarified, and interesting future
directions based on the ideas in this paper. In this final section, we comment on some
of them.

All the supports that we found in section 3 are for the spectrum of BPS excitations.
The vacuum `energy' that is captured as an overall factor could also reveal an interesting
information on the 6d theory. As this quantity is closely related but nonetheless
somewhat different from the conventional Casimir energy of the QFT on
$S^5\times\mathbb{R}$, we first elaborate on its definition with a concrete
illustration from the Abelian theory in section 5.1. We think there appears a subtlety
in calculating the subleading corrections of this quantity from our QFT on
$\mathbb{CP}^2\times\mathbb{R}$. We explain this as well, comparing the situation
with a similar calculation done from a QFT on $S^5$. The two different
5d QFT's provide complementary viewpoints on the 6d physics.

In section 5.2, we further discuss the precise relation between the two 5d QFT's on $S^5$
and on $\mathbb{CP}^2\times S^1$, related to the S-duality of the 5d maximal SYM
on $\mathbb{R}^4\times S^1$ \cite{Tachikawa:2011ch}.

In section 5.3, we comment on the application of the ideas in this paper to the
6d $(1,0)$ SCFT's and their indices. Some general discussions are made,
with possible new aspects compared to the $(2,0)$ theory discussed in this paper.

\subsection{The Casimir energy and related issues}

In section 3, we saw that the index captures a multiplicative factor of
$e^{\beta\frac{N(N^2-1)}{6}}$ from the $s$ fluxes, which can be understood as a
contribution to the `vacuum energy' $\epsilon_0\leftarrow-\frac{N(N^2-1)}{6}$ or
more precisely the vacuum expectation value of $E-R_1$. The interpretation of this
multiplicative factor is given in \cite{Kim:2012av}, generalizing \cite{Aharony:2003sx}
in the index version. Let us review this first, with a concrete example given by
the Abelian 6d theory.

The naive and abstract definition of this quantity is given by the expectation value
of the charge $E-R_1$ for the vacuum on $S^5\times\mathbb{R}$,
\begin{equation}\label{vacuum-energy}
  \langle E-R_1\rangle=-\left.\frac{\partial}{\partial\beta}\log Z[S^5\times S^1]
  \right|_{\beta\rightarrow\infty}\ ,
\end{equation}
where $Z[S^5\times S^1]$ in our case is the 6d index, with $\beta$ conjugate
to $E-R_1$. This quantity has to be carefully defined. To concretely illustrate
the subtleties, let us discuss the free QFT as an example, e.g.
the 6d Abelian $(2,0)$ theory. Then, just like the normal Casimir energy,
(\ref{vacuum-energy}) is given by the alternating sum of the zero point values
of $E-R_1$ carried by the free particle oscillators:
\begin{equation}\label{free-casimir-index}
  (\epsilon_0)_{\rm index}\equiv{\rm tr}\left[(-1)^F\frac{E-R_1}{2}\right]=
  \sum_{\rm bosonic\ modes}\frac{E-R_1}{2}
  -\sum_{\rm fermionic\ modes}\frac{E-R_1}{2}\ .
\end{equation}
The trace is over the infinitely many free particle modes. This is somewhat similar to
the ordinary Casimir energy defined by
\begin{equation}\label{free-casimir-proper}
  \epsilon_0\equiv{\rm tr}\left[(-1)^F\frac{E}{2}\right]=\sum_{\rm bosonic\ modes}\frac{E}{2}
  -\sum_{\rm fermionic\ modes}\frac{E}{2}\ ,
\end{equation}
which appears while one computes the partition function of a QFT on $S^n\times\mathbb{R}$
with inverse-temperature $\beta$ conjugate to the energy $E$ \cite{Aharony:2003sx}.
Both expressions are formal, and should be supplemented by a suitable regularization
of the infinite sums. As in \cite{Aharony:2003sx} for the latter quantity, one can use
the charges carried by the summed-over states to provide regularizations.
The charges that can be used in the regulator are constrained by the symmetries of
the problem under considerations, which are different between
(\ref{free-casimir-index}) and (\ref{free-casimir-proper}).

For (\ref{free-casimir-proper}), the only charge that one can use to regularize the sum
is energy $E$ \cite{Aharony:2003sx}. This is because the symmetry of the path integral
for the partition function on $S^n\times S^1$ includes all the internal symmetry of the
theory, together with the rotation symmetry on $S^n$. Firstly, non-Abelian rotation
symmetries forbid nonzero vacuum expectation values of angular momenta on $S^n$. Also,
there are no sources which will give nonzero values for the internal charges: its
expectation value is zero either if the internal symmetry is non-Abelian, or
if there are sign flip symmetries of the Abelian internal symmetries. On the other hand,
energy $E$ can be used in the regulator function, as $E$ has a definite sign.
The remaining procedure of properly defining (\ref{free-casimir-proper}) is explained
in \cite{Aharony:2003sx}. One introduces a regulator function $f(E/\Lambda)$ with
a UV cut-off $\Lambda$ (to be sent back to infinity at the final stage) which satisfies
the properties $f(0)=1$, $f(\infty)=0$ and is sufficiently flat at $E/\Lambda=0$: $f^\prime(0)=0$,
$f^{\prime\prime}(0)=0$, etc. The rigorous definition replacing
(\ref{free-casimir-proper}) is given by
\begin{equation}
  {\rm tr}\left[(-1)^F\frac{E}{2}f(E/\Lambda)\right]\ .
\end{equation}
When energy level $E$ has an integer-spacing, $E=\frac{m}{R}$ with $m=1,2,3,\cdots$,
and the degeneracy for given $m$ is a polynomial of $m$ (as in \cite{Aharony:2003sx}),
one can easily show that this definition is the same as
\begin{equation}\label{reg-casimir-proper}
  {\rm tr}\left[(-1)^F\frac{E}{2}e^{-\beta^\prime E}\right]
  =-\frac{1}{2}\frac{d}{d\beta^\prime}{\rm tr}\left[(-1)^Fe^{-\beta^\prime E}\right]\ ,
\end{equation}
where we take $\beta^\prime\rightarrow 0$ at the final stage.

On the other hand, the correct regularization of (\ref{free-casimir-index}) is
constrained by different symmetries. Let us first explain this quantity at the
special point $m=\pm\frac{1}{2}$, $a=b=c=0$ where some SUSY enhancement is expected.
From the SYM on $S^5$, enhancement to 16 SUSY contained in the $SU(4|2)_\pm$ supergroup
is explicitly visible \cite{Kim:2012av}. $SU(4|2)_\pm$ is defined to be a subgroup of
$OSp(8^\ast|4)$ which commutes with $E-R_1$ for $+$ subscript, and with $E-R_2$ for
$-$ subscript. From the QFT on $\mathbb{CP}^2\times S^1$, only a subset of
this group is manifestly visible, but still one expects $16$ SUSY and $SU(4|2)$
enhanced symmetry by having some conserved currents carrying nonzero instanton charges.
Thus, with $SU(4|2)_+$ symmetry in the path integral at $m=\frac{1}{2}$, $a=b=c=0$,
the only charge which commutes with this symmetry group is $E-R_1$, and only this
can be used to regularize the sum (\ref{free-casimir-index}). So this sum can be
regularized as ${\rm tr}\left[(-1)^F\frac{E-R_1}{2}f(\frac{E-R_1}{\Lambda})\right]$,
or equivalently as
\begin{equation}\label{reg-casimir-index}
  {\rm tr}\left[(-1)^F\frac{E-R_1}{2}e^{-\beta^\prime(E-R_1)}\right]=
  -\frac{1}{2}\frac{d}{d\beta^\prime}{\rm tr}\left[(-1)^Fe^{-\beta^\prime(E-R_1)}\right]\ .
\end{equation}
The quantities
\begin{equation}
  {\rm tr}\left[(-1)^F e^{-\beta^\prime(E-R_1)}\right]\ ,\ \
  {\rm tr}\left[(-1)^F e^{-\beta^\prime E}\right]
\end{equation}
appearing in (\ref{reg-casimir-index}) and (\ref{reg-casimir-proper}) are the
letter index, and the letter partition function with minus sign inserted for fermions,
respectively, with $\beta^\prime$ playing the role of chemical potentials. (The latter
quantity is not an index because the measure in ${\rm tr}$ breaks all SUSY.) So once
we know the letter index or the letter partition function, one can calculate
$(\epsilon_0)_{\rm index}$ or $\epsilon_0$ for the Abelian 6d $(2,0)$ theory.

Let us calculate the two letter partition functions and the two Casimir energies
for the Abelian theory, to get a better understanding. The Abelian theory has
$5$ real scalars, chiral fermions with $4\times 4$ components (first and second $4$ from
the $SO(5)_R$ spinor and $SO(6)$ chiral spinor), a 3-form flux subject to the self-duality
condition and the Bianchi identity. Letters are defined as a single field dressed by
many derivatives on $\mathbb{R}^6$ acting on them. The charges of these fields and
derivatives under $E$ and $R_1$ are given in Table \ref{abelian-letter}.
\begin{table}[t!]
$$
\begin{array}{c|cccc}
  \hline {\rm field}&\phi^I&\Psi^i_\alpha&H_{MNP}&\partial_M\\
  \hline E&2&\frac{5}{2}&3&1\\
  R_1&(+1,-1,0,0,0)&(+\frac{1}{2},-\frac{1}{2})&0&0\\
  \hline
\end{array}
$$
\caption{The scale dimension $E$ and R-charge $R_1$ of fields and derivatives}\label{abelian-letter}
\end{table}
Let us first calculate the letter partition function
${\rm tr}\left[(-1)^Fe^{-\beta^\prime E}\right]$. This is given by
\begin{equation}
  f(x)=\frac{5x^2(1-x^2)-16x^{\frac{5}{2}}(1-x)+(10x^3-15x^4+6x^5-x^6)}{(1-x)^6}\ ,
\end{equation}
where $x\equiv e^{-\frac{\beta^\prime}{r}}$ with the $S^5$ radius $r$.
The factor $\frac{1}{(1-x)^6}$ is
given by multiplying six kinds of derivatives to the single fields, whose
partition function is given by the numerator. The first term $5x^2(1-x^2)$ is
the partition function of $5$ scalars $\phi^I$ subtracted by the constraint
$-5x^4$ from the equation of motion $\partial^2\phi^I=0$. The second term
$-16x^{\frac{5}{2}}(1-x)$ comes from $16$ fermions subtracted by the constraint
coming from their Dirac equation. The last term $10x^3-15x^4+6x^5-x^6$ comes
from the 2-form potential $B_{MN}$ for the self-dual 3-form flux. Most rigorously,
we have derived it using the tensor spherical harmonics. These can also be
understood more intuitively from the self-dual 3-form subject to the Bianchi
identity constraint, as follows. Firstly, $+10x^3$ comes from $10$ components
of unconstrained self-dual 3-form $H_{MNP}$. This is subject to the Bianchi
identity constraint $\partial_{[M}H_{NPQ]}=0$. The number of these constraints
at scale dimension $E=4$ is $15$, accounting for $-15x^4$. However, this is
an over-subtraction in the sense that an anti-symmetrized derivative acting on
this 4-form $\partial_{[M}\partial_NH_{PQR]}$ is automatically zero. So one has to add
the contribution from the over-subtracted part at dimension $E=5$, which is
$+6x^5$. Finally, the added part is again over-adding the over-subtraction at $E=4$,
as $\partial_{[M}\partial_N\partial_PH_{QRS]}$ need not be added (being automatically zero).
So this explains the last term $-x^6$. From this expression, one obtains
\begin{equation}
  -\frac{1}{2}\frac{d}{d\beta^\prime}f(x)
  =\frac{5r}{16(\beta^\prime)^2}-\frac{25}{384r}+r^{-3}\mathcal{O}(\beta^\prime)^2
\end{equation}
as $\beta^\prime\rightarrow 0$. As explained in \cite{Aharony:2003sx}, the first term
$\frac{5r}{16(\beta^\prime)^2}\sim\frac{5}{16}r\Lambda^2$ should be canceled by a counterterm
to be zero. This is because the vacuum value of $E$ has to be zero in the flat
space limit $r\rightarrow\infty$ from the conformal symmetry. A counterterm of the form
$\Lambda^2\int_{S^5\times S^1} d^6x\sqrt{g}\ R^2$ can cancel this divergence.
After this subtraction and the removal $\beta^\prime\rightarrow 0$ of the regulator,
the second term will appear as the prefactor of the $S^5\times S^1$ partition function
(not an index) as $e^{\frac{25}{384r}\cdot 2\pi r_1}=e^{\frac{25\beta}{384}}$.

The index version of it can be calculated from the letter index
${\rm tr}[(-1)^Fe^{-\beta^\prime(E-R_1)}]$. From Table \ref{abelian-letter},
it suffices to consider the contributions from BPS letters which saturate a BPS
energy bound with a pair of $Q,S$ supercharges. For instance, one can consider
the bound $E_{BPS}=2(R_1+R_2)+j_1+j_2+j_2$. The BPS letters are two complex scalars
with $(R_1,R_2)=(1,0), (0,1)$, three fermions all with $(R_1,R_2)=(+\frac{1}{2},+\frac{1}{2})$ and $(j_1,j_2,j_3)=(-\frac{1}{2},+\frac{1}{2},+\frac{1}{2})$,
$(+\frac{1}{2},-\frac{1}{2},+\frac{1}{2})$ or $(+\frac{1}{2},+\frac{1}{2},-\frac{1}{2})$.
No 3-forms are BPS. The BPS derivatives are three holomorphic ones.
The scalar equation of motion does not act within the BPS sector, while one component of
Dirac equation $\slash\hspace{-.255cm}\partial\Psi=0$ with $(j_1,j_2,j_3)
=(+\frac{1}{2},+\frac{1}{2},+\frac{1}{2})$ and $(R_1,R_2)=(+\frac{1}{2},+\frac{1}{2})$
remains to be BPS. Thus, one obtains the following letter index:
\begin{equation}
  f_{\rm index}(x)=\frac{(x+x^2)-3x^2+x^3}{(1-x)^3}=\frac{x}{1-x}\ ,
\end{equation}
where $x\equiv e^{-\frac{\beta^\prime}{r}}$. From this, one obtains
\begin{equation}\label{abelian-casimir-index}
  -\frac{1}{2}\frac{d}{d\beta^\prime}f_{\rm index}(x)=
  \frac{r}{2(\beta^\prime)^2}-\frac{1}{24r}+r^{-3}\mathcal{O}(\beta^\prime)^2\ .
\end{equation}
Again, the first term has to be canceled by a counterterm, e.g. taking the
form of $\Lambda^2\int d^6x\sqrt{g} R^2$. This is because the vacuum value of
$E-R_1$ has to vanish in the flat space limit, required by the superconformal
symmetry. The second term in the limit $\beta^\prime\rightarrow 0$ becomes the
`index version' of the vacuum energy, providing a weight $e^{\frac{2\pi r_1}{24r}}
=e^{\frac{\beta}{24}}$ for the Abelian index at $m=\frac{1}{2}$, $a=b=c=0$.

Although conceptually closely related, the two Casimir charges defined above are
quantitatively different observables. The difference is perhaps clearly illustrated
above with the Abelian $(2,0)$ theory. With a non-Abelian
gauge group, our index should be capturing the `index version' of the Casimir energy
of the interacting $(2,0)$ theory.

So far we have been explaining the Casimir charge at the maximal SUSY enhancement
point $m=\frac{1}{2}$, $a=b=c=0$. The same argument can be given at another
maximal SUSY point $m=-\frac{1}{2}$, $a=b=c=0$, with unique charge $E-R_2$ to
regulate the infinite sum over the modes. In general, with all four chemical potentials
turned on, it is not clear what kind of regulator one has to turn on. This is because
introducing more chemical potentials reduces the symmetry of the path integral, making
the regulator less constrained. Thus, the `vacuum charge' with generic values of
chemical potentials will be somewhat ambiguous and regulator dependent. This fact can
be concretely checked again by considering the Abelian $(2,0)$ theory. With generic
chemical potentials turned on, the symmetry of the path integral is just $SU(1|1)\times U(1)^3$, where a $U(1)$ comes from $R_1-R_2$ and $U(1)^2$ comes from $j_1-j_3$ and $j_2-j_3$.
The most general regulator one can introduce is
\begin{equation}
  e^{-\beta^\prime\left(E-\frac{R_1+R_2}{2}\right)}e^{\beta^\prime m^\prime(R_1-R_2)}
  e^{-\beta^\prime a_i^\prime j_i}\ ,
\end{equation}
with $a_1^\prime+a_2^\prime+a_3^\prime=0$. $\beta^\prime$ is the regulator which
should be sent to zero at the final stage. The other three parameters $m^\prime$,
$a^\prime,b^\prime$ parametrize the ambiguities of the regulator. Let us fix
$m^\prime$, $a_i^\prime$ and calculate
\begin{equation}\label{casimir-general}
  (\epsilon_0)_{\rm index}=
  \frac{1}{2}{\rm tr}\left[(-1)^F\left(E-\frac{R_1+R_2}{2}-m(R_1-R_2)+a_ij_i\right)
  e^{-\beta^\prime\left(E-\frac{R_1+R_2}{2}\right)}e^{\beta^\prime m^\prime(R_1-R_2)}
  e^{-\beta^\prime a_i^\prime j_i}\right]\ .
\end{equation}
Again defining
\begin{eqnarray}
  f(\beta^\prime,\beta^\prime m^\prime,\beta^\prime a_i^\prime)&=&{\rm tr}
  \left[(-1)^Fe^{-\beta^\prime\left(E-\frac{R_1+R_2}{2}\right)}e^{\beta^\prime m^\prime(R_1-R_2)}e^{-\beta^\prime a_i^\prime j_i}\right]\\
  &=&\frac{e^{-\frac{3\beta^\prime}{2}}(e^{\beta^\prime m^\prime}+e^{-\beta^\prime m^\prime})
  -e^{-2\beta^\prime}(e^{\beta^\prime a^\prime}+e^{\beta^\prime b^\prime}+
  e^{\beta^\prime c^\prime})+e^{-3\beta^\prime}}{(1-e^{-\beta^\prime(1+a^\prime)})
  (1-e^{-\beta^\prime(1+b^\prime)})(1-e^{-\beta^\prime(1+c^\prime)})}\ ,\nonumber
\end{eqnarray}
one finds
\begin{equation}
  (\epsilon_0)_{\rm index}=\frac{1}{2}\left(-\frac{\partial}{\partial\beta^\prime}
  +\frac{m^\prime-m}{\beta^\prime}\frac{\partial}{\partial m^\prime}
  +\frac{a_i^\prime-a_i}{\beta^\prime}\frac{\partial}{\partial a_i^\prime}\right)f\ .
\end{equation}
The expression shows a $\frac{r}{(\beta^\prime)^2}$ divergence in the
$\beta^\prime\rightarrow 0$ limit. Again, this part has to be canceled by
a counterterm for the sensible flat space limit $r\rightarrow\infty$. We have not
identified the correct nature of the counterterms which could possibly provide
the correct $a,b,c$ dependence: this could be an interesting problem. After this
subtraction, $(\epsilon_0)_{\rm index}$ in the $\beta^\prime\rightarrow 0$ limit
is a very complicated expression depending on $m,a,b$ as well as the ambiguous
parameters $m^\prime,a^\prime,b^\prime$ in the regulator.
One of the arbitrary choices of $m^\prime,a^\prime,b^\prime$ is simply
$m^\prime=m$, $a^\prime=a$, $b^\prime=b$. Then one obtains \cite{Kim:2012qf}
\begin{equation}\label{abelian-S5-casimir}
  (\epsilon_0)_{\rm index}=-\frac{1}{24r}\left(1+
  \frac{2abc+(1-ab-bc-ca)\left(\frac{1}{4}-m^2\right)+
  \left(\frac{1}{4}-m^2\right)^2}{(1+a)(1+b)(1+c)}\right)\ .
\end{equation}
This particular regulator will be discussed in more detail below.

So far we explained the basic properties of the vacuum charges appearing in the index, with
concrete illustrations from the 6d Abelian theory. Now let us discuss these charges
appearing in our 5d SYM calculations, both from $S^5$ and $\mathbb{CP}^2\times S^1$.

Let us first discuss the maximal SUSY point $m=\frac{1}{2}$, $a=b=c=0$, when the
meaning of $(\epsilon_0)_{\rm index}$ seems unambiguous. From the partition function
on $S^5$, one finds \cite{Kim:2012av,Kim:2012qf} (see also \cite{Minahan:2013jwa})
\begin{equation}\label{casimir-index-unrefined}
  r(\epsilon_0)_{\rm index}=-\frac{N(N^2-1)}{6}-\frac{N}{24}\ ,
\end{equation}
as explained in section 3.2. We study the comparable quantity
from our QFT on $\mathbb{CP}^2\times S^1$. Let us first study this quantity
from the QFT at $n=-\frac{1}{2}$, and comment on the same quantity viewed from
a QFT at another value of $n$. At $m=\frac{1}{2}$ or $\hat{m}=m+n=0$,
and $a=b=c=0$, the result from section 3.2 is
\begin{equation}\label{unrefined-again}
  e^{\beta\frac{N(N^2-1)}{6}}\prod_{n=1}^{N-1}(1-e^{-n\beta})^{N-n}\cdot
  \prod_{n=1}^\infty\frac{1}{(1-e^{-n\beta})^N}\ .
\end{equation}
Apparently, even at this point we do not get
the result for $(\epsilon_0)_{\rm index}$ obtained from $S^5$ calculation, by missing
the additional $-\frac{N}{24}$ term. Due to the absence of the last term, we do not
even get the $\epsilon_0=-\frac{1}{24}$ at $N=1$, which we independently explained
from the 6d calculation in this subsection. In particular, the last apparent mismatch
says that there should be some elementary factor missing from our
$\mathbb{CP}^2\times S^1$ QFT approach so far.

It is not difficult to find what we are missing.
Let us first recall how we obtained each factor in the above expression.
The first factor, $e^{\beta\frac{N(N^2-1)}{6}}$ and the finite
product of $n$, comes by combining the contributions depending on the anti-self-dual
flux, and the perturbative measure, and doing the contour integral and $s$ flux sum.
The last infinite product of $n$ is the partition function of self-dual $U(N) $
instantons at the point $\hat{m}=0$, $a_i=0$, which is basically the contribution
from $U(1)^N$ self-dual instantons. The last contribution from self-dual instantons
can be written as
\begin{equation}\label{U(1)N-instanton}
  \exp\left[\sum_{n=1}^\infty\frac{1}{n}\frac{Ne^{-n\beta}}{1-e^{-n\beta}}\right]
  \equiv PE\left[\frac{Ne^{-\beta}}{1-e^{-\beta}}\right]\ ,
\end{equation}
so this is the `multi-particle' excitations coming from the 5d instanton `modes'
whose letter index is $\frac{Ne^{-\beta}}{1-e^{-\beta}}$. With the discussions
around (\ref{PE-literature}), the above $PE$ only accounts for the `excitations' coming
from BPS modes. Had the BPS modes been coming from the 5d `elementary fields,' like
those in the perturbative measure, the correct $PE$ to be used is (\ref{PE-precise}) or
(\ref{PE-final}), which include the vacuum energy contribution from these modes.
While we were calculating $(\epsilon_0)_{\rm letter}$ for the 6d free Abelian theory
in this subsection, we included all the ground state energy contributions coming
from `5d perturbative' modes (carrying $k=0$) and the `5d instantonic' modes
(carrying $k\neq 0$). However, from the 5d QFT approach, we are treating the latter
as solitonic configurations, and rely on some semi-classical analysis (which is given
a precise sense via supersymmetric localization). For solitonic modes, the vacuum
energy factor as in (\ref{PE-final}) is not automatically
provided during the path integral.

We can naturally supplement the analysis we did so far, inserting the vacuum energy
contribution from self-dual instantons, by trying to treat the instanton part of the
index on the same footing as the perturbative part. This starts from writing
$Z_{\rm inst}^{(1)}Z_{\rm inst}^{(2)}Z_{\rm inst}^{(3)}$ in a Plethystic form,
\begin{equation}\label{self-dual-letter}
  Z_{\rm inst}^{(1)}Z_{\rm inst}^{(2)}Z_{\rm inst}^{(3)}=
  PE\left[f(\beta,m,a_i)_{\rm inst}\right]\equiv PE\left[\sum_{n=1}^\infty
  \frac{1}{n}f(n\beta,m,a_i)_{\rm inst}\right]\ ,
\end{equation}
with some `letter index' $f$ for `instantonic modes.' Treating this in the same
footing as the perturbative modes, we promote the $PE$ in (\ref{PE-literature}) to
(\ref{PE-final}), thus including an extra vacuum energy contribution
\begin{equation}\label{extra-casimir}
  (\Delta\epsilon_0)_{\rm index}=
  \frac{1}{2}\left(-\frac{\partial}{\partial\beta^\prime}
  +\frac{m^\prime-m}{\beta}\frac{\partial}{\partial m^\prime}+\frac{a_i^\prime-a_i}{\beta^\prime}
  \frac{\partial}{\partial a_i^\prime}\right)f(\beta^\prime,m^\prime,a_i^\prime)
\end{equation}
which in general is subject to all the ambiguities that we already explained
with the Abelian 6d theory. However, at $m=\frac{1}{2}$, $a_i=0$, the letter index
$f(\beta)=\frac{Ne^{-\beta}}{1-e^{-\beta}}$ and big enough symmetry in the path integral
uniquely fix this contribution $(\Delta\epsilon_0)_{\rm index}$ to be $-\frac{N}{24}$.
This follows from exactly the same arguments around (\ref{abelian-casimir-index})
for the 6d Abelian theory. Thus, one can reproduce the Casimir energy
(\ref{casimir-index-unrefined}) at $m=\frac{1}{2}$ by including the self-dual
instanton modes' contribution.

Although this sounds heuristic, at least to us, it certainly seems to
be a `guesswork' without a rigorous justification. In a conservative attitude, one might
say we just added an extra factor of $e^{-\beta(\Delta\epsilon_0)_{\rm index}}$
by hand, no matter what kind of words we put on it. This might be a genuine limitation
of the 5d QFT approach from $\mathbb{CP}^2\times\mathbb{R}$, unless there is a
more convincing way of fixing it.

At this point, note that the partition function
of the 5d QFT on $S^5$ also has an ambiguity of similar sort, which we should fix
in a particular way to view it as a 6d index \cite{Kim:2012av,Kim:2012qf}.
Namely, the $S^5$ partition function presented in
\cite{Lockhart:2012vp,Kim:2012qf} all takes the form of a weak-coupling
expansion at $\beta\ll 1$, taking the form of a small correction to $1$ in either
perturbative or non-perturbative manner (power series in $\beta$ or
$e^{-\frac{4\pi^2}{\beta}}$). This could be regarded as a high temperature
expansion from the 6d index viewpoint. However, just like the ordinary partition
function which counts the number of bosonic states plus that of fermionic states,
the index also usually
develops a high temperature divergence. The divergence of the true free energy
of the 6d $(2,0)$ theory should scale like $T^6\sim\beta^{-6}$ at high temperature,
but the index version of the free energy (minus of the logarithm of the index) may
develop a milder divergence. This divergence takes the form of a factor
\begin{equation}\label{high-T}
  \exp\left[\sum_{n>0}\frac{a_n}{\beta^n}\right]
\end{equation}
multiplied to the partition function calculated in \cite{Lockhart:2012vp,Kim:2012qf}.
From the 6d BPS physics, we expect that the maximal allowed value of $n$ is $3$, since
there are only $3$ BPS derivatives. So presumably the summation over $n$ in the exponent
is finite. The finitely many coefficients $a_n$ do not seem be fixed from a 5d consideration
only, as far as we are aware of at the moment.\footnote{Using $d-1$ dimensional QFT to
study the $d$ dimensional high temperature physics is common. However, the low
dimensional ambiguity comparable to what we discuss here should be a small input from
the $d$ dimensional physics. In our case, since we do not know an independent 6d
formulation of the theory, the high temperature asymptotics of the free energy seems
to manifest itself as an ambiguity.}

A pragmatic prescription for resolving this ambiguity from 5d was proposed as
follows \cite{Kim:2012av,Kim:2012qf}. We can in principle make the strong coupling
re-expansion of the $S^5$ partition function at $\beta\gg 1$, in a power series of
$e^{-\beta}$. With a multiplicative factor like (\ref{high-T}), the partition function
at strong coupling (or low temperature) would take the `index form'
\begin{equation}\label{index-schematic}
  e^{-\beta(\epsilon_0)_{\rm index}}\left(1+\sum_{E_i>0}n_ie^{-\beta E_i}\right)
\end{equation}
with integer coefficients $n_i$. This only happens with a very particular form of
(\ref{high-T}). So we can tune the values of finitely many parameters $a_n$, by
requiring the strong coupling expansion to be an index.
For instance, had we been multiplying (\ref{high-T}) to the partition function
of \cite{Lockhart:2012vp,Kim:2012qf} with wrong coefficients $a_n$, the strong
coupling expansion will be (\ref{index-schematic}) multiplied by a factor of the
form (\ref{high-T}), which is not an index. This prescription was successfully
applied to some special cases in \cite{Kim:2012qf}, in which a strong coupling
re-expansion could be concretely made.

It is curious to find that the approaches from $S^5$ and $\mathbb{CP}^2\times S^1$
exhibit their ambiguities at very different places. Once the factor of
(\ref{high-T}) is fixed in the way explained above, the former partition function
provides a definite value of $(\epsilon_0)_{\rm index}$. On the other hand,
since the latter partition function is manifestly taking an index form, the high
temperature asymptotics like (\ref{high-T}) should be in principle derivable from it.
The ambiguity we find is at $(\epsilon_0)_{\rm index}$ as explained above, although
we have a heuristic conjecture to fix it. Thus, the two different 5d QFT approaches
could be somewhat complementary to each other. See the next subsection for
more discussions on possible future studies on the $S^5$ partition function.

Even if one accepts our prescription to calculate $(\Delta\epsilon_0)_{\rm index}$
from self-dual instantons using (\ref{extra-casimir}), there are further issues
that should be clarified. This is because both points $m=\pm\frac{1}{2}$, $a_i=0$
exhibit enhanced symmetry to $SU(4|2)$, making $(\epsilon_0)_{\rm index}$ to
be an unambiguous quantity. We have calculated $(\epsilon_0)_{\rm index}$ at
$m=\frac{1}{2}$ from the 5d QFT at $n=-\frac{1}{2}$. Similarly, exactly the same
calculation can be done at $m=-\frac{1}{2}$ using the QFT at $n=+\frac{1}{2}$
to obtain (\ref{casimir-index-unrefined}), using (\ref{extra-casimir}). However,
a complication arises if one tries to calculate (\ref{casimir-index-unrefined})
at $m=-\frac{1}{2}$ from the QFT at $n=-\frac{1}{2}$. This is because, all the
extra $14$ SUSY in $SU(4|2)$ at $m=-\frac{1}{2}$, apart from the universal
$Q^{++}_{---}$ and $Q^{--}_{+++}$, are not manifestly
visible from the 5d QFT at $n=-\frac{1}{2}$ at $k=0$. Since the extra SUSY are all
carrying nonzero instanton charges, the simple final result (\ref{unrefined-again})
would come after complicated cancelations across various sectors with different
values of $k$. Note that, from the discussions at the end of section 2.2, $\beta$
is conjugate to $k$ only when we use $\beta\hat{m}=\beta(m+n)$ as the independent
chemical potential. At $n=-\frac{1}{2}$ and $m=-\frac{1}{2}$, $\beta\hat{m}=-\beta$
so that $e^{-\beta}$ factor appears not from instanton expansions only, but also
from other charges. In fact, the contribution of $(\epsilon_0)_{\rm index}$ from
the anti-self-dual fluxes and perturbative measures (without self-dual instantons'
contribution in this subsection) is
\begin{equation}
  e^{\beta\frac{N(N^2-1)}{6}(1-\hat{m})}\ \rightarrow\ e^{\beta\frac{N(N^2-1)}{3}}
\end{equation}
at $m=-\frac{1}{2}$, $n=-\frac{1}{2}$, from the result of section 3.3.
If our conjecture around (\ref{extra-casimir}) is correct, the self-dual instanton
modes should provide the extra contribution of $+\frac{N(N^2-1)}{6}-\frac{N}{24}$
to get (\ref{casimir-index-unrefined}). Unfortunately, to confirm this, one has to
know the letter index $f$ appearing in (\ref{self-dual-letter}) exactly, to all order
in the instanton
expansion. This is because (\ref{extra-casimir}) is a regularized sum of infinite
series, and knowing a finite order truncation of it does not help. Apart from
several simple points, like that leading to (\ref{U(1)N-instanton}), the exact form
of $f$ is difficult to know from the instanton partition functions of \cite{Nekrasov:2002qd}.
We hope we can clarify this issue in the near future.

Finally, let us comment on $(\epsilon_0)_{\rm index}$ at general values of
$m,a_i$. As explained with the Abelian example, it seems that this quantity may
not be unambiguously well defined. However, calculating the index from the $S^5$ QFT
or the $\mathbb{CP}^2\times\mathbb{R}$, we obtain certain value of $(\epsilon_0)_{\rm index}$. We think that this might be because we have implicitly fixed a small ambiguity (comparable
to the $m^\prime, a_i^\prime$ in (\ref{casimir-general})) when
we do the 5d supersymmetric path integrals. In fact, the $S^5$ partition function
with $U(1)$ gauge group calculates $(\epsilon_0)_{\rm index}$ given by
(\ref{abelian-S5-casimir}),
using the regularization with $m^\prime=m$, $a_i^\prime=a_i$. So this might be
a concrete fixing of this ambiguity that we could have implicitly made for general
QFT path integral on $S^5$. For the non-Abelian theory, the $S^5$ partition function
yields a complicated expression (\ref{unrefined-S5}) for $(\epsilon_0)_{\rm index}$ at the
unrefinement point $m=\frac{1}{2}-c$ with $4$ SUSY. We are inclined not to seek for
any physical meaning of (\ref{abelian-S5-casimir}) or $(\epsilon_0)_{\rm index}$ in
(\ref{unrefined-S5}), as we are not sure if these quantities are physically well-defined
and/or useful. (However, see \cite{Kallen:2012zn,Minahan:2013jwa} for the
calculation of this quantity at general $m$ and $a_i=0$, in the large $N$ limit.)

\subsection{S-duality on $\mathbb{R}^4\times S^1$ and related issues}

In this subsection, we comment on the 6d index calculated from $S^5$ partition
function, and some related issues. As explained in the introduction, the expression
for the partition function on $S^5$ that was obtained in the literatures takes
a weak-coupling expansion form. At least so far, this partition function as an index
has been studied only in some special cases \cite{Kim:2012av,Lockhart:2012vp,Kim:2012qf}.
An important building block for this partition function is the partition function of
the 5d $\mathcal{N}=1^\ast$ gauge theory on Omega-deformed $\mathbb{R}^4\times S^1$.
If one can obtain a strong coupling re-expansion of the last quantity at $\beta\gg 1$,
the full re-expansion of the $S^5$ partition function would be much easier to do.
In the context of the last theory on $\mathbb{R}^4\times S^1$, one should understand
$\beta=\frac{g_{YM}^2}{2\pi r}$ with $r$ given by the radius of $S^1$.

There is a good reason to expect that the strong coupling re-expansion of
the partition function of the 5d $\mathcal{N}=1^\ast$ SYM on $\mathbb{R}^4\times S^1$
should take a simple form, at $\beta\gg 1$. This is because this partition function
acquires contribution from instanton particles, and can be understood as the
6d $(2,0)$ theory's partition function on $\mathbb{R}^4\times T^2$
\cite{Kim:2011mv}. The BPS partition function only depends on the complex structure
$\tau=\frac{2\pi i}{\beta}$ of $T^2$, and is independent of its volume. So one
expects the 5d partition function should transform in a simple manner under
the $SL(2,\mathbb{Z})$ action on $T^2$. This would be a special realization of the
S-duality of 5d maximal SYM on $\mathbb{R}^4\times S^1$, studied in \cite{Tachikawa:2011ch}.
In particular, the S-duality would be basically flipping a circle which is visible in 5d,
with another circle in the 6'th direction which is non-perturbatively probed by instantons.

Very recently, the partition function on Omega-deformed $\mathbb{R}^4\times S^1$
was rewritten in a way that this $SL(2,\mathbb{Z})$ transformation property is
clearly visible \cite{vafa}. Let us briefly review the results of this paper.
Nekrasov's partition function (or equivalently a suitable topological string partition
function as studied in \cite{vafa}) of the 5d $\mathcal{N}=1^\ast$ theory captures
the BPS degeneracies of the maximal SYM on $\mathbb{R}^{4,1}$ in its Coulomb phase,
for the states carrying electric charge and/or the instanton charge. The 6d
uplifts of these charges are the charge of the self-dual strings wrapping the 6'th
circle, and the momentum charge along the same circle, respectively.
So Nekrasov's partition function is a linear combination of the 2 dimensional
indices for the wrapped self-dual strings with various electric charges, with a
multiplication of an index for BPS momenta with zero electric charge. However,
in the expression presented in \cite{Nekrasov:2002qd}, this 2d index nature
is very difficult to see, which is manifestly visible only after one can sum
over the infinite instanton series. \cite{vafa} essentially did this sum at
nonzero self-dual string charges, with a closed form expressions for the
2d indices given by ratios of Jacobi's theta functions.
The prefactor index which comes from the momentum bound states with zero
electric charge is not for self-dual strings, and this part is left untouched
in \cite{vafa}. In our appendix A, we did a strong coupling re-expansion of
this part, namely the partition function for pure momenta in the $U(1)$ theory.
The S-dual transformation is a bit more involved than the Jacobi theta functions,
which are simple Jacobi forms: see appendix A for the details. So our appendix A
could be understood as a supplement to \cite{vafa}, making S-duality properties
of the whole partition function on $\mathbb{R}^4\times S^1$ to be more
transparent.

With these results, it would be much easier to clearly check if the $S^5$ partition
function is indeed the 6d index, by confirming that it admits a proper fugacity
expansion with integer coefficients. This would be obtained by first taking the
strong-coupling expansions of the three $\mathbb{R}^4\times S^1$ partition functions
in the integral for the $S^5$ partition function \cite{Lockhart:2012vp,Kim:2012qf}, and
then carefully performing the integral. With our studies in this paper, we
expect that the result is the same as our $\mathbb{CP}^2\times S^1$ index.
As the last result exhibits some delicate structures, like the choice of
the integration contour, etc., it will be very interesting to perform the integral
of \cite{Lockhart:2012vp,Kim:2012qf} and make an alternative study of (hopefully) the
same 6d index. Especially, as we commented in the previous subsection, two different
approaches show different subtleties which could complement each other.
The last motivation could make this study even more interesting.

One novel aspect of the result from our $\mathbb{CP}^2\times\mathbb{R}$ QFT,
apparently absent in the $S^5$ QFT, is the anti-self-dual flux $s$
on $\mathbb{CP}^2$ proportional to the K\"ahler 2-form $J$. We think that $s$ could be
interpreted as a combination of the self-dual strings' electric charges coming from the
three fixed points, which appear while one expands the $S^5$ partition function
following \cite{vafa}. This interpretation is natural already on
$\mathbb{CP}^2\times\mathbb{R}$, as the $A\wedge F\wedge J$
interaction provides electric charges to anti-self-dual fluxes.

\subsection{6d $(1,0)$ superconformal indices}

We can try to apply the ideas and techniques developed in this paper to study
6d $(1,0)$ superconformal field theories. Now with reduced symmetry, there is a variety
of interesting $(1,0)$ theories predicted from string theory. For instance, see
\cite{Seiberg:1996vs,Seiberg:1996qx} and many references thereof.

One interesting 6d $(1,0)$ theory would be that living on $N$ M5-branes probing a
Horava-Witten 9-plane with $E_8$ global symmetry \cite{Horava:1995qa}.
The 5 dimensional reduction of this theory (at least on $\mathbb{R}^{4,1}$) is well
understood. It is a 5d $\mathcal{N}=1$ SYM theory with $Sp(N)$ gauge group, coupled to
one hypermultiplet in the anti-symmetric representation and $N_f=8$ hypermultiplets
in the fundamental representation. The manifestly visible flavor symmetry is only
$SO(2N_f)=SO(16)\subset E_8$ in 5d. This theory is expected to have a 6d UV fixed point.
Perhaps an interesting issue is to see if a global symmetry enhancement is visible after
including instantons (the 6d KK modes). Similar studies were done with 5d gauge theories
with $N_f\leq 7$ fundamental and one anti-fundamental hypermultiplets, which are supposed
to have 5d UV fixed points \cite{Seiberg:1996bd,Morrison:1996xf} with enhanced
$E_{N_f+1}$ symmetry. More recently, the symmetry enhancement of these 5d SCFT's
was studied using the superconformal index, or the $S^4\times S^1$ partition function
\cite{Kim:2012gu}. There should be more interesting 6d $(1,0)$ theories that one could
try to study from the 5d SYM approach.

In this paper, we make some preliminary observations (mostly of kinematic nature) of
our approach applied to the $(1,0)$ theories in general, and try to point out some
new aspects compared to the $(2,0)$ theories studied in this paper. The 6d symmetry
is $OSp(8^\ast|2)$ with $SU(2)_R$ R-symmetry. The supercharges take the form of
\begin{equation}
  Q^R_{j_1,j_2,j_3}=Q^{\pm\frac{1}{2}}_{\pm\frac{1}{2},\pm\frac{1}{2},\pm\frac{1}{2}}\ ,
\end{equation}
where $R$ is the Cartan of $SU(2)_R$, with $\pm\frac{1}{2}$ eigenvalues on supercharges.
Without losing generality, we would like to take $Q\equiv Q^+_{---}$ and its conjugate
$S$, and study the related superconformal index. We would like to study this
index from a QFT on $\mathbb{CP}^2\times\mathbb{R}$, again obtained by
a circle reduction of the 6d theory on $S^5\times\mathbb{R}$
after a $\mathbb{Z}_K$ orbifold along the Hopf fiber of $S^5$.
We want the $\mathbb{Z}_K$ orbifold to preserve the above $Q,S$.
The only combination of charges that one can use to make such a
supersymmetric $\mathbb{Z}_K$ orbifold is
\begin{equation}
  Kk\equiv j_1+j_2+j_3+3R\ .
\end{equation}
We would again like to realize $k$ as the topological instanton charge
in the 5d QFT viewpoint.

Let us investigate some properties of $k$, especially at $K=1$.
Firstly, the supercharges $Q^R_{j_1,j_2,j_3}$ all take integral eigenvalues
of $k$. All other generators of $OSp(8^*|2)$ also take integral eigenvalues.
Let us also study the eigenvalues of $k$ on the fields. The fields in
the 6d vector multiplet or the tensor multiplet have integral eigenvalues.
Firstly, the vector and tensor fields have integral $k$ (with $R=0$).
Other fields in these multiplets are connected to them using the
$OSp(8^*|2)$ generators, which all assume integral values of $k$.
On the other hand, the fields in the hypermultiplet would have half-an-odd
integer eigenvalues of $k$. This is because the two complex scalars in
the hypermultiplet from a doublet under $SU(2)$ R-symmetry.

So a question is how one could realize half-integral eigenvalues for
the topological charge in SYM on $\mathbb{CP}^2\times\mathbb{R}$.
It could be that a half-integral instanton charge on $\mathbb{CP}^2$ could be
playing the role, coming from the fact that $\mathbb{CP}^2$ is not a spin manifold
\cite{Witten:2003ya}. In fact, the instanton charge $-\frac{1}{2}\sum_{i=1}^Ns_i^2$
coming from our anti-self-dual instanton flux is half-integral in
(\ref{saddle-action-asd-scalar}). However, in
our $(2,0)$ index, the presence of an overall decoupled $U(1)$ imposes the
Gauss law constraint $\sum_{i=1}^Ns_i=0$, making it to be integral. In
the presence of fundamental matters, such as the $(1,0)$ theory made by
putting M5-branes near the Horava-Witten plane, there are no such constraints
and one could have half-integral $k$'s. We are not sure if this mechanism
would provide all the necessary half-integral $k$'s in various $(1,0)$ theories.
A more detailed study on this issue, with various examples of $(1,0)$ theories
in mind, would be necessary.

\vskip 0.5cm

\hspace*{-0.8cm} {\bf\large Acknowledgements}

\vskip 0.2cm

\hspace*{-0.75cm} We thank Frederik Denef, Rajesh Gopakumar, Amihay Hanany, Sungjay Lee,
Jan Manschot, Shiraz Minwalla, Satoshi Nawata, Costis Papageorgakis, Jaemo Park,
Sara Pasquetti, Shlomo Razamat, David Tong, Martin Wolf and especially Davide Gaiotto
and Jaume Gomis for helpful discussions. This work is supported by the National Research
Foundation of Korea (NRF) Grants No. 2006-0093850 (KL), 2009-0084601 (KL), 2010-0007512 (SK),
2012R1A1A2042474 (SK), 2012R1A2A2A02046739 (SK) and 2005-0049409 through the Center for
Quantum Spacetime (CQUeST) of Sogang University (SK,KL).  The work of SK at the Perimeter
Institute is also supported by the Government of Canada through Industry Canada, and by the
Province of Ontario through the Ministry of Economic Development \& Innovation. This work (KL) is supported in part by the National Science Foundation under Grant No. PHYS-1066293 and the hospitality of the Aspen Center for Physics.

\appendix

\section{S-duality of Abelian partition function on $\mathbb{R}^4\times S^1$}

The S-duality that we expect for maximal SYM (or its mass deformation) on $\mathbb{R}^4_{\epsilon}\times S^1$, or the 6d theory on $\mathbb{R}_\epsilon^4\times T^2$
has to be understood with care, depending on the signs and ranges of the Omega deformation
parameters. To illustrate
this, let us first consider the partition function of the Abelian maximal SYM, with
$m=0$, in various Omega backgrounds. When $\epsilon_1=\epsilon_2$, one finds that
the instanton partition function is simply $1$, as $I_-$ appearing as the instanton
center-of-mass factor is zero. Also, the perturbative letter index becomes
$I_+=I_--1\rightarrow-1$. The last number simply comes from the ghost zero mode,
which we discard to make the partition function well-defined. (In all calculus on
$S^5$ or $\mathbb{CP}^2\times S^1$, this cancels out with a positive constant in $PE$,
coming from a bosonic zero mode in the ghost multiplet.)
So the perturbative partition function is also $1$. The net partition function
$1$ is trivially self S-dual in this limit. (We shall shortly extend this trivial result
to a more interesting case.) On the other hand, when $\epsilon_1=-\epsilon_2$, the Abelian
instanton partition function is $\eta(e^{-\frac{4\pi^2}{\beta}})^{-1}$, and the perturbative
part is $1$ since $I_+=0$. So the partition function is a nontrivial modular form. So we
should have different S-duality transformation rules in different Omega backgrounds.
We want to classify them, at least when they appear as building blocks of the $S^5$
partition function.

We start from the weak coupling partition function at $\beta\ll 1$,
multiplying the instanton and perturbative contributions together, while
taking the Omega backgrounds to be pure imaginary. We start from
\begin{eqnarray}\label{weak-abelian}
  &&PE\left[\frac{1}{2}\frac{\sin\pi(m+\epsilon_+)\sin\pi(m-\epsilon_+)}
  {\sin\pi\epsilon_1\sin\pi\epsilon_2}+\frac{1}{2}
  +\frac{\sin\pi(m+\epsilon_-)\sin\pi(m-\epsilon_-)}
  {\sin\pi\epsilon_1\sin\pi\epsilon_2}\frac{e^{-\beta_D}}{1-e^{-\beta_D}}\right]\nonumber\\
  &&\equiv PE\left[\frac{1}{2}I_++\frac{1}{2}+I_-\frac{e^{-\beta_D}}{1-e^{-\beta_D}}\right]\ .
\end{eqnarray}
The expression is assumed with small ${\rm Im}(\epsilon_{1,2})=\varepsilon>0$,
namely with the $\epsilon_{1,2}+i\varepsilon$ prescription. The addition of
$\frac{1}{2}$ term is because, expanding the first perturbative part with
this prescription, one obtains a contribution $-\frac{1}{2}$ from a fermionic
zero mode (`zero-form' from ghost in BRST setting), making the $PE$
operation meaningless. $\beta_D\equiv\frac{4\pi^2}{\beta}$ is the S-dual coupling.
Using $I_--I_+=1$, and $\frac{e^{-\beta_D}}{1-e^{-\beta_D}}=
\frac{1}{2}\coth\frac{\beta_D}{2}-\frac{1}{2}$, the above expression can be
rearranged as
\begin{equation}\label{+dominated-start}
  PE\left[\frac{1}{2}I_-\coth\frac{\beta_D}{2}\right]\ .
\end{equation}
To make an S-dual expansion of this quantity at $\beta_D\ll 1$, we employ
a trick used in \cite{Kim:2012qf}, which is inspired by the techniques of
Gopakumar-Vafa to  rewrite the topological string partition function
at strong coupling \cite{Gopakumar:1998ii}.
We first change the exponent of $PE$ into an integral, using
\begin{equation}\label{delta}
  \beta_D\sum_{n=-\infty}^\infty\delta(s-\beta_D n)=\sum_{n=-\infty}^\infty
  e^{\frac{2\pi ins}{\beta_D}}\ .
\end{equation}
Using this, one obtains
\begin{equation}\label{integral}
  \sum_{n=-\infty}^\infty\int_\epsilon^\infty\frac{ds}{s}\
  \frac{\sin\frac{\pi(m+\epsilon_-)s}{\beta_D}\sin\frac{\pi(m-\epsilon_-)s}{\beta_D}}
  {2\sin\frac{\pi\epsilon_1s}{\beta_D}\sin\frac{\pi\epsilon_2s}{\beta_D}}
  \ \coth\frac{s}{2}\ e^{\frac{2\pi i ns}{\beta_D}}\ .
\end{equation}
One has to understand this integral carefully, as the integrand has poles
at $s=\frac{\beta_D p}{\epsilon_{1,2}}$ with $p=1,2,3,\cdots$, yielding an infinite
sum over $p$ with an oscillating measure from $e^{\frac{2\pi ins}{\beta_D}}$. One possible
way to make this sum `safer' is to introduce a `small exponential damping factor' in
$e^{\frac{2\pi ins}{\beta_D}}$,
\begin{equation}
  e^{\frac{2\pi ins}{\beta_D}}\rightarrow
  e^{\frac{2\pi in(s\pm i\varepsilon)}{\beta_D}}
\end{equation}
with a small positive number $\varepsilon$ in the definition of the periodic
delta function, where the $\pm$ signs are for $n>0$ and $n<0$, respectively.
As for the term with $n=0$, deformations with
$\pm$ signs will yield the same result. Let us comment on the most convenient
prescription later, for this $n=0$ sector.

We first make the integrand finite at $s=0$ by a suitable subtraction:
the finite integral is given by
\begin{equation}
  \hspace*{-1.6cm}\sum_{n=-\infty}^\infty\int_0^\infty\frac{ds}{s}\
  \frac{\sin\frac{\pi(m+\epsilon_-)s}{\beta_D}\sin\frac{\pi(m-\epsilon_-)s}{\beta_D}}
  {\sin\frac{\pi\epsilon_1s}{\beta_D}\sin\frac{\pi\epsilon_2s}{\beta_D}}
  \left(\frac{1}{2}\coth\frac{s}{2}-\frac{1}{s}\right)e^{\frac{2\pi i ns}{\beta_D}}+
  \int_0^\infty\frac{ds}{s^2}\left(
  \frac{\sin\frac{\pi(m+\epsilon_-)s}{\beta_D}\sin\frac{\pi(m-\epsilon_-)s}{\beta_D}}
  {\sin\frac{\pi\epsilon_1s}{\beta_D}\sin\frac{\pi\epsilon_2s}{\beta_D}}
  -\frac{m^2-\epsilon_-^2}{\epsilon_1\epsilon_2}\right)
  e^{\frac{2\pi i ns}{\beta_D}}
\end{equation}
and the remaining divergent part is simply
\begin{equation}\label{divergent}
  \frac{m^2-\epsilon_-^2}{\epsilon_1\epsilon_2}\sum_{n=-\infty}^\infty
  \int_\epsilon^\infty\frac{ds}{s^2}e^{\frac{2\pi ins}{\beta_D}}=
  \frac{\pi^2(m^2-\epsilon_-^2)}{6\epsilon_1\epsilon_2\beta_D}=
  \frac{m^2-\epsilon_-^2}{24\epsilon_1\epsilon_2}\ \beta\ .
\end{equation}
As the integrand apart from $e^{\frac{2\pi ins}{\beta_D}}$ is an even function
in $s$, the finite integral can be understood as $\int_{-\infty}^\infty$ by combining
a pair of integrals associated with $n$, $-n$. For the case with $n=0$, one can
replace the integral by $\frac{1}{2}\int_{-\infty}^\infty$ from the evenness of
the whole integrand. The summation of integrals for $n\neq 0$ is now restricted to
$n=1,2,3,\cdots$, with the measure $e^{\frac{2\pi ins}{\beta_D}}$.
In fact, the $\pm i\varepsilon$ prescriptions for $n>0$ and $n<0$ that we introduced
above patch nicely, and demands that the contour of $s$ along the real line is slightly
deformed to the positive imaginary direction. Note that deforming it to the negative
imaginary direction is not allowed, as the summation of $n$ in (\ref{delta})
would then diverge. On the other hand, the integral with $n=0$ admits any deformation
$\pm i\varepsilon$. One possible way will be thus deforming it above the real axis,
and another will be deforming below the real axis. From our argument above, the result
should be the same for $n=0$. The difference between the two prescriptions is whether
one includes the residues for poles on the real axis or not. Since the poles are
distributed evenly around $s=0$, and since the series of residues will be an even
function in $p$ (running over the whole integer) as the integrand is even in $s$,
the summation over all residues vanishes. So it makes sense to conveniently
assume the $s+i\varepsilon$ prescription for the $n=0$ integral, once we make $s$
run over the whole real axis. The above integrals can be made into closed contour
integrals by adding the large upper half circle on the complex $s$ plane.
The addition of half circle does not give extra contribution if one assumes
\begin{equation}\label{semi-circle-condition}
  2|\epsilon_+|-|\epsilon_-+m|-|\epsilon_--m|>0\ .
\end{equation}
We assume so, which we take as the first case of our study in the parameter
space $\epsilon_\pm,m$.
For instance, this condition is fulfilled when ${\rm \epsilon_{1,2}}\gg 1$,
$\epsilon_+\gg\epsilon_-,m$. To be more precise, in order to take the integral
domain to cross the point $s=0$, one should have to subtract half of the
values of the integrand at $s=0$, multiplied by $\beta_D/2$. This is because
we originally excluded the delta function contribution at $s=0$ in (\ref{delta})
to get (\ref{integral}), by restricting the integral to start from $s=\epsilon>0$.
But now we want to extend it over $s=0$ to use contour integral techniques.
After this, the extra contribution one finds in the exponent is
\begin{equation}\label{prefactor}
  -\frac{\pi^2(m^2-\epsilon_-^2)}{6\epsilon_1\epsilon_2\beta}+
  \beta\frac{m^2-\epsilon_-^2}{24\epsilon_1\epsilon_2}\left(m^2+\epsilon_-^2
  -\frac{\epsilon_1^2+\epsilon_2^2}{2}\right)=
  -\frac{\pi^2(m^2-\epsilon_-^2)}{6\epsilon_1\epsilon_2\beta}+
  \beta\frac{(m^2-\epsilon_-^2)(m^2-\epsilon_+^2)}{24\epsilon_1\epsilon_2}\ .
\end{equation}
So apart from the finite integral to be discussed shortly, the extra
exponent from (\ref{divergent}) and (\ref{prefactor}) is
\begin{equation}
  \frac{\epsilon_-^2-m^2}{24\epsilon_1\epsilon_2}\left(\frac{4\pi^2}{\beta}-\beta\right)+
  \beta\frac{(m^2-\epsilon_-^2)(m^2-\epsilon_+^2)}{24\epsilon_1\epsilon_2}\ .
\end{equation}
The first term will be absorbed into the definition of the two
mutually dual partition functions.

The finite integral can now be written as
\begin{equation}
  \hspace*{-1cm}\sum_{n=1}^\infty\oint\frac{ds}{s}\
  \frac{\sin\frac{\pi(m+\epsilon_-)s}{\beta_D}\sin\frac{\pi(m-\epsilon_-)s}{\beta_D}}
  {2\sin\frac{\pi\epsilon_1s}{\beta_D}\sin\frac{\pi\epsilon_2s}{\beta_D}}
  \left(\coth\frac{s}{2}-\frac{1}{s}\right)e^{\frac{2\pi i ns}{\beta_D}}+
  \oint\frac{ds}{s^2}\left(
  \frac{\sin\frac{\pi(m+\epsilon_-)s}{\beta_D}\sin\frac{\pi(m-\epsilon_-)s}{\beta_D}}
  {2\sin\frac{\pi\epsilon_1s}{\beta_D}\sin\frac{\pi\epsilon_2s}{\beta_D}}
  -\frac{m^2-\epsilon_-^2}{2\epsilon_1\epsilon_2}\right)
  e^{\frac{2\pi i ns}{\beta_D}}
\end{equation}
plus
\begin{equation}
  \frac{1}{2}\oint\frac{ds}{s}\
  \frac{\sin\frac{\pi(m+\epsilon_-)s}{\beta_D}\sin\frac{\pi(m-\epsilon_-)s}{\beta_D}}
  {2\sin\frac{\pi\epsilon_1s}{\beta_D}\sin\frac{\pi\epsilon_2s}{\beta_D}}
  \left(\coth\frac{s}{2}-\frac{1}{s}\right)+
  \frac{1}{2}\oint\frac{ds}{s^2}\left(
  \frac{\sin\frac{\pi(m+\epsilon_-)s}{\beta_D}\sin\frac{\pi(m-\epsilon_-)s}{\beta_D}}
  {2\sin\frac{\pi\epsilon_1s}{\beta_D}\sin\frac{\pi\epsilon_2s}{\beta_D}}
  -\frac{m^2-\epsilon_-^2}{2\epsilon_1\epsilon_2}\right)\ .
\end{equation}
The integrals can now be evaluated as sums of residues. As
${\rm Im}(\epsilon_{1,2})>0$, we only have to consider poles coming from
$\coth\frac{s}{2}$ in the first integral.
The poles are at $s=2\pi ip$ for $p=1,2,3,\cdots$. The first integral provides
a factor of
\begin{equation}
  \exp\left[\sum_{p=1}^\infty\frac{1}{p}\
  \frac{\sin\frac{p\beta(m+\epsilon_-)}{2}\sin\frac{p\beta(m-\epsilon_-)}{2}}
  {\sinh\frac{p\beta\epsilon_1}{2}\sin\frac{p\beta\epsilon_2}{2}}\
  \frac{e^{-\beta}}{1-e^{-\beta}}\right]=
  PE\left[I_-(\beta\epsilon_1,\beta\epsilon_2,\beta m)
  \frac{e^{-\beta}}{1-e^{-\beta}}\right]
\end{equation}
to the partition function,
where we inserted $\beta_D=\frac{4\pi^2}{\beta}$. This is the instanton part
of the strong-coupling index. The second integral provides a factor of
\begin{equation}
  PE\left[\frac{1}{2}I_-(\beta\epsilon_1,\beta\epsilon_2,\beta m)\right]
  =PE\left[\frac{1}{2}I_+(\beta\epsilon_1,\beta\epsilon_2,\beta m)+\frac{1}{2}\right]\ .
\end{equation}
This can be identified as the perturbative part of the strong-coupling index,
again with an elimination of the real fermion zero mode. Now let us define
\begin{equation}
  Z\equiv e^{\frac{m^2-\epsilon_-^2}{24\epsilon_1\epsilon_2}
  \beta}PE\left[\frac{1}{2}I_+(\beta\epsilon_{1,2},\beta m)
  +\frac{1}{2}+I_-(\beta\epsilon_{1,2},\beta m)\frac{e^{-\beta}}{1-e^{-\beta}}\right]
\end{equation}
and $Z_D$ to take the same form as above, with replacements of
$\beta,\beta \epsilon_{1,2},\beta m$ by
$\beta_D=\frac{4\pi^2}{\beta}$, $\beta_D\epsilon^D_{1,2}\equiv 2\pi i\epsilon_{1,2}$,
$\beta_D m_D\equiv 2\pi im$. The last two equations are defining $\epsilon_{1,2}^D$
and $m_D$. Then, one finds the following S-duality transformation:
\begin{equation}\label{+dominated-S-dual}
  Z_D\equiv Z(\beta_D,\epsilon^D_{1,2},m_D)=
  \exp\left[\frac{\beta(m^2-\epsilon_-^2)(m^2-\epsilon_+^2)}
  {24\epsilon_1\epsilon_2}\right]Z(\beta,\epsilon_{1,2},m)\ .
\end{equation}
This establishes a version of S-duality of the Abelian partition function
on $\mathbb{R}^4_\epsilon\times T^2$, at least in certain range of Omega
background parameters. In particular, this case contains
the trivial case with $m=0$, $\epsilon_1=\epsilon_2$ in which case $Z=Z_D=1$.

Now we are interested in other regions on the $\epsilon_{1,2}$ plane. It is
useful to consider the perturbative part $\frac{1}{2}I_+$ in some detail, as
the nature of its series expansion subtly depends on the parameter ranges.
One finds
\begin{equation}\label{pert-letter-expand}
  I_+=\frac{\sinh\frac{\beta(m+\epsilon_+)}{2}\sinh\frac{\beta(m-\epsilon_+)}{2}}
  {\sinh\frac{\beta\epsilon_1}{2}\sinh\frac{\beta\epsilon_2}{2}}=\pm
  \frac{e^{-\beta\epsilon_\pm}(e^{\beta m}+e^{-\beta m}-e^{\beta\epsilon_+}
  -e^{-\beta\epsilon_+})}{(1-e^{-\beta\epsilon_1})(1-e^{\mp\beta\epsilon_2})}\ .
\end{equation}
We have set $\epsilon_1>0$ without losing generality, but since the relative sign
between $\epsilon_{1,2}$ matters, we consider two possible expansions of denominator
in $e^{\mp\beta\epsilon_2}$.
When both $\epsilon_{1,2}$ are positive (or more generally when they have same signs),
there exists a fermion zero mode as discussed above (coming from the 3rd term in
the numerator). Other three terms in the numerator yield stable modes supposing
that $\epsilon_+$ is large enough compared to $|m|$.
On the other hand, when $\epsilon_1>0>\epsilon_2$, the numerator
becomes
\begin{equation}\label{pert-letter-num}
  e^{\beta(-\epsilon_-+m)}+e^{-\beta(\epsilon_-+m)}-e^{\beta\epsilon_2}
  -e^{-\beta\epsilon_1}\ .
\end{equation}
The last two terms are for nonzero modes, and supposing the $m>0$ without losing
generality, the 2nd term is also for a nonzero mode (since $\epsilon_->0$).
Now the first term is for a nonzero mode when $\epsilon_->m$.

At this point, let us consider the $\mathbb{R}^4\times S^1$ partition function as
a building block of the $S^5$ partition function, as one motivation of this
appendix is to develop an S-duality rule (supplementing \cite{vafa}) and use it
to study the $S^5$ partition function. At the three fixed points on the
$\mathbb{CP}^2$ base of $S^5$, the parameters are given by \cite{Kim:2012qf}
\begin{eqnarray}
  {\rm 1st\ fixed\ point}&:& (\epsilon_1,\epsilon_2,m)\rightarrow\left(\frac{b-a}{1+a},
  \frac{c-a}{1+a},\frac{m}{1+a}+\frac{3}{2}\right)\nonumber\\
  {\rm 2nd\ fixed\ point}&:& (\epsilon_1,\epsilon_2,m)\rightarrow\left(\frac{c-b}{1+b},
  \frac{a-b}{1+b},\frac{m}{1+b}+\frac{3}{2}\right)\nonumber\\
  {\rm 3rd\ fixed\ point}&:& (\epsilon_1,\epsilon_2,m)\rightarrow\left(\frac{a-c}{1+c},
  \frac{b-c}{1+c},\frac{m}{1+c}+\frac{3}{2}\right)\ .
\end{eqnarray}
Of course the last $\frac{3}{2}$ can be replaced by any half an odd integer,
as $\pi$ times this is an argument of the periodic sine function.
Since the three factors became particularly simple around
$m=\frac{1}{2}$ and $|c|\ll |a|,|b|$ (near the maximal SUSY point), we try to
expand the 3 partition functions around this point. In this case, the $\epsilon_\pm,m$
parameters around the third fixed point are given by
\begin{equation}
  (\epsilon_+,\epsilon_-,m)\rightarrow\left(-\frac{3c}{2(1+c)},
  \frac{a-b}{2(1+c)},\frac{\delta}{1+c}+\frac{\epsilon_+}{3}\right)
\end{equation}
where $\delta\equiv m-\frac{1}{2}$, and we used $\frac{m}{1+c}-\frac{1}{2}$
(replacing $+\frac{3}{2}$ shift by $-\frac{1}{2}$)
for the mass parameter using the periodicity of $\sin$ functions.
To be definite, let us choose an order of parameters $a,b,c,\delta$.
We take $b\ll c<0\ll a$, so that we could use $b\approx -a$ (coming from $a+b+c=0$)
for the purpose
of evaluating inequalities. (The above relation automatically implies $|c|\ll 1$).
Also, we take $\delta$ to be $|\delta|\ll |c|\ll |a|,|b|$. Thus, at all three fixed
points, the effective mass parameters are smaller than the effective $\epsilon_+$
parameters, since
\begin{equation}
  |\epsilon_+|-\left|\frac{\delta}{1+a_i}+\frac{\epsilon_+}{3}\right|
  \approx\frac{2}{3}|\epsilon_+|>0\ .
\end{equation}
This makes it possible for us to safely apply the $\epsilon_+$ dominated
S-duality transformation (\ref{+dominated-S-dual}) at the first and second fixed
points.  This is because in these cases, one additionally finds $|\epsilon_+|\approx\frac{3|a_i|}{2(1+a_i)}>
\frac{|a_i|}{2(1+a_i)}\approx|\epsilon_-|$, making $\epsilon_+$ to be the largest
parameter. Thus the numerator of (\ref{pert-letter-expand}) with upper signs
has a fermion zero mode that we already treated, without any other dangerous modes.

For the 3rd fixed point, where $|\epsilon_+|\approx\frac{3|c|}{2}\ll
a\approx\epsilon_-$, $\epsilon_+$ is not the largest parameter so that one cannot use
(\ref{+dominated-S-dual}). Comparing $|\epsilon_-|$ and $|m|$ at the third fixed point,
one finds
\begin{equation}
  |\epsilon_-|-|m|\approx a-\frac{|\epsilon_+|}{3}>0\ .
\end{equation}
So $I_+$ (\ref{pert-letter-expand})m (\ref{pert-letter-num}) has no zero modes.
When $\epsilon_-$ is the largest parameter, we start from a weak-coupling
expression similar to (\ref{weak-abelian}), but without adding $+\frac{1}{2}$
as there are no fermion zero modes in $I_+$:
\begin{equation}\label{weak-abelian-2}
  PE\left[\frac{1}{2}I_++I_-\frac{e^{-\beta_D}}{1-e^{-\beta_D}}\right]
  =PE\left[\frac{1}{2}I_+\coth\frac{\beta_D}{2}
  +\frac{e^{-\beta_D}}{1-e^{-\beta_D}}\right]\ .
\end{equation}
Here we replaced $I_-=I_++1$, since now $I_-$ has a bosonic `zero mode.'
The first term in the $PE$ is obtained by taking (\ref{+dominated-start})
and then replacing $\epsilon_-$ in $I_-$ by $\epsilon_+$. Thus, all the arguments
which led us to (\ref{+dominated-S-dual}) go through, after renaming parameters
in a minor way. In particular, the condition (\ref{semi-circle-condition})
\begin{equation}
  2|\epsilon_+|-|\epsilon_-+m|-|\epsilon_--m|\approx
  2|\epsilon_+|-2|\epsilon_-|>0
\end{equation}
which was important for adding the upper half circle is again satisfied
with the roles of $\epsilon_\pm$ changed. So the $PE$ of the first exponent
in (\ref{weak-abelian-2}) is essentially self-dual, with a prefactor like that
in (\ref{+dominated-S-dual}). In particular, all $\epsilon_-$ on the left hand
side of (\ref{prefactor}) are replaced by $\epsilon_+$.
Combining this with a term like (\ref{divergent}), again with $\epsilon_-$
replaced by $\epsilon_+$,  one obtains
\begin{equation}
  \frac{\epsilon_+^2-m^2}{24\epsilon_1\epsilon_2}
  \left(\frac{4\pi^2}{\beta}-\beta\right)+
  \beta\frac{(m^2-\epsilon_-^2)(m^2-\epsilon_+^2)}{24\epsilon_1\epsilon_2}
\end{equation}
in this case (i.e. third fixed point) as an extra exponent.
The second term appearing in the  $PE$ of (\ref{weak-abelian-2}) yields the
inverse of the Dedekind eta function, if one multiplies a factor of $e^{\frac{\beta}{24}}$
into the definition of the partition function. The modular property of this
function is well known. So in this case, suited for the third
fixed point, we should put an extra factor of
\begin{equation}
  \exp\left[-\beta\frac{\epsilon_+^2-m^2}{24\epsilon_1\epsilon_2}
  +\frac{\beta}{24}\right]
\end{equation}
into the definition of $Z$. Then, one obtains
\begin{equation}\label{-dominated-S-dual}
  Z_D\equiv Z(\beta_D,\epsilon_{1,2}^D,m_D)=\left(\frac{2\pi}{\beta}\right)^{1/2}
  \exp\left[\frac{\beta(m^2-\epsilon_-^2)(m^2-\epsilon_+^2)}
  {24\epsilon_1\epsilon_2}\right]Z(\beta,\epsilon_{1,2},m)\ .
\end{equation}
So apart from the second factor which is the same as the S-duality prefactor
in (\ref{+dominated-S-dual}), one has an extra modular factor in the S-duality
when $\epsilon_-$ is the largest parameter. This includes the case with
$m=0$, $\epsilon_1=-\epsilon_2$ mentioned at the beginning of this appendix.

Now as a check of the two S-dual transformations (\ref{+dominated-S-dual}),
(\ref{-dominated-S-dual}), we combine all the three expressions to see the S-duality
transformation of $Z^{(1)}Z^{(2)}Z^{(3)}$. This will reproduce the strong coupling expansion
of the $S^5$ partition function studied in \cite{Lockhart:2012vp,Kim:2012qf}, but now
in a way which makes the role of S-duality on $\mathbb{R}^4\times S^1$ clear.
The subtractions of $-\frac{1}{2}$ from the fermion zero modes in the perturbative parts of
$Z^{(1)}$ and $Z^{(2)}$ amount to be $+1$ inside $PE$: this is provided by the
bosonic zero mode's contribution in the ghost multiplet of gauge fixed action on
$S^5$. Firstly, the prefactors we multiply to define $Z$ at weak
coupling is $\frac{4\pi^2}{\beta}$ times
\begin{equation}
  -\frac{1}{1+a}\ \frac{\frac{(b-c)^2}{4(1+a)^2}-
  \frac{(2m-1-a)^2}{4(1+a)^2}}{24\frac{b-a}{1+a}\cdot\frac{c-a}{1+a}}
  -\frac{1}{1+b}\ \frac{\frac{(c-a)^2}{4(1+b)^2}-
  \frac{(2m-1-b)^2}{4(1+b)^2}}{24\frac{c-b}{1+b}\cdot\frac{a-b}{1+b}}
  +\frac{1}{1+c}\left(\frac{1}{24}-\frac{\frac{9c^2}{4(1+c)^2}-
  \frac{(2m-1-c)^2}{4(1+c)^2}}{24\frac{a-c}{1+c}\cdot\frac{b-c}{1+c}}\right)
\end{equation}
in the exponent. The factors of $\frac{1}{1+a_i}$ arise because the effective
$\beta_D$ at the three fixed points is $\frac{4\pi^2}{\beta(1+a_i)}$.
One can show that this exactly equals
\begin{equation}
  \frac{1}{24}\left(\frac{1}{1+c}-\frac{\left(\frac{1}{2}-c\right)^2-m^2}
  {(1+a)(1+b)(1+c)}\right)\ ,
\end{equation}
which is the $\mathcal{O}(\beta^{-1})$ term in the weak coupling
that we had to add as the constant shift of the action on $S^5$ \cite{Kim:2012qf}.
We also collect all the $\mathcal{O}(\beta)$ terms in the exponent that
are absorbed into the definition of $Z_D$'s at strong coupling. This should be
identified as the Casimir energy. In the exponent, we find $\beta$ times
\begin{eqnarray}
  \hspace*{-1cm}&&
  (1+a)\frac{(m-\frac{1+a}{2})^2-\frac{(b-c)^2}{4}}{24(b-a)(c-a)}
  +(1+b)\frac{(m-\frac{1+b}{2})^2-\frac{(c-a)^2}{4}}{24(c-b)(a-b)}+
  \frac{1+c}{24}\left(1+\frac{(m-\frac{1+c}{2})^2-\frac{9c^2}{4}}
  {(a-c)(b-c)}\right)\nonumber\\
  &&+\left[\frac{\left((m-\frac{1+a}{2})^2-
  \frac{(b-c)^2}{4}\right)\left((m-\frac{1+a}{2})^2-\frac{9a^2}{4}\right)}
  {24(1+a)(b-a)(c-a)}+(a,b,c\ {\rm perm.})\right]\ ,
\end{eqnarray}
where the first line comes from the $\mathcal{O}(\beta)$ terms similar to
the $\mathcal{O}(\beta^{-1})$ terms discussed above, and the second line
comes from the extra $\frac{\beta(m^2-\epsilon_-^2)(m^2-\epsilon_+^2)}
{24\epsilon_1\epsilon_2}$ terms appearing in the dualization.
One can show that these all add up to
\begin{equation}
  \frac{1}{24}\left(1+\frac{2abc+(1-ab-bc-ca)(\frac{1}{4}-m^2)
  +(\frac{1}{4}-m^2)^2}{(1+a)(1+b)(1+c)}\right)\ ,
\end{equation}
which is exactly the Casimir energy contribution (\ref{abelian-S5-casimir})
that one finds on $S^5$ with a particular regulator \cite{Kim:2012qf}.

Finally, we combine the three pieces of indices at strong coupling.
It suffices to consider the sum of three `letter indices' appearing in $PE$.
One obtains
\begin{equation}
  \left[\frac{1}{2}I_-\left(\beta(b-a),\beta(c-a),
  \beta\big(m-\frac{1+a}{2}\big)\right)\coth\frac{\beta(1+a)}{2}
  +(a,b,c\ {\rm permutations})\right]-\frac{1}{2}\ ,
\end{equation}
where we rewrote the letter index from the third fixed point as
\begin{equation}
  \frac{1}{2}I_+\coth\frac{\beta}{2}+\frac{e^{-\beta}}{1-e^{-\beta}}
  =\frac{1}{2}I_-\coth\frac{\beta}{2}-\frac{1}{2}\ .
\end{equation}
One can show that the above sum of three letter indices completely agrees with
the letter index of the Abelian 6d (2,0) theory:
\begin{equation}
  \frac{e^{-\frac{3\beta}{2}}(e^{\beta m}+e^{-\beta m})
  -e^{-2\beta}(e^{\beta a}+e^{\beta b}+e^{\beta c})+e^{-3\beta}}
  {(1-e^{-\beta(1+a)})(1-e^{-\beta(1+b)})(1-e^{-\beta(1+c)})}\ ,
\end{equation}
completing the derivation of the 6d Abelian index from the $S^5$ partition function
using S-duality on $\mathbb{R}^4\times S^1$.

One comment for the S-duality transformation (\ref{+dominated-S-dual}),
(\ref{-dominated-S-dual}) is that the prefactors look more complicated than
those appearing for the ordinary Jacobi forms. Namely, the Jacobi form
$\phi(\tau,z)$ with weight $k$ and index $m$ transform under S-duality as
\begin{equation}
  \phi\left(-\frac{1}{\tau},\frac{z}{\tau}\right)=\tau^k
  e^{\frac{2\pi imz^2}{\tau}}\phi(\tau,z)\ .
\end{equation}
The Jacobi theta functions, which were the building blocks for the self-dual
string part of the $\mathbb{R}^4\times S^1$ partition function \cite{vafa},
are Jacobi forms. Maybe (\ref{+dominated-S-dual}) and (\ref{-dominated-S-dual})
can be understood as suitable generalized Jacobi forms, or equivalently Fourier
coefficients of generalized Siegel modular forms.

\section{Spinors on $\mathbb{CP}^2\times\mathbb{R}$}

In this section, we explain the spinors on $\mathbb{CP}^2\times \mathbb{R}$ used
in this paper, and especially the Killing spinors which correspond to the
supersymmetry preserved by our 5d gauge theories.
For our purpose, we will obtain the 5d Killing spinors from the dimensional reduction of 6d Killing spinors
on $S^5\times \mathbb{R}$.
It is convenient to first fix our convention for 6d and 5d spinor calculus.
All the computations below will be carried out in Euclidean signature. The Lorentzian
spinors can be similarly studied by carefully taking the Wick rotation
into account.

We take $8\times 8$ gamma matrices $\Gamma^M$ in 6d  as
\begin{eqnarray}
	\Gamma^{1,2,3,4} = \gamma^{1,2,3,4}\otimes \sigma^1 \,, \ \ \ \Gamma^\tau = \gamma^\tau\otimes\sigma^1 \,, \ \ \ \Gamma^{5} = -{\bf 1}_4\otimes \sigma^2\ ,
\end{eqnarray}
where the 5th direction is chosen to be the Hopf fiber direction of $S^5$,
along which we will dimensionally  reduce the 6d theory.
In 6 dimensions, the chirality matrix is $i\Gamma^{\tau12345}={\bf 1}_4 \otimes\sigma^3$ and the charge conjugation
matrix is chosen as $C=\Gamma^{13}$ which acts on the gamma matrices as follows:
\begin{equation}
	C\Gamma_\tau C^{-1} = (\Gamma_\tau)^* \,, \ \ \
	C\Gamma_MC^{-1} = -(\Gamma_M)^* \ \ (M=1,2,3,4,5)\ .
\end{equation}
The $4\times 4$ gamma matrices $\gamma^\mu$ for 5d spacetime are given by %in Euclidean signature are in the representation
\begin{eqnarray}
	\gamma^{1,2,3} = \sigma^{1,2,3}\otimes \sigma^1 \,, \ \ \ \gamma^4 = {\bf 1}_2\otimes\sigma^2 \,,
	\ \ \  \gamma^\tau = -{\bf 1}_2\otimes \sigma^3\ ,
\end{eqnarray}
while the $4\times 4$ gamma matrices $\hat\gamma^I$ for the internal $SO(5)_R$ symmetry are chosen so that %take the following forms
%and the $4\times 4$ gamma matrices $\hat\gamma^I$ for the internal $SO(5)_R$ spinors take the following forms
\begin{eqnarray}
	\hat\gamma^1 = \sigma^1\otimes\sigma^1 \,, \ \ \ \hat\gamma^2 = \sigma^2\otimes\sigma^1
	\,, \ \ \ \hat\gamma^4 = \sigma^3\otimes\sigma^1 \,, \ \ \ \hat\gamma^5={\bf 1}_2\otimes\sigma^2
	\,, \ \ \ \hat\gamma^3 =\hat\gamma^{1245} = -{\bf 1}_2\otimes \sigma^3 \,.\ \
\end{eqnarray}
We note that this internal R-symmetry is, in general, broken down to $U(1)\times U(1)$ under 5d reduction.

There are 32 Killing spinors on $S^5\times \mathbb{R}$  satisfying
\begin{equation}\label{32-killing}
	\nabla_M\epsilon_\pm = \pm\frac{1}{2r}\Gamma_M\Gamma_\tau\epsilon_\pm\ .
\end{equation}
They are 6d chiral spinors, $i\Gamma^{\tau12345}\epsilon=\epsilon$. We impose 6d symplectic Majorana condition on them as
\begin{equation}
	(\epsilon_-)^* = C\otimes \hat{C}\epsilon_+\ ,
\end{equation}
where $\hat{C}$ is the invariant tensor for $USp(4)_R\,(\approx SO(5)_R)$ internal symmetry.

The 6d metric on $S^5\times \mathbb{R}$ is given by
\begin{eqnarray}
	ds^2_{S^5\times \mathbb{R}} &=& ds_{S^5}^2 + d\tau^2 \,, \\ \nonumber
	ds^2_{S^5} &=& r^2 ds_{\mathbb{CP}^2}^2+r^2\left(dy+\theta_\mu dx^\mu\right)^2\ ,
\end{eqnarray}
where $d\theta=2J$ and $J$ is the K\"ahler form on $\mathbb{CP}^2$ base.
We perform dimensional reduction along the Hopf fiber circle $S^1$, parametrized by coordinate $y$. %, and want to have some amount of supersymmetries preserved in the 5d theory on $\mathbb{CP}^2\times \mathbb{R}$.
The reduction leads to the 5d metric
\begin{eqnarray}
	ds^2_{\mathbb{CP}^2\times \mathbb{R}} &=& r^2ds_{\mathbb{CP}^2}^2 + d\tau^2 \,, \\ \nonumber
	ds^2_{\mathbb{CP}^2} &=& d\rho^2 +\frac{1}{4}\sin^2\rho\,(\boldsymbol{\sigma_1}^2+\boldsymbol{\sigma_2}^2)
	+\frac{1}{4}\sin^2\rho\,\cos^2\rho\,\boldsymbol{\sigma_3}^2\ ,
\end{eqnarray}
where the 1-forms $\boldsymbol{\sigma_i}$ are given by
\begin{eqnarray}
	\boldsymbol{\sigma_1} &=& \sin\psi d\theta-\cos\psi\sin\theta d\phi \,, \nonumber \\
	\boldsymbol{\sigma_2} &=& \cos\psi d\theta +\sin\psi\sin\theta d\phi \,, \nonumber \\
	\boldsymbol{\sigma_3} &=& d\phi + \cos\theta d\phi \,.
\end{eqnarray}
It is convenient to choose a $\mathbb{CP}^2$ frame, $ds_{\mathbb{CP}^2}^2 = e^a e^a$, with
the vierbein basis
\begin{eqnarray}
	e^1 = d\rho \,, \quad e^2 = \frac{1}{2}\sin\rho\cos\rho \,\boldsymbol{\sigma_3} \,, \quad
	e^3 = \frac{1}{2}\sin\rho\,\boldsymbol{\sigma_1} \,, \quad e^4 = \frac{1}{2}\sin\rho \,\boldsymbol{\sigma_2} \,.
\end{eqnarray}
In terms of these these vierbeins, the K\"ahler 2-form $J$ of $\mathbb{CP}^2$ is expressed as
%Using these vierbeins, the K\"ahler 2-form $J$ of $\mathbb{CP}^2$ is given by
$J=e^1\wedge e^2-e^3\wedge e^4$.

We want to have some amount of supersymmetries preserved in the 5d theory on $\mathbb{CP}^2\times \mathbb{R}$.
Naive reduction will break all the supersymmetries since all of them carry non-zero charges for the circle $S^1$ rotation. Thus it is necessary to twist the translation along $y$
with the internal R-symmetry before reduction. This amounts to redefining all
the R-charged fields by suitably multiplying $y$-dependent phases. We shall choose
a twisting such that $i\partial_y$ changes to
\begin{equation}\label{twisted-shift}
 	i\partial_y \rightarrow i\partial_y  + \frac{3}{2}(R_1+R_2)+n(R_1-R_2)
 	\equiv -Kk + M_n\ .
\end{equation}
in the new frame. Here, $Kk$ is the same as \eqref{twisting} defined as
\begin{equation}
 	Kk \equiv j_1+j_2+j_3+\frac{3}{2}(R_1+R_2)+n(R_1-R_2) \,,
\end{equation}
in terms of the charges before we redefine the R-charged fields. It is simply
$-i\partial_y$ as in the right hand side of (\ref{twisted-shift}) after the
redefinition. We introduce the $\mathbb{Z}_K$ quotient acting on the newly defined $y$
by $y\rightarrow y+\frac{2\pi}{K}$. At large $K$, the 6d theory reduces to 5d theory
that contains fields and supercharges at $k=0$ only.

The Killing spinor equation (\ref{32-killing}) can be written in terms of 5d derivatives and gamma matrices as
\begin{equation}
	 \left[\partial_y-\frac{1}{4}J_{\mu\nu}\gamma^{\mu\nu}\right]\epsilon_\pm
   =\pm\frac{i}{2}\gamma_\tau\epsilon_\pm\,, \ \ \
	\left[\nabla_\mu-\frac{1}{r}\ \theta_\mu\partial_y -\frac{i}{2r}J_{\mu\nu}\gamma^\nu\right]\epsilon_\pm = \pm\frac{1}{2r}\gamma_\mu\gamma_\tau\epsilon_\pm\ ,
\end{equation}
where $\mu=1,2,3,4,\tau$ labels 5d coordinates %denotes the index for 5d coordinates
and $\nabla_\mu$ is the 5d
covariant derivative. There are 32 Killing spinors solving these equations.
Let us consider the twisting explained above and reduce to 5 dimensions by keeping the
modes with $k=0$ only. At the level of Killing spinors, we are focussing on those
which are independent of $y$. After twisting and setting $\partial_y=0$,
one obtains the following 5d Killing spinor equations
\begin{equation}\label{Killing-spinor-eq3}
	 \left[-iM_n-\frac{1}{4}J_{\mu\nu}\gamma^{\mu\nu}\right]\epsilon_\pm
=\pm\frac{i}{2}\gamma_\tau\epsilon_\pm\,, \ \ \
	\left[\nabla_\mu+\frac{i}{r}\ \theta_\mu M_n -\frac{i}{2r}J_{\mu\nu}\gamma^\nu\right]\epsilon_\pm = \pm\frac{1}{2r}\gamma_\mu\gamma_\tau\epsilon_\pm\ .
\end{equation}

The number of Killing spinors (or supercharges) in 5d which are the solutions to (\ref{Killing-spinor-eq3}), depends on the twisting parameter $n$.
For generic value of $n$, only two Killing spinors (corresponding to $Q_{---}^{++}$ and the conjugate $S_{+++}^{--}$) remain to be the solutions to the 5d Killing spinor equations, which satisfy
\begin{equation}\label{Killing-spinor-eq4}
	D_\mu\epsilon_\pm = \frac{i}{2r}J_{\mu\nu}\gamma^\nu\epsilon_\pm\pm\frac{1}{2r}\gamma_\mu\gamma_\tau\epsilon_\pm\,,
	\ \ \ M_n\epsilon_\pm = \mp\frac{3}{2}\epsilon_\pm \,, \ \ \
	i\gamma^{12}\epsilon_\pm = -i\gamma^{34}\epsilon_\pm = \mp\epsilon_\pm\ ,
\end{equation}
where $D_\mu = \nabla_\mu +\frac{i}{r}\ \theta_\mu M_n$.
The first equation implies that two Killing spinors are covariantly constant on $\mathbb{CP}^2$
with the `charged' covariant derivative $D_\mu$. Hence they are singlets under the $SU(3)$
isometry. These two Killing spinors are the ones that we have used to define the index in this
paper.

\section{Off-shell QFT for vector multiplet from supergravity}\label{app:OffshellQFT}

In this appendix, we construct the off-shell action (\ref{vector-action})
on $\mathbb{CP}^2\times \mathbb{R}$ for the vector multiplet, from the off-shell
supergravity of \cite{Kugo:2000af,Hanaki:2006pj}. We follow the notation of
\cite{Kim:2012qf}.

The vector multiplet Lagrangian is governed by the background scalar and vector fields in the Weyl multiplet and the gravi-photon multiplet coupled to our fields in the vector multiplet.
We first need to identify nonzero background fields of the Weyl multiplet in 5d supergravity.
This can be achieved by comparing (\ref{Killing-spinor-eq2}) with the gravitino SUSY variation and solving the condition of vanishing dilatino SUSY variation
in the Weyl multiplet. The first equation in (\ref{Killing-spinor-eq2}),
or (\ref{Killing-spinor-eq4}), can be rewritten as
\begin{equation} \label{C1}
	D_\mu\epsilon_\pm +\frac{i}{4r}\gamma_{\mu\nu\lambda}J^{\nu\lambda}\epsilon_\pm = \gamma_\mu\eta_\pm \,, \ \ \
	\eta_\pm =\frac{i}{4r}J_{\mu\nu}\gamma^{\mu\nu}\epsilon_\pm \pm\frac{1}{2r}\epsilon_\pm =\mp\frac{1}{2r}\epsilon_\pm \ .
\end{equation}
This will be compared with the gravitino SUSY variation of 5d supergravity given by
\begin{equation} \label{C2}
	\delta\psi_\mu = \mathcal{D}_\mu\epsilon + \frac{1}{2}\gamma_{\mu\nu\lambda}v^{\nu\lambda}\epsilon - \gamma_\mu\eta=0\ ,
\end{equation}
where $\mathcal{D}_\mu\epsilon = (\partial_\mu+\frac{1}{4}\omega_{\mu ab}\gamma^{ab}+\frac{1}{2}b_\mu)\epsilon - \tilde{V}_\mu\epsilon$.
We see that these two equations \eqref{C1} and \eqref{C2} are identical when the background fields and the Killing spinors of 5d supergravity take the values
\begin{equation}
	b_\mu = 0 \,, \ \ \ \tilde{V}_\mu = -\frac{i}{r}\theta_\mu M_n \,, \ \ \ v = \frac{i}{2r}J \,, \ \ \  \epsilon = \epsilon_\pm \,, \ \ \ \eta=\eta_\pm\ .
\end{equation}
The background auxiliary field $\rm{D}$ is determined to be ${\rm D}=12/r^2$ by gaugino SUSY variation
\begin{eqnarray}
	\delta \chi &=& {\rm D}\epsilon - 2\gamma^c\gamma^{ab}\epsilon D_av_{bc} + \gamma^{\mu\nu}\tilde{F}_{\mu\nu}\epsilon - 2\gamma^a\epsilon \, \epsilon_{abcde}v^{bc}v^{de}+4\gamma^{\mu\nu}v_{\mu\nu}\eta \nonumber \\
	&=& {\rm D}\epsilon -\frac{12}{r^2}\epsilon = 0\ .
\end{eqnarray}
This fixes all the background values for bosonic fields in the Weyl multiplet.

Our dynamical fields in the vector multiplets also couple to the background
gravi-photon vector multiplet. The SUSY transformation for all vector multiplets is
just (\ref{offshell-SUSY-vector}). The gravi-photon multiplet involves a scalar
dilaton $\phi^0$, and the `RR 1-form' potential $A^0_\mu$, and a triplet of
$D$-term fields $D^{0I}$. As we require SUSY invariance, these fields should
assume to take background values which make the gaugino SUSY variation to vanish.
We find the following solution
\begin{equation}
	\mathcal{V}^0=(A^0_\mu, \phi^0, D^0) = (\theta_\mu,\alpha,
    \frac{4-\alpha}{r}\delta^{I}_3)\ ,
\end{equation}
where $\alpha$ is a constant, and the 1-form $\theta$ satisfies $d\theta=2J$.

We finally choose the cubic function $\mathcal{N}=C_{IJK}\mathcal{V}^I\mathcal{V}^J\mathcal{V}^k
=\frac{1}{2}\phi^0{\rm tr}(\phi^2)$
where $\phi$ is the adjoint scalar in the dynamical vector multiplet.
Altogether, the bosonic action of the vector multiplet from the off-shell supergravity becomes
\begin{eqnarray}\label{sugra-general}
	\tilde{g}^2_{YM}\mathcal{L}_B &=& {\rm tr}\, \bigg[ \alpha\left( \frac{1}{4}F_{\mu\nu}F^{\mu\nu}+\frac{1}{2}D_\mu\phi D^\mu\phi-\frac{1}{2}D^ID^I \right)+\frac{1}{r^2}\left(\frac{3}{2}\alpha-4\right)\phi^2 \\ \nonumber
	&& +\frac{1-\alpha}{r}J^{\mu\nu}F_{\mu\nu}\phi - \frac{4-\alpha}{r}D^3\phi
	-\frac{i}{2re}\epsilon^{\mu\nu\lambda\rho\sigma}\left(A_\mu\partial_\nu A_\lambda-\frac{2i}{3}A_\mu A_\nu A_\lambda\right)J_{\rho\sigma}\bigg] \ .
\end{eqnarray}
In all the analysis, we required only 2 off-shell supercharges. The off-shell Lagrangian corresponding to our on-shell action (\ref{onshell-action}) is obtained by setting
$\alpha=1$. At $\alpha=1$,
\begin{eqnarray}
	\tilde{g}^2_{YM}\mathcal{L}_B &=& {\rm tr}\, \bigg[\frac{1}{4}F_{\mu\nu}F^{\mu\nu}+\frac{1}{2}D_\mu\phi D^\mu\phi-\frac{1}{2}D^ID^I -\frac{5}{2r^2}\phi^2
	- \frac{3}{r}D^3\phi \\ \nonumber
	&&-\frac{i}{2re}\epsilon^{\mu\nu\lambda\rho\sigma}\left(A_\mu\partial_\nu A_\lambda-\frac{2i}{3}A_\mu A_\nu A_\lambda\right)J_{\rho\sigma}\bigg]\ ,
\end{eqnarray}
which is the bosonic part of the off-shell action in section 2.
It is straightforward to write fermionic part from the 5d supergravity and it completes the derivation of the off-shell action (\ref{vector-action}).

We fixed $\alpha=1$ basically from the reduction of the 6d Abelian theory on
$S^5/\mathbb{Z}_K\times\mathbb{R}$. Requiring more SUSY at $n=\pm\frac{1}{2}$ or
$\pm\frac{3}{2}$ should freeze $\alpha=1$. For instance, the theories at $n=\pm\frac{3}{2}$
should have an internal $SU(2)$ symmetry, rotating the vector multiplet scalar $\phi$ and
a complex scalar in the hypermultiplet as a triplet. On the other hand, nonzero
$J^{\mu\nu} F_{\mu\nu}\phi$ term in (\ref{sugra-general}) is incompatible with this symmetry. This leads to $\alpha=1$ at $n=\pm\frac{3}{2}$. However, forgetting
the concrete setting of the 6d theory on $S^5/\mathbb{Z}_K$, one might find a physical realization
for the QFT at general $\alpha$. %the QFT at general $\alpha$ might find a physical realization.
For instance, it might be obtained by
starting from a 6d $(2,0)$ theory on $S^5\times\mathbb{R}$ in which $S^5$ is
metrically squashed with a continuous parameter $\alpha$, rather than
$\mathbb{Z}_K$.\footnote{We thank Costis Papageorgakis for discussions on this
possibility.}

\end{document}